\documentclass[showpacs,showkeys,preprintnumbers,amsmath,amssymb,times,graphicx]{revtex4}

\usepackage{graphicx} 
\usepackage{dcolumn}
\usepackage{bm}

\usepackage{color}
\usepackage{amsmath}
\usepackage{amsfonts}
\usepackage{amssymb}
\usepackage{amsbsy}
\usepackage[figuresright]{rotating}
\usepackage[font=small,labelfont=bf]{caption}
\usepackage{natbib}

\def\3{\ss}

\def\q0{\phantom{1}}

\def\vht{{\bf V}_{\rm HT}}

\newcommand{\bff}{\mathbf}

\def\be{\beta}

\def\thetabn{\Theta_{Bn}}

\def\parp2{\frac{\partial^{2}}{\partial p^{2} }}

\def\m21{$2^{\circ}\times 1^{\circ}$}

\def\ts{\thinspace}

\def\ne8{Ne\ts{$\scriptstyle {\rm VIII}$} }

\newcommand{\asr}{{Adv. \ Space. \ Res.}}

\newcommand{\ag}{{Ann.\ Geo\-phy\-si\-c\ae}}

\newcommand{\grl}{{Geo\-phys.\ Res.\ Lett.}}

\newcommand{\jgr}{{J.\ Geo\-phys.\ Res.}}

\newcommand{\ssr}{{Space Sci.\,Rev.}}

\newcommand{\pf}{{Phys. Fluids}}

\newcommand{\pop}{{Phys.\ Plasmas}}

\def\ion[#1 #2]{#1\,{\sc #2}}
\def\lamb[#1]{#1\,{\AA}}
\def\serts89{SERTS-89}
\def\fe12{Fe\,{\sc xii}}

\def\mb[#1]{\makebox[0.15cm][l]{#1}}

\begin{document}

\title{Fundamentals of  Non-relativistic Collisionless Shock Physics: \\  III. Quasi-Perpendicular Supercritical Shocks}

\author{R. A. Treumann$^\dag$ and C. H. Jaroschek$^{*}$}\email{treumann@issibern.ch}
\affiliation{$^\dag$ Department of Geophysics and Environmental Sciences, Munich University, D-80333 Munich, Germany  \\ 
Department of Physics and Astronomy, Dartmouth College, Hanover, 03755 NH, USA \\ 
$^{*}$Department Earth \& Planetary Science, University of Tokyo, Tokyo, Japan
}%

\begin{abstract}The theory and simulations of quasi-perpendicular and strictly perpendicular collisionless shocks are reviewed. The text is structured into the following sections and subsections: 1. Setting the frame, where the quasi-perpendicular shock problem is formulated, reflected particle dynamics is described in theoretical terms, foot formation and foot ion acceleration discussed, and the shock potential explained. 2. Shock structure, where the observational evidence is given, and where the simulation studies of quasi-perpendicular shocks are described as far as they deal with shock structure, i.e. the discussion of the different physical shock scales, and their investigation in one-dimensional simulations for small mass-ratios, determination of the shock-transition scale in two ways, from experiment and from simulations which identifies two scales: the foot scale and the ramp scale, the latter being determined from the overshoot magnetic field, and shock reformation is described in one- and two-dimensional simulations, showing that there are regimes when no shock reformation occurs even in supercritical quasi-perpendicular shocks, when the upstream plasma $\beta_i$ is high or when oblique whistlers stabilise the shock in two dimensions; but high Mach number shocks will always become non-stationary. 3. Ion dynamics, describing its role in shock reformation and the various ion-excited instabilities. 4. Electron dynamics, describing electron instabilities in the foot: Buneman and modified two-stream instabilities, generation of electron tails and heating, generation of phase-space holes, and discussion of various wave properties, Weibel instability, 5. The problem of stationarity, posing the theoretical reasons for shocks being non-stationary, discussing nonlinear whistler mediated variability, two-stream and modified two-stream variability, formation of ripples in two-dimensions, 6. Summary and conclusions: The possibility of shock breaking.
\end{abstract}
\pacs{}
\keywords{}
\maketitle

\section{Setting the Frame}\noindent
As long as the shocks are subcritical with Mach numbers ${\cal M}<{\cal M}_c$ the distinction between quasi-perpendicular and quasi-parallel shocks is not overwhelmingly important, at least as long as the shock normal angle is far from zero. The mechanisms of dissipation in such sub-critical (or laminar) shocks have been discussed in the previous chapter. However, when the Mach number increases and finally exceeds the critical Mach number, ${\cal M}>{\cal M}_c$, the distinction becomes very important.\index{Mach number!critical} 

We speak of quasi-perpendicular super-critical shocks\index{shocks!quasi-perpendicular} when the shock-normal angles $\thetabn<45^\circ$, and this because of good reasons. First, super-critical shocks cannot be maintained by dissipation alone. This has been clarified in Chapters 1 and 2. The inflow of matter into a supercritical shock is so fast that the time scales on which dissipation would take place are too long for dissipating the excess energy and lowering the inflow velocity below the downstream magnetosonic velocity. Hence, the condition for criticality, as we have shown in Chapter 1, is that the downstream flow velocity becomes equal to the downstream magnetosonic speed, which yielded the critical Mach number, ${\cal M}_c\lesssim 2.76$. We have also shown that ${\cal M}_c(\thetabn)$ is a function of the shock normal angle and can become quite small, even though of course ${\cal M}_c(\thetabn)\gtrsim 1$ for existence of a shock. 

In order to help maintaining a shock in the supercritical case the shock must forbid an increasing number of ions to pass across its ramp. This is done by reflecting some particles back upstream. This is not a direct dissipation process, rather it is an emergency act of the shock. It throws a fraction of the incoming ions back upstream and by this reduces both the inflow momentum and energy densities. Clearly, this reflection process slows the shock down by attributing a negative momentum to the shock itself. The shock slips back and thus in the shock frame also reduces the difference velocity to the inflow, i.e. it reduces the Mach number. In addition, however, the reflected ions form an unexpected obstacle for the inflow and in this way reduce the Mach number a second time. 

These processes are very difficult to understand, and we will go into more detail of them in this chapter. However, we must ask first, what the reason is for this rigid limit in $\thetabn$ for calling a shock a quasi-perpendicular supercritical shock. The answer is that a shock as long belongs to the class of quasi-perpendicular shocks as reflected particles cannot escape from it upstream along the upstream magnetic field. After having performed half a gyro-circle back upstream they return to the shock ramp and ultimately traverse it to become members of the downstream plasma population. 
\begin{figure}[t!]
\hspace{0.0cm}\centerline{\includegraphics[width=0.6\textwidth,clip=]{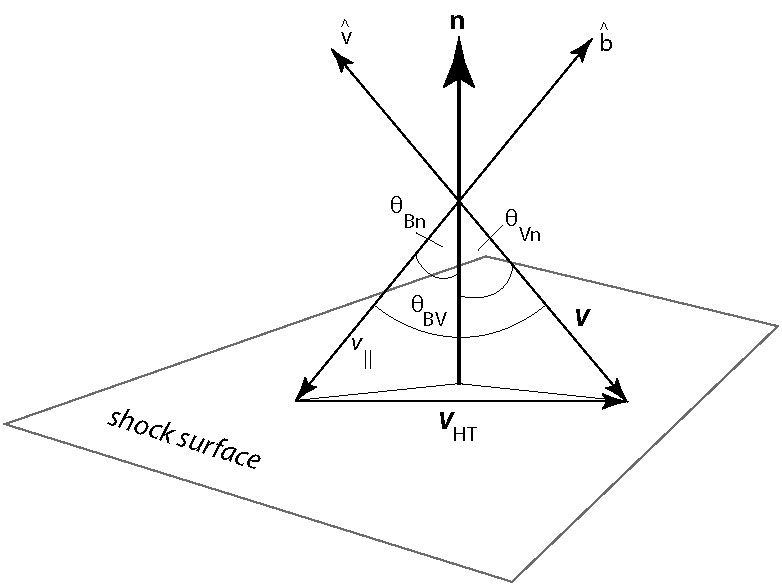} }
\caption[The shock co-ordinate system]
{\footnotesize The shock coordinate system showing the shock normal ${\bf n}$, velocity and magnetic field directions $\hat v, \hat b$, the three angles $\thetabn,\theta_{Vn},\theta_{BV}$ between $\hat b$ and ${\bf n}$, velocity ${\bf V}$ and ${\bf n}$, and velocity ${\bf V}$ and $\hat b$, respectively. The velocity $\vht$ in the shock plane is the de Hoffmann-Teller velocity.}\label{chap4-fig-vhtgeom}
\end{figure}

\subsection{Particle dynamics}\noindent
To see this we must return to the orbit a particle performs in interaction with a supercritical shock when it becomes reflected from the shock.\index{particles!shock reflection} In the simplest possible model one assumes the shock to be a plane surface, and the reflection being specular turning the component $v_n$ of the instantaneous particle velocity ${\bf v}$ normal to the shock by 180$^\circ$, i.e. simply inflecting it. In a very simplified version we have already considered this problem in Chapter 2. Here we follow the explicit calculation for these idealized conditions as given by \cite{Schwartz1983} who treated this problem in the most general way. One should, however, keep in mind that the assumption of ideal specular reflection is the extreme limit of what happens in reality. In fact, reflection must by no means be specular because of many reasons.  One of the reasons is that the shock ramp is not a rigid wall; the particles penetrate into it at least over a distance of a fraction of their gyroradius. In addition, they interact with waves and even excite waves during this interaction and during their approach of the shock. Altogether, it must be stressed again that the very mechanisms by which they become reflected are poorly known, indeed. Specular reflection is no more than a convenient assumption. Nevertheless, observations suggest that assuming specular reflection seems to be quite a good approximation to reality.\index{process!specular reflection}

Figure\,\ref{chap4-fig-vhtgeom} shows the coordinate frame\index{frame!shock normal} used at the planar shock, with shock normal ${\bf n}$, magnetic $\hat b$ and velocity $\hat v$ unit vectors, respectively. Shown are the angles $\thetabn, \theta_{Vn}, \theta_{BV}$. The velocity vector ${\bf V}_{\rm HT}$ is the de Hoffmann-Teller velocity which lies in the shock plane and is defined in such a way that in the coordinate system moving along the shock plane with velocity $\vht$ the plasma flow is along the magnetic field. ${\bf V}-\vht=-v_\|\,\hat b$. Because of the latter reason it is convenient to consider the motion of particles in the de Hoffmann-Teller frame.\index{frame!de Hoffmann-Teller} The guiding centres of the particles in this frame move all along the magnetic field. Hence,\index{velocity!de Hoffmann-Teller}
\begin{equation}
v_\|=V\frac{\cos\theta_{Vn}}{\cos\thetabn}, \qquad \vht=V\left(-\hat v +\frac{\cos\theta_{Vn}}{\cos\thetabn}\hat b\right) \equiv \frac{{\bf n}\times{\bf V}\times{\bf B}}{{\bf n\cdot B}}
\end{equation}
The de Hoffmann-Teller velocity is the same to both sides of the shock ramp, simply because the normal component $B_n$ of ${\bf B}$ and the tangential electric field. Thus, in the de Hoffmann-Teller frame there is no induction electric field ${\bf E}=-{\bf n\times V\times B}$ are both continuous. The remaining problem is thus two-dimensional (because trivially $\hat n, \hat b$ and $-v_\|\hat b$ are coplanar, which is nothing else but the coplanarity theorem holding under these undisturbed idealized conditions).

In the de Hoffmann-Teller (primed) frame the particle velocity is described by the motion along the magnetic field $\hat b$ plus the gyromotion of the particle in the plane perpendicular to $\hat b$:
\begin{equation}\label{chap4-eq-eqmotion}
{\bf v}'(t)=v_\|'\hat b +v_\perp[\hat x \cos\,(\omega_{ci}\,t+\phi_{\,0})\mp \hat y \sin\,(\omega_{ci}\,t+\phi_{\,0})] 
\end{equation}
The unit vectors $\hat x, \hat y$ are along the orthogonal coordinates in the gyration plane of the ion, the phase $\phi_{\,0}$ accounts for the initial gyro-phase of the ion, and $\pm$ accounts for the direction of the upstream magnetic field being parallel (+) or antiparallel to $\hat b$. 

In specular reflection the velocity component along ${\bf n}$ is reversed, and hence (for cold ions) the velocity becomes
\begin{equation}
{\bf v}'= -v_\|\,\hat b + 2v_\|\cos\,\thetabn\, \hat n \nonumber
\end{equation}
which (with $\phi_{\,0}=0$) yields for the components of the velocity
\begin{equation}
\frac{v_\|'}{V}=\frac{\cos\theta_{Vn}}{\cos\thetabn}(2\cos^2\thetabn - 1), \qquad \frac{v_\perp}{V} =2\sin\thetabn\cos\theta_{Vn}
\end{equation}
These expressions can be transformed back into the observer's frame\index{frame!observer's} by using $\vht$. It is, however, of grater interests to see, under which conditions a reflected particle turns around in its upstream motion towards the shock. This happens when the upstream component of the velocity $v_x=0$ of the reflected ion vanishes. For this we need to integrate Eq.\,(\ref{chap4-eq-eqmotion}) which for $\phi_{\,0}={\,0}$ yields 
\begin{equation}
{\bff x}'(t)=v_\|'t\,\hat b +\frac{v_\perp}{\omega_{ci}}\{[\sin\,\omega_{ci}t]\hat x \pm[\cos\,\omega_{ci}t -1]\hat y\}
\end{equation}
Scalar multiplication of this expression with ${\bf n}$ yields the ion displacement normal to the shock in upstream direction. The resulting expression 
\begin{equation}\label{chap4-eq-reflectdist}
{\bf x}_n'(t^*)=v_\|'t^*\cos\thetabn +\frac{v_\perp}{\omega_{ci}}\sin\thetabn \sin\omega_{ci}t^* =0
\end{equation}
vanishes at time $t^*$ when the ion reencounters the shock with normal velocity $v_n(t^*)=v_\|'\cos\thetabn +v_\perp \sin\thetabn\cos\omega_{ci}t^*$. The maximum displacement away from the shock in normal direction is obtained when setting this velocity to zero, obtaining for the maximum displacement time
\begin{equation}\label{chap4-eq-cosmin1}
\omega_{ci}t_m=\cos^{-1}\left(\frac{1-2\cos^2\thetabn}{2\sin^2\thetabn}\right)
\end{equation}
This expression must be inserted in ${\bf x}_n$ yielding for the distance a reflected ion with gyro-radius $r_{ci}=V/\omega_{ci}$ can achieve in upstream direction
\begin{equation}\label{chap4-eq-upstdist}
\Delta x_n=r_{ci}\cos\theta_{Vn}[\omega_{ci}t_m(2\cos^2\thetabn -1)+2\sin^2\thetabn \sin\omega_{ci}t_m]
\end{equation}
For a perpendicular shock $\thetabn=90^\circ$ this distance is $\Delta x_n\simeq 0.7 r_{ci}\cos\theta_{Vn}$ which is less than an ion gyro radius. The distance depends strongly on the shock normal angle. Note that the argument of $\cos^{-1}$ in Eq.\,(\ref{chap4-eq-cosmin1}) exceeds unity for $\thetabn\leq 45^\circ$. Hence there are no solution for such angles. This is related to the fact that Equation (\ref{chap4-eq-reflectdist}) has solutions only for shock normal angles $\thetabn>45^\circ$. Reflected ions thus return to the shock only when the magnetic field makes an angle with the shock normal larger than this value. For less inclined shock normal angles the reflected ions escape along the magnetic field upstream of the shock and do not return. This sharp distinction between shock normal angles $\thetabn<45^\circ$ and $\thetabn>45^\circ$ thus provides the clear natural discrimination between quasi-perpendicular and quasi-parallel shocks we were looking for.\index{shocks!distinction of Q$_\perp$/Q$_\parallel$ }
\begin{figure}[t!]
\hspace{0.0cm}\centerline{\includegraphics[width=0.75\textwidth,clip=]{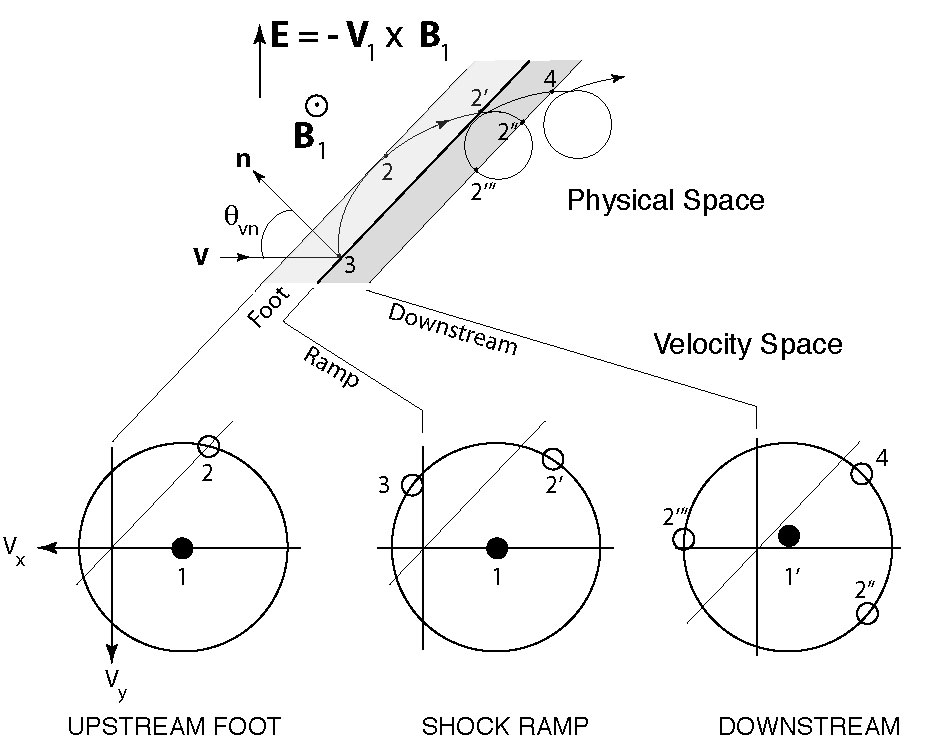} }
\caption[Reflected ion orbits in the foot]
{\footnotesize {\it Top}: Reflected ion orbits in the foot of a quasi-perpendicular shock in real space. The ion impacts under an instantaneous angle $\theta_{vn}$, is reflected from the infinitely thin shock, performs a further partial gyration in the upstream field ${\bf B}_1$ where it is exposed to the upstream convection electric field ${\bf E}=-{\bf V}_1\times {\bf B}_1$ in which it is accelerated as is seen from the non-circular section of its orbit in the shock foot. It hits the shock ramp a second time now at energy high enough to overcome the shock potential, passing the ramp and arriving in the compressed downstream magnetic field behind the shock where it performs gyrations of reduced gyro-radius. {\it Bottom}: The ion distribution function mapped into velocity space $v_x,v_y$ for the indicated regions in real space, upstream in the foot, at the ramp, and downstream of the shock ramp. Upstream the distribution consists of the incoming dense plasma flow (population 1, dark circle at $v_y=0$) and the reflected distribution 2 at large negative $v_y$. At the ramp in addition to the incoming flow 1 and the accelerated distribution 2' there is the newly reflected distribution 3. Behind the ramp in the downstream region the inflow is decelerated 1' and slightly deflected toward non-zero $v_y$, and the energized passing ions exhibit gyration motions in different instantaneous phases, two of them (2", 4) directed downstream, one of them (2'") directed upstream. \citep[redrawn after][]{Sckopke1983}.}\label{chap4-fig-vhtgeom}
\end{figure}

\subsection{Foot formation and acceleration}\noindent\index{shocks!foot}
Shock reflected ions in a quasi-perpendicular shock cannot escape far upstream. Their penetration into the upstream plasma is severely restricted by formula (\ref{chap4-eq-upstdist}). Within this distance the ions perform a gyrational orbit before returning to the shock. 

Since the reflected ions are about at rest with respect to the inflowing plasma they are sensitive to the inductive convection electric field ${\bf E}=-{\bf V}_1\times{\bf B}_1$ behaving very similar to pick-up ions\index{particles!pick-up ions} and becoming accelerated in the direction of this field to achieve a higher energy \citep{Schwartz1983}. When returning to the shock their maximum (minimum) achievable energy is 
\begin{equation}
{\cal E}_{\rm max}=\frac{m_i}{2}\left[(v_\|'+V_{\rm HT\|})^2 +(V_{\rm HT\perp}\pm v_\perp)^2\right]
\end{equation}
This energy is larger than their initial energy with that they have initially met the shock ramp and, under favorable conditions, they now might overcome the shock ramp potential and escape downstream. Otherwise, when becoming reflected again, they gain energy in a second round until having picked up sufficient energy for passing the shock ramp.
\begin{figure}[t!]
\hspace{0.0cm}\centerline{\includegraphics[width=0.9\textwidth,clip=]{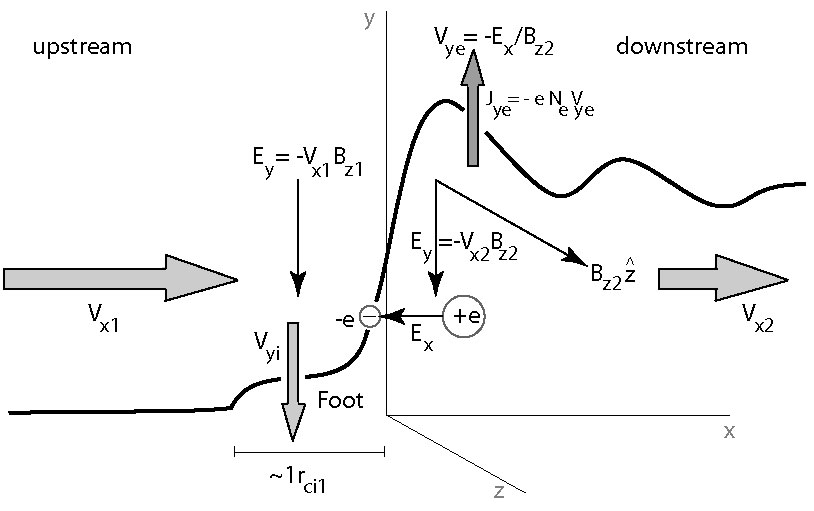} }
\caption[Geometry of an ideal quasi-perpendicular shock]
{\footnotesize Geometry of an ideally perpendicular supercritical shock showing the field structure and sources of free energy.  The shock is a compressive structure. The profile of the shock thus stands for the compressed profile of the magnetic field strength $|{\bf B}|$, the density $N$, temperature $T$, and pressure $NT$ of the various components of the plasma. The inflow of velocity $V_1$ and outflow of velocity $V_2$ is in $x$ direction, and the magnetic field is in $z$ direction. Charge separation over an ion gyroradius $r_{ci}$ in the shock ramp magnetic field generates a charge separation electric field $E_x$ along the shock normal which reflects the low-energy ions back upstream. These ions see the convection electric field $E_y$ of the inflow, which is along the shock front, and become accelerated.The magnetic field of the current carried by the accelerated back-streaming ions causes the magnetic foot in front of the shock ramp. The shock electrons are accelerated antiparallel to $E_x$ perpendicular to the magnetic field. The shock electrons also perform an electric field drift in $y$-direction in the crossed $E_x$ and compressed $B_{z2}$ fields which leads to an electron current $j_y$ along the shock.  These different currents are sources of free energy which drives various instabilities in different regions of the perpendicular shock.}\label{chap2-fig-perpsh}
\end{figure}

In addition to this energization of reflected ions which in the first place have not made it across the shock, the reflected ions when gyrating and being accelerated in the convection electric field constitute a current layer just in front of the shock ramp of current density $j_y\sim eN_{i,{\rm refl}}v_{y,{\rm refl}}$ which gives rise to a foot magnetic field of magnitude $B_{z,{\rm foot}}\sim \mu_0j_y\Delta x_n$. It is clear that this foot ion current, which is essentially a drift current in which only the reflected newly energized ion component participates, constitutes a source of free energy as it violates the energetic minimum state of the inflowing plasma in its frame. Being the source of free energy it can serve as a source for excitation of waves via which it will contribute to the lack of dissipation. However, in a quasi-perpendicular shock there are other sources of free energy as well which are not restricted to the foot region. 

Figure\,\ref{chap2-fig-perpsh} shows  a sketch of some of the different free-energy sources\index{instability!free energy source} and processes across the quasi-perpendicular shock. In addition to the shock-foot current and the presence of the fast cross-magnetic field ion beam there, the shock ramp is of finite thickness. It contains a charge separation electric field $E_x$ which in the supercritical shock is strong enough to reflect the lower energy ions. In addition it accelerates electrons downstream thereby deforming the electron distribution function. 

The presence of this field, which has a substantial component perpendicular to the magnetic field, implies that the magnetized electrons with their gyro radii being smaller than the shock-ramp width experience an electric drift $V_{ye}=-E_x/B_{z2}$ along the shock in the ramp which can be quite substantial giving rise to an electron drift current $j_{ye}= -eN_{e,{\rm ramp}}V_{ye}=eN_{e,{\rm ramp}}E_x/B_{z2}$ in $y$-direction. This current has again its own contribution to the magnetic field, which at maximum is roughly given by $B_z\sim \mu_0 j_{ye}\Delta x_n$. Here we use the width of the shock ramp. The electron current region might be narrower, of the order of the electron skin depth $c/\omega_{pe}$. However, as long as we do not know the number of magnetized electrons which are involved into this current nor the width of the electric field region (which must be less than an ion gyro-radius because of ambipolar effects) the above estimate is good enough. 

The magnetic field of the electron drift current causes an overshoot in the magnetic field in the shock ramp on the downstream side and a depletion of the field on the upstream side. When this current becomes strong it contributes to current-driven cross-field instabilities like the modified two-stream instability. 

Finally, the mutual interaction of the different particle populations present in the shock at its ramp and behind provide other sources of free energy. A wealth of instabilities and waves is thus expected to be generated inside the shock. To these micro-instabilities add the longer wavelength instabilities which are caused by the plasma and field gradients in this region. These are usually believed to be less important as the crossing time of the shock is shorter than their growth time. However, some of them propagate along the shock and have therefore substantial time to grow and modify the shock profile. In the following we will turn to the discussion of numerical investigations of some of these processes reviewing their current state and provide comparison with observations.

\subsection{Shock potential drop}\noindent  \index{shocks!electric potential drop}One of the important shock parameters is the electric potential drop across the shock ramp -- or if it exists also across the shock foot. This potential drop is not necessarily a constant but changes with location along the shock normal. We have already noted that it is due to the different dynamical responses of the inflowing ions and electrons over the scale of the foot an ramp regions. Its theoretical determination is difficult, however when going to the de Hoffmann-Teller frame the bulk motion of the particles is only along the magnetic field, and in the stationary electron equation of motion the ${\bf V}_e\times{\bf B}$-term drops out and to first approximation the cross shock potential is simply given by the pressure gradient (when neglecting any contributions from wave fields). The expression is then simply
\begin{equation}
\Delta \Phi(x)=\int_0^x\frac{1}{eN_e(n)}[\nabla\cdot{\textsf{P}_e(n)}]\cdot{\rm d}{\bf n}
\end{equation}
Integration is over $n$ along the shock normal ${\bf n}$. For a gyrotropic electron pressure, valid for length scales longer than an electron gyroradius, ${\textsf P}_e=P_{e\perp}{\textsf I}+(P_{e\|}-P_{e\perp}){\bf BB}/BB$ one obtains  \citep{Goodrich1984}, taking into account that ${\bf E\cdot B}$ is invariant,
\begin{equation}
\frac{{\rm d} }{{\rm d}n}\Phi(n)=-\frac{E_\|}{\cos\thetabn}=\frac{1}{eN_e}\left[\frac{\rm d}{{\rm d}n}P_{e\|}-(P_{e\|}-P_{e\perp})\frac{\rm d}{{\rm d}n}(\ln B)\right]
\end{equation}
which, when used in the above expression, yields
\begin{equation}
e\Delta\Phi(x)=\int_0^x{\rm d}n\left\{\frac{{\rm d}T_{e\|}}{{\rm d}n}+T_{e\|}\frac{{\rm d}}{{\rm d}n}\ln\left[\frac{N(n)}{N_1}\frac{B_1}{B(n)}\right]+T_{e\perp}\frac{{\rm d}}{{\rm d}n}\ln\left[\frac{B(n)}{B_1}\right]\right\}
\end{equation}
This expression can be written approximately in terms of the gradient in the electron magnetic moment $\mu_e=T_{e\perp}/B$ as follows:
\begin{equation}
e\Delta\Phi(x)\simeq \Delta(T_{e\|}+T_{e\perp})-\int_0^x{\rm d}n\frac{{\rm d}\mu_e(n)}{{\rm d}n}B(n)
\end{equation}
with $T_e$ in energy units. When the electron magnetic moment is conserved, the last term disappears, yielding a simple relation for the potential drop $e\Delta\Phi(x)\simeq \Delta(T_{e\|}+T_{e\perp})$ as the sum of the changes in electron temperature. The perpendicular temperature change can be expressed as $\Delta T_{e\perp}=T_{e\perp,1}\Delta B/B_1$ which is in terms of the compression of the magnetic field. 

The parallel change in temperature is more difficult to express. One could do it in terms of the temperature anisotropy $A=T_{e\|}/T_{e\perp}$ as has been done by \cite{Kuncic2002}, and then vary the anisotropy. But this is a question of the particular model. It is more important to note that this adiabatic estimate of the potential drop does not account for any dynamical process which generates waves and substructures in the shock. It thus gives only a hint on the order of magnitude of the potential drop across the foot-ramp region in quasi-perpendicular shocks. 

\begin{figure}[t!]
\hspace{0.0cm}\centerline{\includegraphics[width=0.6\textwidth,clip=]{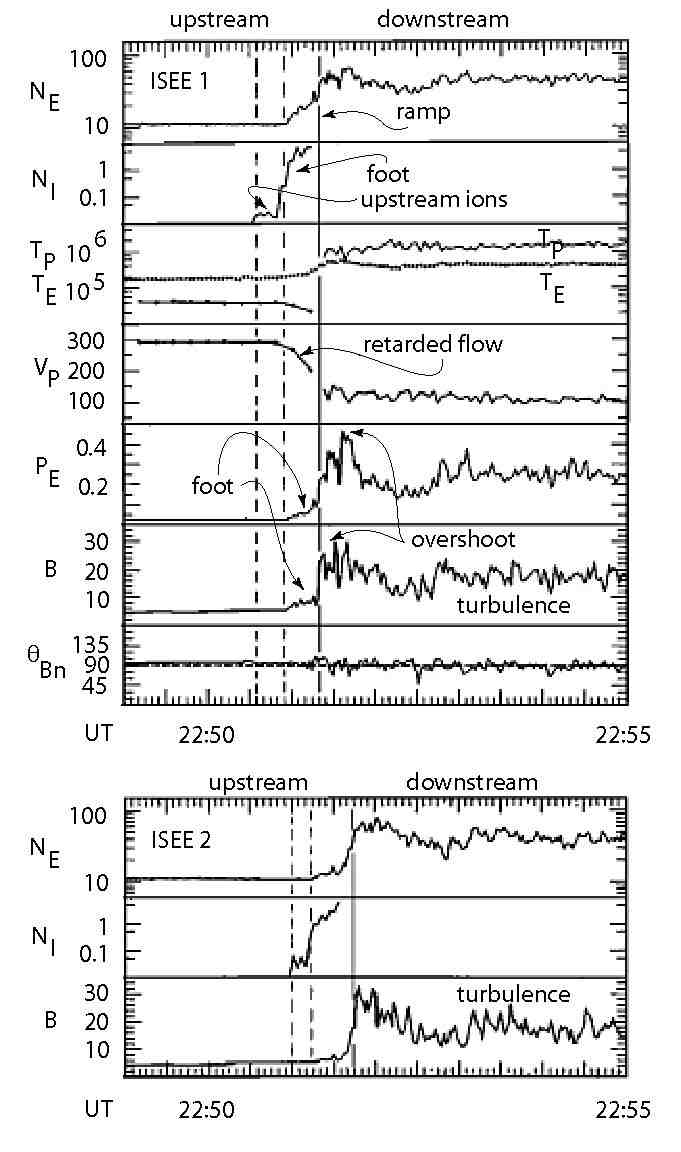} }
\caption[Sckopke et al.'s 1983 detection of the foot and shock ramp structure]
{\footnotesize Time profiles of plasma and magnetic field parameters across a real quasi-perpendicular shock that had been crossed by the ISEE 1 and 2 spacecraft on November 7, 1977 in near-Earth space \citep[after][]{Sckopke1983}. The shock in question is the Earth's bow shock wave which will be described in detail Chapter 8. Here the measurement serve as typical for a quasi-perpendicular shock. $N_E$ is the elecron density, $N_I$ the reflected ion density, both in cm$^{-3}$, $T_p,T_E$ are proton and electron in K. $V_P$ is the proton (plasma) bulk velocity in km\,s${-1}$, $P_E$ electron pressure in 10$^{-9}$ N\,m$^{-2}$, $B$ the magnitude of the magnetic field in nT,  and $\thetabn$. The vertical lines mark the first appearance of reflected ion, the outer edge of the foot in the mgnetic profile, and the ramp in the field magnitude, respectively. The abscissa is the Universal Time UT referring to the measurements. The upper block are observations from ISEE 1, the lower block observations from ISEE 2.}\label{chap4-fig-sckopkesh}
\end{figure}

\section{Shock Structure}\label{structure4-2}\noindent
Figure\,\ref{chap4-fig-sckopkesh} shows observations from one of the first unambiguous satellite crossings of a quasi-perpendicular supercritical (magnetosonic Mach number ${\cal M}_{ms}\sim 4.2$) shock in near Earth space. The crossing occurred at the Earth's bow shock, the best investigated shock in the entire cosmos! A complete discussion of its properties will be given in Chapter 7. Here it should mainly serve for visualization of the properties of a real collisionless shock how it appears in the data. The shock crossing shown in the figure is indeed a textbook example.

\subsection{Observational evidence}\noindent
The crossing occurred on an inbound path of the two spacecraft ISEE\,1 (upper block of the figure) and ISEE\,2 (lower block of the figure) from upstream to downstream in short sequence only minutes apart. In spite of some differences occurring on the short time scale the two shock crossings are about identical, identifying the main shock transition as a spatial and not as a temporal structure. Temporal variations are nevertheless visible on the scale of a fraction of a minute. 

From top to bottom the figure shows the electron density ($N_E$), energetic ion density ($N_I$), proton and electron temperatures ($T_P,T_E$), bulk flow velocity ($V_P$), electron pressure ($P_E$), magnetic field ($B$), and $\thetabn$. The latter is close to 90$^\circ$ prior to shock crossing (in the average $\thetabn\sim 85^\circ$), and fluctuates afterwards around 90$^\circ$ identifying the shock as quasi-perpendicular. Accordingly, the shock develops a foot in front of the shock ramp as can be seen from the slightly enhanced magnetic field after 22:51\,UT in ISEE\,1 and similar in ISEE\,2, and most interestingly also in the electron pressure. At the same time the bulk flow velocity starts decreasing already, as the result of interaction and retardation in the shock foot region. The foot is also visible in the electron density which increases throughout the foot region, indicating the presence of electrons which, as is suggested by the increase in pressure, must have been heated or accelerated. \index{shocks!shock profile}

The best indication of the presence of the foot is, however, the measurement of energetic ions (second panel from top). These ions are observed first some distance away from the shock but increase drastically in intensity when entering the foot. These are the shock-reflected ions which have been accelerated in the convection electric field in front of the shock ramp. Their occurrence before entrance into the foot is understood when realizing that the shock is not perfectly perpendicular. Rather it is quasi-perpendicular such that part of the reflected ions having sufficiently large parallel upstream velocities can escape along the magnetic field a distance larger than the average upstream extension of the foot. For nearly perpendicular shocks, this percentage is small. 

The shock ramp\index{shocks!ramp} in Figure\,\ref{chap4-fig-sckopkesh} is a steep wall in $B$ and $P_E$, respectively. The electron temperature $T_E$ increases only moderately across the shock while the ion temperature $T_P$ jumps up by more than one magnitude, exceeding $T_E$ downstream behind the shock. This behaviour is due to the accelerated returning foot-ions which pass the shock. $P_E$, $B$, and $N_E$ exhibit overshoots behind the shock ramp proper. Farther away from the shock they merge into the highly fluctuating state of lesser density, pressure, and magnetic field that can be described as some kind of turbulence. Clearly, this region is strongly affected by the presence of the shock which forms one of its boundaries, the other boundary being the obstacle which is the main responsible for the formation of the shock.

The evidence provided by the described measurements suggests that the quasi-perpen\-di\-cu\-lar shock is a quasi-stationary entity. This should, however, not been taken as apodictive. Stationarity depends on the spatial scales as well as the time scales. A shock is a very inhomogeneous subject containing all kinds of spatial scales. Being stationary on one scale does not imply that it is stationary on another scale. For a shock like the Earth's bow shock considered over times of days, weeks or years the shock is of course a stationary subject. However on shorter time scales of the order of flow transition times this may not be the case. A subcritical shock of the kind discussed in Chapter 3 may well be stationary on long and short time scales. However, for a supercritical shock the conditions for forming a stationary state are quite subtle. From a single spacecraft passage like that described above it cannot be concluded to what extent, i.e. on which time scale and on which spatial scale and under which external conditions (Mach number, angle, shock potential, plasma-$\beta$, \dots) the observed shock can be considered to be stationary \citep[a discussion of the various scales has been given, e.g., by][]{Galeev1988}. Comparison between the two ISEE spacecraft already shows that the small-scale details as have been detected by both spacecraft are very different. This suggests that -- in this case -- on time scales less than a minute variations in the shock structure must be expected.\index{shocks!scales} \index{Galeev, Alexander A.}

Generally spoken, one must be prepared to consider the shock as a non-stationary phenomenon \citep[this has been realized first by][]{Morse1972} which depends on many competing processes and, most important, even as a whole is not in thermal equilibrium. It will thus be very sensitive to small changes in the external parameters and will permanently try to escape the non-equilibrium state and to approach equilibrium. Since its non-equilibrium is maintained by the conditions in the flow, it is these conditions which determine the time scales over which a shock evolves, re-evolves and changes its state. In the following we will refrain from analytical theory and, forced by the complexity of the problem, mainly discuss numerical experiments on shocks. However, at a later stage we will return to the problem of non-stationarity again. Real supercritical shocks, whether quasi-perpendicular or quasi-parallel are in a permanently evolving state and thus are intrinsically nonstationary. 

\subsection{Simulation studies of quasi-perpendicular shock structure}\noindent
Perpendicular or quasi-perpendicular collisionless shocks are relatively easy to treat in numerical simulations. Simulations of this kind have been mostly one-dimensional. Only more recently they have begun to become treated in two dimensions. 

Already from the first one-dimensional numerical experiments on collisionless shocks \citep[for an early review cf., e.g.,][]{Biskamp1973} if became clear that such shocks have a very particular structure. This structure, which we have describe in simplified version in Figure\,\ref{chap2-fig-perpsh} and which could to some extent also be inferred from the observations of Figure\,\ref{chap4-fig-sckopkesh}, becomes ever more pronounced the more refined the resolution becomes and the better the shorter scales can be resolved. \index{Biskamp, Dieter}

As already mentioned, collisionless shocks are in thermodynamic non-equilibrium and therefore can only evolve if a free energy source exists and if the processes are violent enough to build up and maiintain a shock. Usually in a freely evolving system the free energy causes fluctuations which serve dissipating and redistributing the free energy towards thermodynamic and thermal equilibria. (Thermal equilibria are characterised by equal temperatures among the different components, e.g. $T_e=T_i$ which is clearly not given in the vicinity of a shock as seen from Figure\,\ref{chap4-fig-sckopkesh}. Thermodynamic equilibria are characterised by Gaussian distributions for all components of the plasma. To check this requires information about the phase space distribution of particles. Shocks contain many differing particle distributions, heated, top-flat, beam distributions, long energetic tails, and truncated as well as gyrating distributions which we will encounter later. Consequently, they are far from thermodynamic equilibrium.) 

For a shock to evolve the amount of free energy needed to dissipate is so large that fluctuations are unable to exercise their duty. This happens at large Mach numbers. The shock itself takes over the duty of providing dissipation. It does it in providing all kinds of scales such short that a multitude of dissipative processes can set on. 

\subsubsection{Scales}\noindent  For a quasi-perpendicular shock propagating and evolving in a high-$\beta$ plasma there is a hierarchy of such scales available (we recall that $\beta =2\mu_0nT/B^2$ refers to the thermal energy of the flow. The kinetic $\beta_{kin\perp}= 2\mu_0Nm_iV_n^2/2B^2\equiv{\cal M}^2>1$  implies that the kinetic energy  in the flow exceeds the magnetic energy. Hence the flow dominates the magnetic field, which is transported by the flow. In plasmas with $\beta_{kin\perp}={\cal M}^2< 1$ the magnetic field dominates the dynamics, and shock waves perpendicular to the magnetic field cannot evolve. Parallel shocks are basically electrostatic in the $\beta_{kin\perp}\ll 1$-case and can evolve when the flow is sufficiently fast along the field, as is observed in the auroral magnetospheres of the magnetized planets in the heliosphere. On the other hand, for large Mach numbers and $\beta\gtrsim 1$ conditions shocks do exist, as the example of the solar wind shows). 

The different scales can be oganized with respect to the different regions of the shock. 
\begin{description}
\item[{\sf 1.}] \noindent The macroscopic scale of the foot region, which determines the width of the foot, is the ion gyroradius based on the inflow velocity $r_{ci,1}=V_1/\omega_{ci,1}$. With the slight modification of replacing the upstream magnetic field with the (inhomogeneous) ramp magnetic field $B_r(x)$ this also becomes approximately the scale of the macroscopic electric potential drop in the ramp, $\Delta_{\phi,r}\sim\,r_{ci,r}\sim V_1/\omega_{ci,r}$. \end{description}

Other scales are
\begin{description}
\item[{\sf 2.}] \noindent the ion inertial length $c/\omega_{pi}$, which is also a function of space inside the ramp because of the steep density increase $N(x)$. It determines the dispersive properties of the fast magnetosonic wave which is locally responsible for steepening and shock ramp formation;
\item[{\sf 3.}] \noindent the thermal ion gyrodradius $r_{ci}=v_i/\omega_{ci}$. It determines the transition from unmagnetized to magnetized ions and from nonadiabatic to adiabatic heating of the ions;
\item[{\sf 4.}] \noindent the density gradient scale $L_P=(\nabla_x\ln P)^{-1}$. It determines the importance of drift waves along the shock which, when excited, structure the shock in the third dimension perpendicular to the shock normal and the magnetic field;
\item[{\sf 5.}] \noindent  the electron inertial length $c/\omega_{pe}$. It is the scale length of whistlers which are excited in front of the shock and are generally believed to play an essential role in shock dynamics; 
\item[{\sf 6.}] \noindent the thermal electron gyroradius $r_{ce}=v_e/\omega_{ce}$. It determines whether electrons behave magnetized or nonmagnetized. In the shock they are usually magnetized under all conditions of interest. However, when nonadiabatic heating becomes important for electrons it takes place on scales comparable to $r_{ce}$; 
\item[{\sf 7.}] \noindent the Debye length $\lambda_D$. It determines the dispersive properties of ion acoustic waves which are responsible for anomalous resistivity and for smaller scale density substructures  in the shock like the phase space holes mentioned earlier which evolve on scales of several Debye lengths. It also determines the scales of the Buneman two-stream (BTS) and modified two-stream (MTS) instabilities which are the two most important instabilities in the shock foot.\index{Buneman, Oscar}
\end{description}

The importance of some of these scales has been discussed by \citep{Kennel1985} assuming that some mostly anomalous resistance has been generated in the plasma. In this case the speed of the fast magnetosonic wave,\index{velocity!magnetosonic} which is responsible for fast shock formation, is written as\index{Kennel, Charles F.}
\begin{equation}
c_{ms}^2=c_{ia}^2+\frac{V_A^2}{1+k^2R^2}, \qquad R=
\left\{
\begin{array}{lr}
 R_\eta= (\eta/\mu_0)(k/\omega),  & \qquad \eta \neq 0 \\
 R_e=c/\omega_{pe},   & \qquad \eta\to 0 \\
\end{array}
\right.
\end{equation}
taking explicitly care of the dispersion of the wave which leads to wave steepening. The macroscopic scale of shock formation enters here through the definition of $R$ which in the collisionless case becomes the electron skin depth. Starting from infinity far away from the shock one seeks for growing solutions of the linear magnetic disturbance $b_z\sim\exp \lambda x$ in the stationary point equation
\begin{displaymath}
R^2_eb_z^{\prime\prime}+R_\eta b_z^{\prime}=Db_z, \qquad\qquad D\equiv\frac{1-{\cal M}^{-2}}{1-c_{ia}^2/V^2}
\end{displaymath}
where the prime $^\prime\equiv\partial/\partial x$ indicates derivation with respect to $x$. With $b_z\to 0$ for $x\to -\infty$ this yields for the growth length 
\begin{equation}
\lambda_>=-\frac{R_\eta}{2R_e^2}+\left[\frac{D}{R_e^2}+\left(\frac{R_\eta}{2R_e^2}\right)^2\right] ^\frac{1}{2}\to\,\,\ \frac{D^\frac{1}{2}}{R_e}\qquad {\rm for}\qquad R_\eta\ll R_e
\end{equation}
which identifies the approximate shock transition scale as proportional to the electron skin depth, $\Delta\simeq c/\omega_{pe}D^\frac{1}{2}$, just what one intuitively would believe to happen for freely moving electrons and ions. Since the upstream sound speed $c_{ia}\ll V$ is small compared with the fast flow $V$, we have $D\approx 1-{\cal M}^{-2}$, and the shock ramp width becomes slightly larger than the electron skin depth
\begin{equation}
\Delta \simeq c{\cal M}/\omega_{pe}({\cal M}^2-1)^\frac{1}{2}
\end{equation}
For large Mach numbers this width approaches the electron inertial length. However, we have already seen that at large Mach numbers the competition between dispersion and dissipation does not hold anymore in this simple way. 

With increasing wave number $k$ the fast magnetosonic mode merges into the whistler \index{waves!whistler}branch\index{dispersion relation!whistler} with its convex dispersion curve. This implies that dispersive whistler waves will outrun the shock becoming precursors of the shock, a problem we have discussed in Chapter 2. Whistlers propagate only outside their resonance cone. The limiting angle between ${\bf k}$ and the magnetic field ${\bf B}$ for which the whistler outruns the shock is given by $\theta_{\rm wh,lim}\lesssim \cos^{-1}[{\cal M}_A(m_e/m_i)^\frac{1}{2}]$, artificially limiting the Alfv\'enic Mach number ${\cal M}_A=V/V_A< 43$.

In one-dimensional simulations with all quantities changing only along the shock normal ${\bf n}$ and  the ${\bf k}$-vectors of waves along ${\bf n}$ as well, one choses angles between (${\bf k, n}$) and ${\bf  B}$ larger than this in order to have clean effects which are not polluted by those whistlers. However, the maximum phase speed of whistlers is the Alfv\'en speed (see Figure\,\ref{chap2-fig-ms-whistler}). Hence, as long as the upstream velocity is less than the Alfv\'en speed, a standing whistler precursor will be attached to the shock in front. When the upstream velocity exceeds the Alfv\'en velocity, phase standing whistlers become impossible. This happens at the critical whistler Mach number given in Chapter 2. The shock structure becomes more complicated then by forming shock substructures \citep{Galeev1988} on scales of $c/\omega_{ce}$, and the shock might become\index{shocks!non-stationary} non-stationary \citep{Krasnoselskikh2002}.

\subsubsection{One-dimensional structure}\noindent
One-dimensional observations as those presented in Figure\,\ref{chap4-fig-sckopkesh} confirm the theoretical prediction of the gross structure of a quasi-perpendicular shock. They can, however, when taken by themselves, not resolve the spatial structure of the shock on smaller scales, nor do they allow to infer about the evolution of the shock. To achieve a clearer picture of both, the structure and the evolution, the observations must be supported by numerical simulations. 

Such simulations have been performed in the past in various forms either as hybrid simulations or as full particle simulations. In hybrid simulations the electrons form a neutralising, massless background with no dynamics, i.e. the electrons react instantaneously while maintaining and merely adjusting their equilibrium Boltzmannian distribution to the locally changing conditions. Such simulations overestimate the role of the ions and neglect the dynamical contribution of the electrons. They nevertheless give a hint on the evolution and gross structure of a shock on the ion scales. Hybrid simulations have the natural advantage that they can be extended over relatively long times $\omega_{ci}t\gg 1$.

On the other hand full particle simulations are usually done for unrealistically small mass ratios $m_i/m_e\ll 1836$ much less than the real mass ratio. The electrons in these simulations are therefore heavy even under non-relativistic conditions. Their reaction is therefore unnaturally slow, the electron plasma and cyclotron frequencies are low, and the electron gyroradius, inertial length, and Debye length are unnaturally large. Under these conditions electrons readily become unmagnetized, nonadiabatic electron heating is prominent, and dispersive effects on ion-acoustic waves are overestimated. 

Moreover, because of the large electron mass the electron thermal speed\index{velocity!electron thermal} is reduced, and the Buneman two-stream instability sets on earlier and grows faster than under realistic conditions. This again should affect electron heating and structuring of the shock. On the other hand, the reduced electron gyroradius also reduces the shock potential, because the differences in ion and electron penetration-depths into the shock are smaller than in reality. This reduces the reflection capability of the shock, reduces the direct electric field heating of the impacting electrons, reduces the electron drift current in the shock ramp and shock transition and thus underestimates the dynamic processes in the shock, its structure, time dependence, formation and reformation and the strength of the foot effect and density of the foot population. 

It is very difficult to separate all these effects, and comparison of different simulations is needed.  
\begin{figure}[t!]
\hspace{0.0cm}\centerline{\includegraphics[width=1.0\textwidth,clip=]{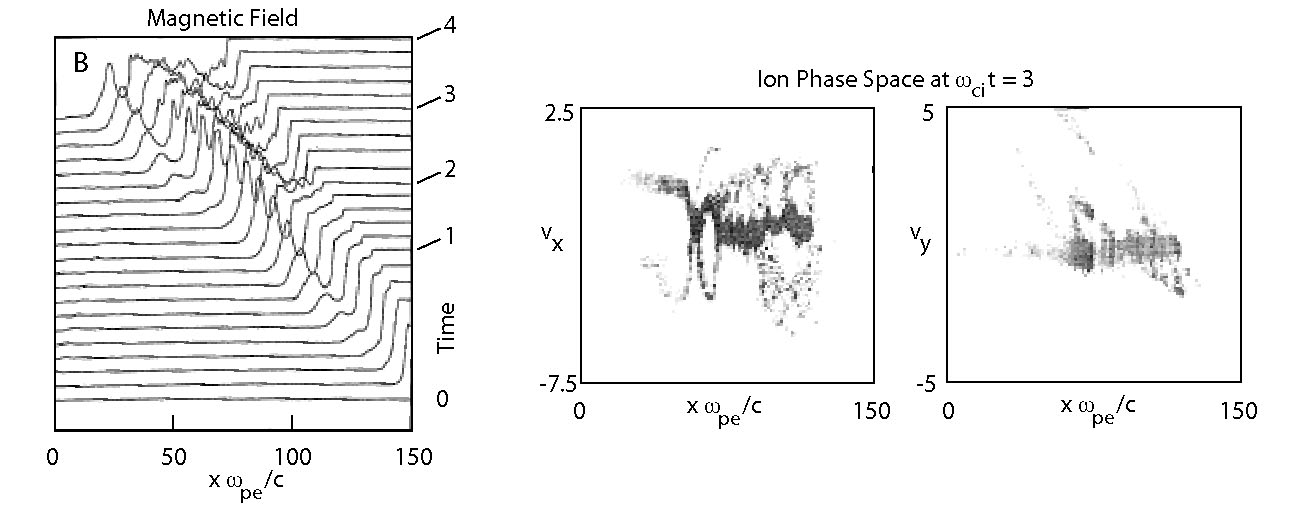} }
\caption[Biskamp \& Welter's first PIC shock simulation]
{\footnotesize One-dimensional full particle-in-cell (PIC) simulations \citep[after][]{Biskamp1972} of shock formation assuming a mass ratio of $m_i/m_e=128$. {\it Left}: Time-evolution of the magnetic field in stack-plot representation. Time is measured in  units of $\omega_{ci}^{-1}$, space in units of (heavy) electron inertial lengths $c/\omega_{pe}$. The simulations are for a supercritical shock with ${\cal M}=2.3$. Note the evolution of the magnetic field and the formation of a ramp, a foot and an overshoot.{\it Right}: Ion phase space plots $(v_x,x)$ and $(v_y,x)$ at time $\omega_{ci}t=3$. Velocities are measured in units of the Alfv\'en velocity $V_A=B/\sqrt{\mu_0Nm_i}$.}\label{chap4-fig-bis1}
\end{figure}
\subsubsection{Low-mass ratio simulations}\noindent  Figure\,\ref{chap4-fig-bis1} shows an early one-dimensional low-mass-ratio perpendicular shock simulation with $m_i/m_e=128$. Simulation times are short, not more than four ion-gyration times when energy conservation starts breaking down. Moreover, only a  very small number of macro-particles (see Chapter 1) per simulation grid cell could be carried along in these simulation. Thus the noise in the simulations is large, not allowing for long simulation times, readily introducing diverging fake modes and fake dissipation/heating. \index{simulations!low-mass ratio}

Nevertheless, the left-hand side of the figure shows the evolution of the magnetic field from the homogeneous state into a shock ramp and further the destruction and, what is known by now from much longer and better resolved simulation studies, the reformation of the shock profile. It should be noted that the shock in this case forms by reflection of the fast initial supercritical flow with Mach number ${\cal M}=2.3$ (which is above critical for the conditions of the simulation), entering the one-dimensional simulation `box' from the left, from a `magnetic piston' located at the right end of the box. This reflection causes a back-streaming ion-beam that interacts with the inflowing ions and drives an electromagnetic ion-ion instability which grows to large amplitude. The system is not current-free. In the interaction region of the two ion components the magnetic field forms a shock ramp. But after a short time of a fraction of an ion gyro-period a new ramp starts growing in the foot of the ramp, which itself evolves into a new ramp while the old ramp becomes eroded. This new ramp has not sufficient time to evolve to a full ramp as another new ramp starts growing in its foot. This causes the shock ramp to jump forward in space in steps from one ramp to the next, leaving behind a downstream compressed but fluctuating magnetic field region. The jump length is about the width of the foot region. It will become clear later why this is so.

Hybrid simulations \citep{Leroy1981,Leroy1982,Leroy1984} with fluid electrons and an artificially introduced anomalous resistivity show similar behaviour even though a number of differences have been found which are related to shock reformation. In particular shock reformation is slow or absent in hybrid simulations if not care is taken on the reaction of the electrons. The responsible instability in the foot region cannot evolve fast enough even though the hybrid simulation which take care of the ion dynamics also find reflection of ions and the evolution of a foot in front of the ramp. These differences must be attributed to the above mentioned lesser reliability of hybrid simulations than full particle codes. 

Extended low-mass ratio full particle simulations in one space dimensions over a wide range of shock-normal angles $\thetabn<45^\circ$ have been performed by \cite{Lembege1987a,Lembege1987b} with the purpose to study plasma heating. These simulations used mass ratios of $m_i/m_e\,=\,100$ and a magnetic-piston generated shock. The simulations were completely collisionless, relatively small Mach number but nevertheless supercritical when taking into account the decrease in critical Mach number with $\thetabn$ \citep{Edmiston1984}. They showed the formation of a foot and overshoot, the generation of an electric charge separation field in the shock transition from the foot across the shock with a highly structured electric field which was present already in the shock foot. Moreover, indications for a periodicity of the electric field structure in the foot region were given which we now understand as standing whistler wave precursors in the shock foot for oblique shock angles and supercritical but moderate Mach numbers \citep{Kennel1985,Balikhin1995}. In addition to the field variations these simulations already demonstrated much of the supercritical particle dynamics related to shock reflection and foot formation which we will discuss separately below.\index{Mach number!whistler critical}

Before discussing the ion phase space plots in Figure\,\ref{chap4-fig-bis1} on the right we are going to describe recent investigations on the effects of the mass ratio dependence of the one-dimensional full particle simulations on the shock structure. Of course, in the end only such simulations can be believed which not only take into account the full mass ratio but which are long enough for following the evolution of the shock from a small disturbance up to a stage where the shock on some time scale has approached kind of a state that in a certain sense  does non further evolve. This state is either stationary or it repeats and restores itself such that it is possible to speak at all of a quasi-perpendicular supercritical shock.

\subsubsection{The shock transition scale} \noindent Determination of the shock foot scale is relatively easy both in simulations as also from observation. From observations, as already mentioned, it has been first determined by \cite{Sckopke1983} who found that the foot scale is slightly less but close ($\sim0.7\,r_{ci,\rm refl}$) to the reflected ion gyroradius in quasi-perpendicular shocks. The reasons for this number have been given by \cite{Schwartz1983} and are related to the reflected ions coupling to the upstream convection electric field in which they are accelerated. This can also be checked in simulations. Of more interest is the determination of the shock transition, i.e. the width of the shock ramp which from theory is not well determined since it depends on several factors which can hardly be taken into account at once. 
\begin{figure}[t!]
\hspace{0.0cm}\centerline{\includegraphics[width=1.0\textwidth,clip=]{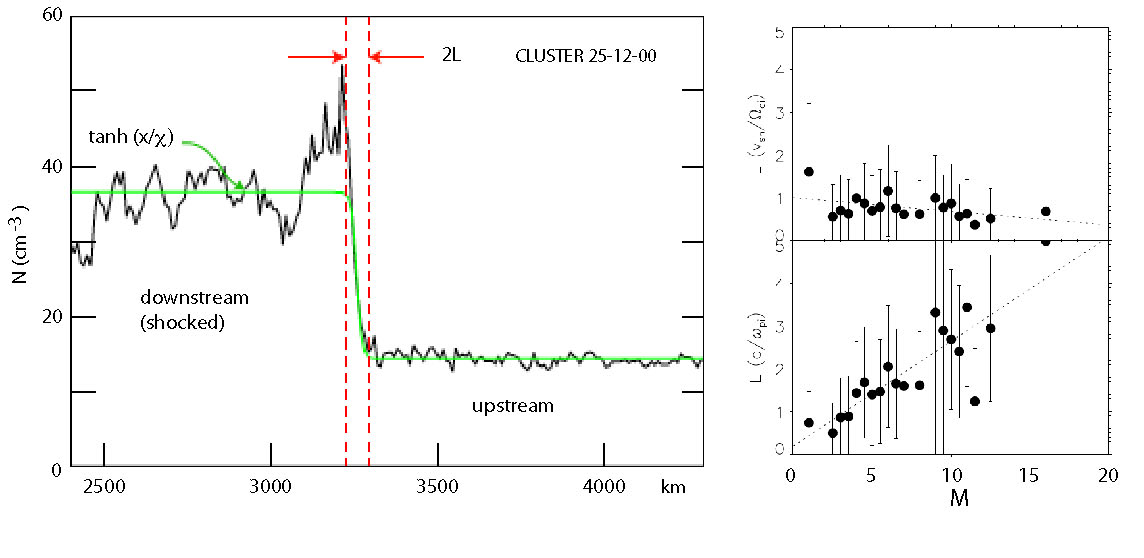} }
\caption[Cluster shock density scale]
{\footnotesize {\it Left}: Shock density transition-fit by a tanh-function in order to determine the shock-ramp transition scale \citep[after][]{Bale2003}. 98 of those shock transitions have been used in order to find a dependence os the shock ramp width from some physical parameter. {\it Right}: The dependence inferred by \cite{Bale2003}. The upper part of the figure scales the dependence of the gyroradius with Mach number, the lower part the dependence of the ion inertial scale. Apparently there is no dependence of gyroradius on Mach number, while there is a clear linear dependence of the inertial scale on Mach number.}\label{chap4-fig-bale-d}
\end{figure}

The width of the shock transition is particularly important in its relation to the width of the electrostatiic potential drop across the shock. There are essentially three transitions scales, the magnetic scale $\Delta_B$, the density scale $\Delta_N$, and the electric potential scale $\Delta_E$. Since the shock is not in pressure equilibrium, the first two scales must not necessarily be proportional to each other. However, the electric field and density gradient might be related, so one expects that $\Delta_N\sim\Delta_E$ even though this is not necessarily so, in particular not when instabilities arise which cause very small scale electric field gradients. In principle one can distinguish three different cases \citep{Lembege1999} which describe different physics:\index{scales!shock transition}
\begin{description}
\item[{\sf 1.} ]  \noindent $\Delta_E\gg\Delta_B$. This is a case that has been reported to have been observed in Bow shock crossings by \cite{Scudder1986,Scudder1995}. The magnetic ramp is much steeper in this case than the structure of the electric field. The latter smears out over the foot and ramp regions. In this case the electrons will behave completely adiabatically, while the ions may be only partially or even non-magnetized.
\item[{\sf 2.} ]  \noindent $\Delta_E\sim\Delta_B$. In this case there will be a significant deviation from adiabatic behaviour of the electrons in the shock transition. Electron heating and motion will not be adiabatic anymore, and the electron distribution will significantly be disturbed \citep[see, cf.,][]{Balikhin1995}. Observations of such cases have been reported by \cite{Formisano1982}.
\item[{\sf 3.} ]  \noindent $\Delta_E\ll\Delta_B$. This case which is also called the `isomagnetic' transition \citep{Eselevich1982,Kennel1985} corresponds to shock transitions with electrostatic substructuring which are sometimes also called subshocks.\index{shocks!ion acoustic subshock}
\end{description}

The most recent {\it experimental} determination of the density transition scale has been provided by \cite{Bale2003} using data from 98 Bow Shock crossings by the Cluster spacecraft quartet. The result is shown in Figure\,\ref{chap4-fig-bale-d} for an example of this fit-determination by fitting a tanh-profile to the shock density transition. The point is that these authors found a dependence of the shock ramp transition on Mach number when the transition is scaled in ion inertial units, while there is no dependence when scaled in ion gyroradii. This caling suggestst that the shock scales with the gyroradius, since $(V/c)(\omega_{pi}/\omega_{ci})\sim {\cal M}_A$.  

\begin{figure}[t!]
\hspace{0.0cm}\centerline{\includegraphics[width=1.0\textwidth,clip=]{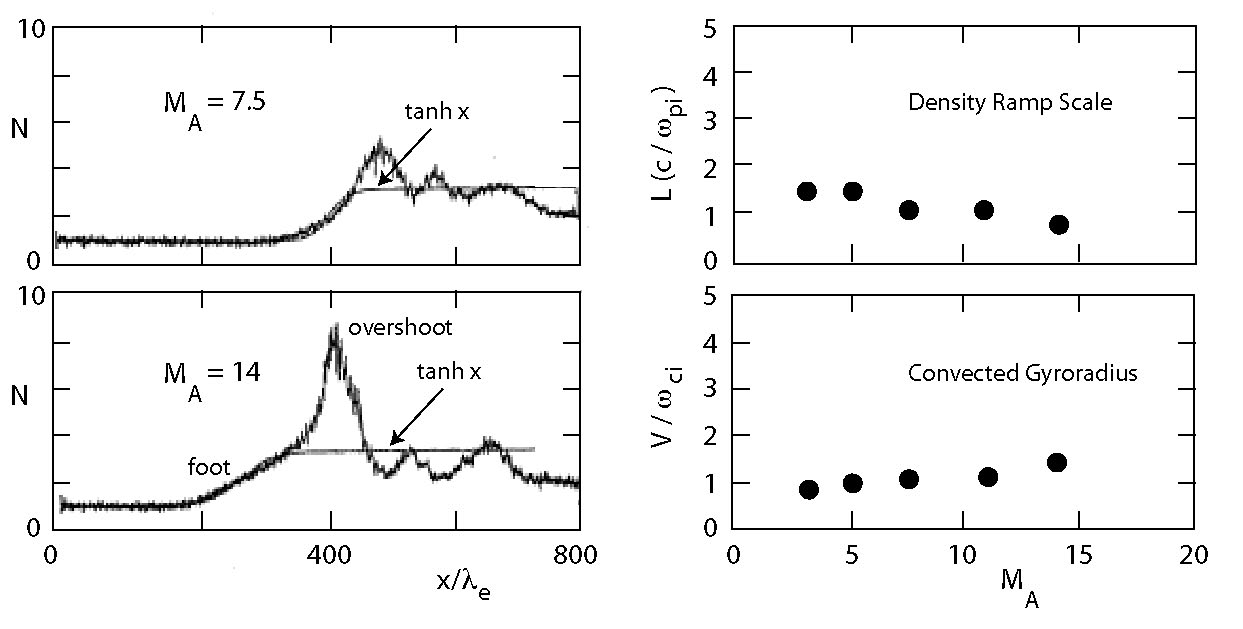} }
\caption[Shock density profile in PIC simulations]
{\footnotesize {\it Left}: Density profile in two full PIC simulations\index{simulations!full particle PIC} of large Mach numbers. Indicated is the pronounced overshoot and the long extended foot. The straight lines are $\tanh x$-fits to the simulations showing the neglect of the overshoot and ramp during such fits which only account for the foot region. From fitting the ramp width the curves on the right are obtained. {\it Right}: Density ramp scales and convected ion gyroradii (in units of upstream inertial length) obtained in one-dimensional full particle PIC simulations of quasi-perpendicular shocks \citep[after][]{Scholer2006} as function of Alfv\'enic Mach number. Use has been made of the full particle mass ratio 1838, $\thetabn=87^\circ$, and $\omega_{pe}/\omega_{ce}=4$. The magnetic field used is that of the overshoot. One observes that the ratio of ion gyroradius to ion inertial length is constant. Also the scale of the ramp is about $\sim1c/\omega_{pi}$, supporting a narrow ramp. The simulations also show that the scale of the ramp sharpens with increasing Mach number.}\label{chap4-fig-bale-s}
\end{figure}

In order to check this behaviour numerically, \cite{Scholer2006} performed a series of one-dimensional full-particle PIC simulations with the correct mass ratio $m_i/m_e=1838$ and for the Alfv\'enic Mach number range $3.2\leq {\cal M}_A\leq 14$ and a shock normal angle $\thetabn=87^\circ$ in order to have a component of $k_\|$ parallel to ${\bf B}$, but with small ratio $\omega_{pe}/\omega_{ce}=4$ to compromise computing reasons. Figure\,\ref{chap4-fig-bale-s} shows the results of these simulations. A $\tanh x$-fit neglects in fact the entire ramp and takes account only of the foot region. Correcting the above described measurements it is thus found that the ramp thickness is just of the order of $\sim 1 \lambda_i=c/\omega_{pi}$ and decreases slightly with increasing Mach number. However, from the form of the density profile it seems clear that the shock ramp is basically determined by the overshoot, and one must take the overshoot magnetic field value in calculating the gyroradius. The convected gyroradius based on the overshoot magnetic field $B_{\rm ov}$ and measured in $\lambda_i$ is about constant very close to unity. Thus the shock ramp scale is given by the convective ion gyroradius based on the overshoot magnetic field. One should, however, note that the computing power in the simulations does not yet allow for larger ratios $\omega_{pe}/\omega_{ce}$ which may affect the result. Moreover, higher dimensional simulations would be required to confirm the general validity of those calculation and conclusions.

Hence, combining the observations of \cite{Bale2003} and the results of the simulation studies of \cite{Scholer2006} we may conclude that  the scale of the shock foot is given by the upstream-convected ion gyroradius $r_{ci}=V/\omega_{ci,1}$ based on the upstream field $B_1$, while the scale of the shock ramp is given by the ramp-convected ion gyroradius $r_{ci,\rm ov}=V/\omega_{ci,\rm ov}$ based on the value of the magnetic field $B_{\rm ov}$ overshoot. This is an important difference which can be taken as a {\it golden rule} for estimates of the structure of quasi-perpendicular shocks even though, of course, these values are dynamical values which change from position to position across the foot and ramp. The scale differences are the reasons for the large upstream extension of the foot and the relative steepness of the shock ramp.
\begin{figure}[t!]
\hspace{0.0cm}\centerline{\includegraphics[width=0.87\textwidth,height=0.38\textheight,clip=]{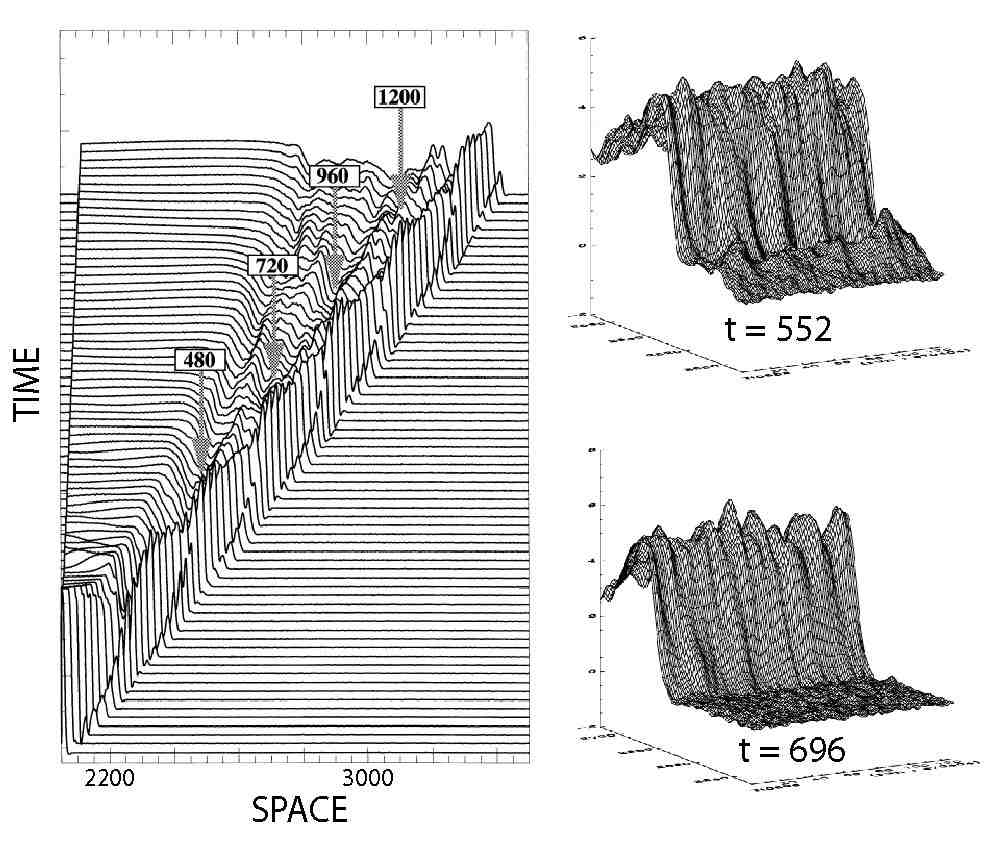} }
\caption[Shock magnetic profile in two-dimensional PIC simulations with reformation]
{\footnotesize Magnetic field from full particle PIC simulations of shock reformation \citep[after][]{Lembege2002}. {\it Left}: Reformation cycles of the magnetic field in the shock. Time is measured in inverse electron plasma frequencies $\omega_{pe}^{-1}$. The reforemation times are indicated by the arrows in the plot with time given when the cycle is complete. {\it Right}: Tow snaphots in time of the view of the shock front in the magnetic field at reformation. The interesting finding is that the front in this two-dimensional view is not a smooth plane but is quite distinctly structured in space.}\label{chap4-fig-lemb1}
\end{figure}

The observed constancy of the overshoot\index{shocks!overshoot} magnetic field-based convective ion gyroradius $r_{ci}\propto V/B_{\rm ov}$ with Mach number ${\cal M}_A\propto V$ can be understood when considering the about linear increase of the overshoot magnetic field $B_{\rm ov}\propto{\cal M}$ with Mach number (or with upstream velocity $V$) which holds for supercritical Mach numbers ${\cal M}>{\cal M}_{\rm crit}$ as long as ${\cal M}$ is not too large. At very large -- but still non-relativistic -- Mach numbers ${\cal M}<{\cal M}_{\rm max}$ the increasing steepness of the shock ramp and the increasing extension of the foot ultimately lead to the excitation of smaller scale structures in the ramp and the foot, which smear out any further increase in the overshoot. 

The generation of these structures by a variety of instabilities might even turn the shock foot and ramp regions into regions where large anomalous collisions and thus resistances are generated as the result of wave-particle interactions. In this case the shock returns to become resistive again due to preventing large numbers of reflected ions from passing across the steep shock ramp and large shock potential, using the kinetic energy of the reflected particle population for the generation of a broad wave spectrum which acts to scatter the particles around in the foot and ramp regions and, possibly, also up to some distance in the transition region behind the ramp. This kind of confinement of reflected particles over long times will then be sufficiently long for providing the heating and dissipation which is required for sustaining a resistive shock which, then, is the result of the combined action of ion viscosity and anomalous resistivity, i.e. anomalous collisions. 
In addition, the scattering of the trapped reflected particle population\index{particles!trapped} necessarily results in plasma heating, and some particles will become accelerated to high velocities in these interactions as well. It is then possible that these particles provide the seed population for energetic particles which have been accelerated to high energies in the well-known shock-Fermi-one and shock-Fermi-two acceleration mechanisms. 

So far the range of Mach numbers ${\cal M}_{\rm max}<{\cal M}<{\cal M}_{\rm rel}$ where this will happen is unknown, as it is hardly accessible to numerical simulations. However, the available simulations seem to point in this direction as long as the Mach numbers remain non-relativistic. In relativistic shocks with ${\cal M}={\cal M}_{\rm rel}$ different effects arise which are not subject to our discussion at this place.

\begin{figure}[t!]
\hspace{0.0cm}\centerline{\includegraphics[width=1.0\textwidth,clip=]{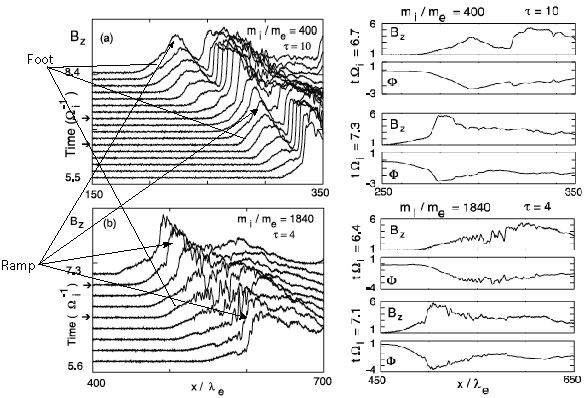} }
\caption[One-dimensional PIC simulations of quasi-perpendicular shocks]
{\footnotesize {\it Left}: One-dimensional PIC simulations \citep[after][]{Scholer2003} of quasi-perpendicular $\thetabn=87^\circ$ shock reformation for two different mass ratios $m_i/m_e=400, 1840$. Time is measured in $\omega_{ci}^{-1}$ (here denoted as $\Omega_i^{-1}$), space in $\lambda_e=c/\omega_{pe}$. The parameter $\tau=\omega_{pe^2}/\omega_{ce}^2$ is taken small in both cases for computational reasons which have to be compromised. The higher mass ratio shows a more violent time evolution because of the lighter electrons and their higher mobility. But both runs show the reformation of the shock from the evolution of their feet. Inspection shows that the original foot region builds up until it becomes itself the shock taking over the role of the ramp. Afterwards a new secondary foot evolves in front of this new ramp. {\it Right}: Two time sections (indicated by the arrows on the left part of the figure) showing the spatial profiles across the shocks. Clearly, the higher mass ratio run shows a more subtle structure of the shock profile in $B_z$ and shock potential $\Phi$, but the gross features are very similar. The potential exists already in the foot but the main drop occurs in the ramp. Moreover, the lower mass ratio has a more concentrated foot region. }\label{chap4-fig-scho03-1}
\end{figure}

\subsection{Shock reformation} \noindent\index{shocks!reformation}\index{process!shock reformation} It has already been mentioned several times that supercritical shocks do under certain conditions reform themselves periodically -- or quasi-periodically--, which is kind of a non-stationarity of the shock that does not destroy the shock but, at the contrary, keeps it intact in a temporarily changing way. We will come later to the problem of real non-stationarity. 
\vspace{-0.3cm}
\subsubsection{Reformation in one dimension: Mass ratio dependence} \noindent Reformation of quasi-per\-pen\-di\-cu\-lar shocks is thus an important shock property which is closely related to highly super-critical shocks and the formation of a foot region, i.e. to the reflection of ions from the shock ramp. 

In fact, reformation was already observed by \cite{Biskamp1972} in the early  short-simulation time PIC simulations shown in Figure\,\ref{chap4-fig-bis1}, where we have noted it explicitly. Reformation of quasi-perpendicular shocks has also been reported, for instance,  by \cite{Lembege1987a}, \cite{Lembege1992, Lembege2002}, \cite{Hellinger2002} and others who all used small ion-to-electron mass ratios. Figure\,\ref{chap4-fig-lemb1} shows an example of shock reformation in a magnetic field stack plot together with the structure of the shock ramp during two reformation times. There is a distinct reformation cycle in this simulation and also a distinct structure of the ramp/shock front which is far from being smooth, a fact  to which we will return during discussion of non-stationarity of shocks.

Full particle electromagnetic PIC simulations with realistically large mass ratios have been performed only very recently \citep{Matsukiyo2003,Scholer2003,Scholer2004} and only in one spatial dimension, showing that reformation at least occurs at small ion-$\beta_i\sim 0.2$. In these simulations the shock is produced by injecting a uniform plasma from $-x$ and letting it reflect from a stationary wall at the right end of the simulation box. The plasma carries a uniform magnetic field in the $(x,z)$-plane, and the plasma is continuously injected in the $+x$-direction. Since the right-hand reflecting boundary is stationary the shock, which is generated via the ion-ion beam instability in the interaction of the incoming and reflected ion beams, moves to the left at velocity given by the supercritical shock Mach number ${\cal M}_A\sim 4.5$. The upstream plasma has $\beta_i=\beta_e=0.05$, and the shock normal angle is $\thetabn=87^\circ$. 

Two runs of these simulations are shown in Figure\,\ref{chap4-fig-scho03-1}, one is for a mass ratio of 400, the other for a mass ratio of 1840. \index{simulations!large-mass ratio}The left-hand side of the Figure shows stack plots of time profiles of the nearly perpendicular magnetic field $B_z$  with time running in equidistant units upward on the ordinate. Since the plasma is injected from the left and reflected at the right boundary the shock is seen to move from the right to the left in this pseudo-threedimensional representation. Time is measured in ion cyclotron periods $\omega_{ci}^{-1}$, while space on the abscissa is in units of the electron inertial length $c/\omega_{pe}$. The magnetic profiles are strikingly similar for both mass ratios. In both cases a relatively flat foot develops in front of the steeper shock ramp caused by the shock reflected ions. The magnetic field of this foot starts itself increasing with time with growth being strongest close to the upstream edge of the foot until the foot field becomes so strong that it replaces the former shock ramp and itself becomes the new and displaced shock ramp.\index{shocks!ramp}  \index{simulations!one-dimensional PIC}

This is seen most cleanly in the upper low mass-ratio part of the figure. The foot takes over, steepens and becomes itself the shock. One can recognise in addition that, even earlier, the intense foot already had started reflecting ions by himself and developing its own flat pre-foot region. This pre-foot evolves readily to become the next foot, while the old ramps become part of the downstream turbulence. 

During this reformation process the shock progresses from right to left. However, this progress is not a continuous motion at constant speed, Both the foot and the ramp jump forward in steps. One such step ahead is seen, for instance, at time $t\omega_{ci}=7.6$ in the upper part on the left. Sitting in the shock frame one would experience some forward acceleration at this time seeing the ramp running downstream as a magnetic wave the source of which seem to be the instantaneous shock ramp, while it is nothing but the edge of the old shock foot. Hence the shock ramp and shock overshoot act as the source of a pulsating magnetic wave that is injected into the downstream turbulence from the shock with periodicity of roughly $\Delta t\sim 1.8\omega_{ci}^{-1}$ for $m_i/m_e=400$. 
\begin{figure}[t!]
\hspace{0.0cm}\centerline{\includegraphics[width=1.0\textwidth,clip=]{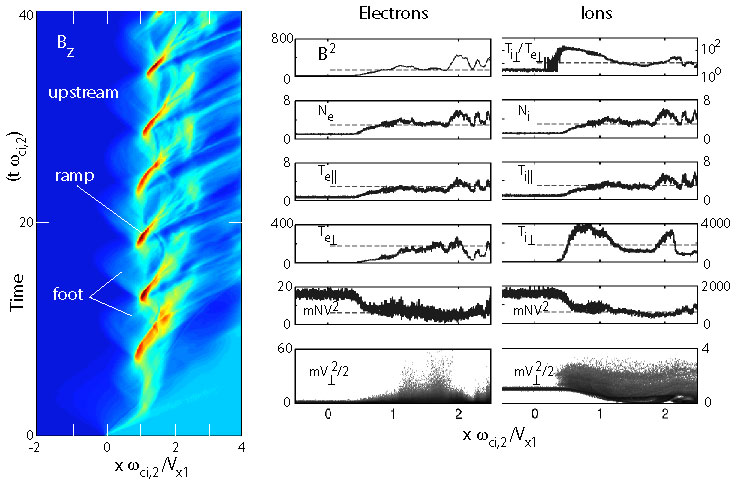} }
\caption[One-dimensional quasi-perpendicular shock simulations by \cite{Umeda2006}]
{\footnotesize {\it Left}: Evolution of the magnetic field in a quasi-perpendicular high Mach number ${\cal M}_A=10$ PIC simulation \citep{Umeda2006}. Here the presentation is in the shock frame of reference, and the shock has been initialised by assuming Rankine-Hugoniot initial jump conditions. The non-physical nature of this assumption is visible in the initial evolution and fast displacement of the shock to the right. After the initial unphysical disturbance has disappeared a selfconsistent physical state is reached in which the shock quasi-periodically reforms itself. The competition between the shock foot and ramp formation is nicely seen in the colour plot of the magnetic field $B_z$. {\it Right}: Electron and ion plasma parameters in computational units. Of interest is only their relative behaviour, not the absolute values. The profiles are taken at time $t\omega_{ci,2}=38.1$. They show the compression of the plasma and heating of electrons and ions. Parallel electron and ion heating is comparable, but ions are heating much stronger than electrons in perpendicular direction causing a large perpendicular temperature anisotropy downstream of the shock.} \label{chap4-fig-umeda}
\end{figure}

The higher mass-ratio run in the lower part on the left also shows reformation of the shock. However there are some differences. First, the magnetic profiles are much stronger disturbed exhibiting much more structuring. Second, the foot region is considerably more extended in upstream direction. Third, the ramp is much steeper, and reformation is faster, happening on a time scale of $\Delta t\sim 1.3 \omega_{ci}^{-1}$, roughly 30\% faster than in the above case.  Reformation is, however, more irregular at the realistic mass ratio with the property of reforming the shock ramp out of a long extended relatively smooth shock foot which exhibits pronounced oscillations.

The right-hand side of the figure shows two shock profiles at constant times for the two different mass-ratio simulations runs. The first profile at $t\omega_{ci}=6.7$ has been taken when a well developed foot and ramp had been formed on the shock, the second profile at $t\omega_{ci}=7.3$ is at the start of the new foot towards the end of the simulations.  At the low mass ratio the foot profile is quite smooth showing that the foot is produced by the accumulation of reflected ions at the upstream edge of the foot where the ions have the largest velocity in direction $y$ along the shock. This is where, during their upstream gyration in the upstream magnetic field, they orbit about parallel to the upstream convection electric field and gain most energy. 

Hence, here, the current density is largest due to the accumulation of the reflected ions, due to the retardation of some ions from the inflow already at this place, and due to the speeding up of the reflected and retarded ions in $y$-direction by the convection electric field $E_y$. All this leads to a maximum in the current density $j_y$ and thus causes a maximum in the magnetic field $B_z$  close to the upstream edge of the foot. 

Most interestingly, the electric potential exhibits its strongest drop right here in the foot region with a second but smaller drop in the ramp itself. It is the electric field that belongs to this potential drop that retards the inflow already before it reaches the shock ramp. At the contrary, when the shock ramp is well developed, the main potential drop is for short time right at the ramp and extends even relatively far into the downstream region.

For the realistic mass ration the foot- and ramp-transitions are both highly structured at $t\omega_{ci}=6.4$ exhibiting fluctuations in both the magnetic field and electric potential, but the electric potential drop extends all over the foot region with nearly no drop in the ramp. When the ramp has been reformed at $t\omega_{ci}=7.1$, the foot region still maintains a substantial potential drop, but 50\% of the total drop is now found in the ramp with the downstream potential recovering. This is interesting as it implies that lower energy electrons will become trapped in the overshoot region, an effect which is much stronger for the large mass-ratio than for small mass-ratios and thus closer to reality.

Some recent one-dimensional full particle PIC simulations by \cite{Umeda2006} at Mach number ${\cal M}_A=10$ and medium mass ratio $m_i/m_e=100$ throw additional light on the reformation process when keeping in mind that reformation is not as strongly dependent on the mass ratio as originally believed. Figure\,\ref{chap4-fig-umeda} shows a collection of their results which this time are represented in the shock frame of reference. 

The simulations have been performed by assuming an initial Rankine-Hugoniot equilibrium in the PIC code. The non-physicality of this initialisation is manifested in the initial evolution over the first few ion cyclotron periods. During this time the simulation adjusts itself to the correct physics, and the non-physical disturbance decays. The shock frame has shifted by this to a new position, which in the shock frame is located farther downstream (which takes into account of the moment transferred to the shock by the reflection of the upstream ions who lower the shock speed). 

The further evolution of the shock shows the quasi-periodic reformation and the play between the foot and the ramp formation. The periodicity is roughly $\sim 10\omega_{ci,2}^{-1}$. When the foot takes over to become the ramp, the ramp jumps ahead in a fraction of this time. Afterwards the formation of the foot retards the ramp motion, and the ramp softens and displaces itself downstream to become a downstream moving spectrum of magnetic oscillations which is injected into the downstream region in the form of wave packets. The various plasma parameters in the left part of the figure show in addition the compression of plasma and field, and the dominance of perpendicular ion heating which is, of course, due to the accelerated foot ions which pass into downstream.

\begin{figure}[t!]
\hspace{0.0cm}\centerline{\includegraphics[width=1.0\textwidth,clip=]{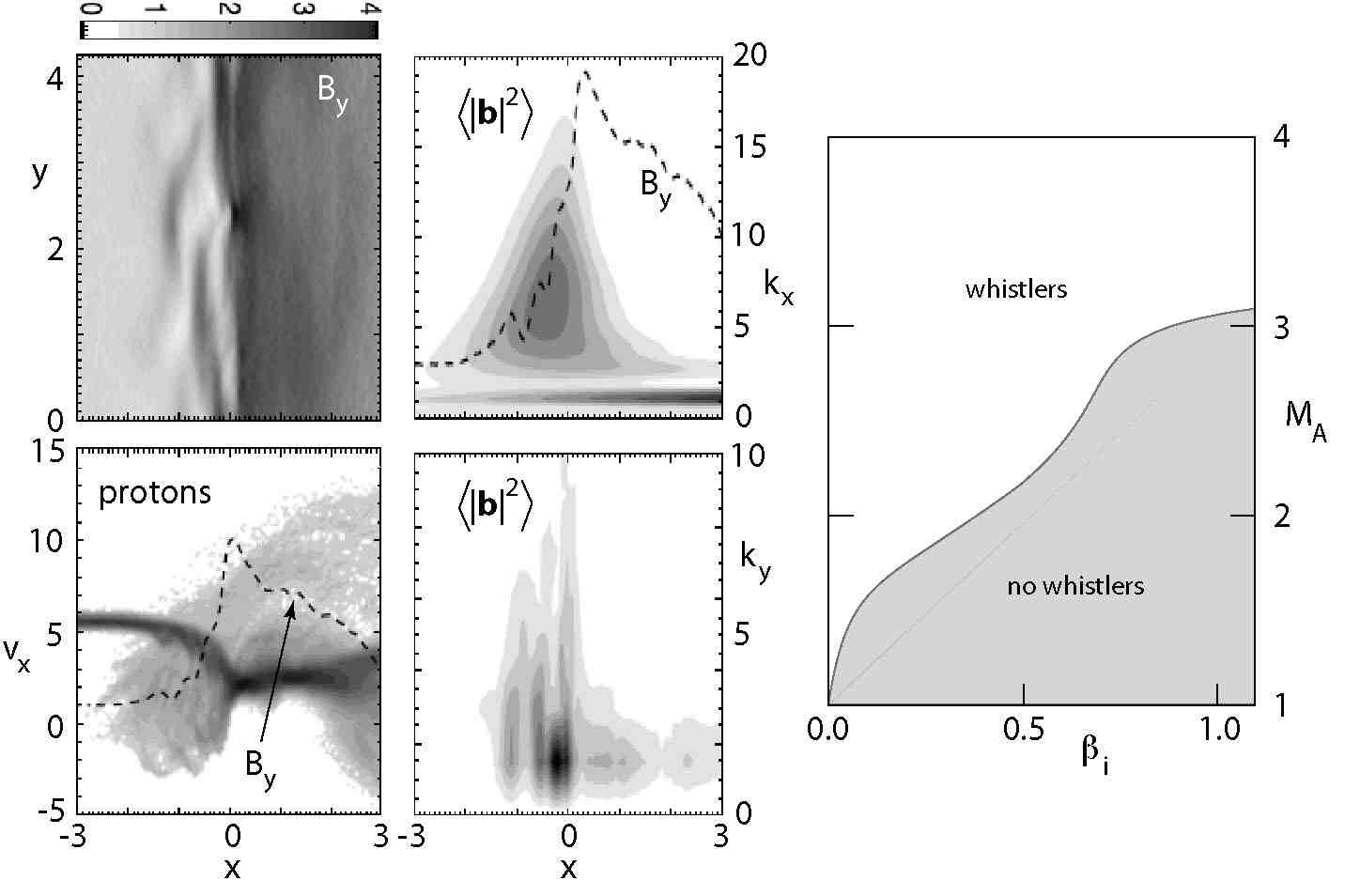} }
\caption[Two-dimensional quasi-perpendicular shock simulations]
{\footnotesize {\it Left}: Two-dimensional PIC simulations \citep[after][]{Hellinger2007} of the end time $\omega_{ci}t=28$ of the evolution of a {\it strictly perpendicular} shock using $m_i/m_e=400$. Shown is the magnetic structure in the $(x,y)$-plane, the proton phase space $(x,v_x)$ and the power of magnetic fluctuations in dependence on space $x$ and wave numbers $(k_y, k_x)$. Lengths are measured in ion inertial lengths $c/\omega_{pi}$, velocities in Alfv\'en speeds $V_A$, wave numbers in inverse ion inertial lengths $\omega_{pi}/c$. Magnetic fields and powers are in relative units (see grey scale bar). No shock reformation is seen in the upper panel of $B_y$ on the left. A periodic foot evolves periodically causing a higher and steeper ramp overshoot when its cycle ends, but the shock ramp does not become exchanged with a new ramp. Note also that the next foot cycle begins before the end of the former cycle, i.e. the shock foot itself reflects ions. The power spectra show a periodic spatial spectrum of whistlers standing in and restricted to the shock foot. Periodicity in $k_y$ is caused by interference between outward and inward moving whistlers. The proton phase space shows the retardation of the incoming flow in the shock foot, the occurrence of reflected ions in the foot and the heating of foot ions. Forward heating is also seen in the overshoot. {\it Right}: A parametric (2D-hybrid simulation) investigation of the evolution of phase locked whistlers in the shock foot in dependence on Mach number ${\cal M}_A$ and $\beta_i$. Large Mach numbers and small $\beta_i$ support the excitation of standing whistlers.}\label{chap4-fig-hell}
\end{figure}

\subsubsection{Two-dimensional reformation: Whistlers and Mach number dependence} \noindent First two-dimensional simulations of a strictly perpendicular $\thetabn=90^\circ$ shock formation have recently been performed by \cite{Hellinger2007}. These simulations were intended to study the reformation process in two dimensions when the  perpendicular shock is supercritical. Since PIC simulations are very computer-time consuming, most of the simulation runs by \cite{Hellinger2007} used a two-dimensional hybrid code with the shock being generated by a magnetic piston as in the case of the simulations by \cite{Lembege1987a}. The interesting result of this simulation study was that no shock reformation was found while phase locked whistlers were detected which formed a characteristic interference pattern in the shock foot regions. This result is surprising as for strictly perpendicular shocks no whistlers should be generated according to the one-dimensional theory \citep[see the above discussion on whistlers and, e.g.,][]{Kennel1985,Balikhin1995}. \index{simulations!two-dimensional PIC}

In order to cross check their hybrid simulation results \cite{Hellinger2007} also performed a two-dimensional PIC simulation \citep{Lembege1992} with  mass ratio $m_i/m_e=400$, Mach number ${\cal M}_A=5.5$, electron plasma-to-cyclotron frequency ratio $\omega_{pe}/\omega_{ce}=2$, upstream $\beta_e=0.24, \beta_i=0.15$ and 4 particles per cell. The results of this simulation have been compiled in Figure\,\ref{chap4-fig-hell}, showing only the PIC simulations and no hybrid simulations. Deviating from the former one-dimensional simulations by \cite{Matsukiyo2003} the magnetic field is in $y$. The block consisting of the four panels on the left in the figure are the simulation results at the end of the simulation run, showing the compression of the magnetic field $B_y$ in the $(x,y)$-plane, proton velocity space $(v_x,x)$ -- only the normal velocity component is shown --, and the average magnetic fluctuation spectra  $\langle b^2\rangle$ as functions of wave number components $k_x, k_y$. As the authors describe, after a short initial time when the shock foot forms and the shock reforms, reformation stops and does not recover again in these two dimensional run. Instead, the shock foot starts exhibiting large magnetic fluctuations. These are seen in the low-frequency magnetic power spectra being confined solely to the shock foot (as recognised from the $B_y$-profile) as seen from the $k_x$ dependence of the magnetic power spectrum and forming an interference pattern in $k_y$. 

These fluctuations are identified as whistler waves propagating obliquely (in $k_x$ and $k_y$) across the foot and the magnetic field. Since $k_x\sim 3 k_y$ their perpendicular wave numbers are large, they are quite oblique, and their parallel wavelengths are long. They are excited in the foot and because of their obliqueness probably propagate close to the resonance cone. Their main effect is to resonantly suppress shock reformation by inhibiting the ions to accumulate in the foot. Hence, under the conditions of these simulations the shock turns out to be stable and does not reform. It maintains its structure thanks to the generation of oblique whistlers in the shock foot which dissipate so much energy that the shock becomes about resistive. In one-dimensional simulations this regime has not been seen and is probably inhibited for strictly perpendicular shocks. In two-dimensional simulations, on the other hand, the additional degree of freedom provided by the introduction of the second spatial dimension allows for the generation of the whistlers which are suppressed in the one-dimensional case (where ${\bf k}$ has only the component $k_x$).

Guided by these simulations \cite{Hellinger2007} have undertaken a parametric study of the regime where whistler excitation and thus presumably stationary shock structures lacking reformation should exist. Their results are given on the right in Figure\,\ref{chap4-fig-hell} in ($\beta_i,{\cal M}_A$)-space. According to this figure, whistlers will not be excited at low Mach numbers. Here the two-dimensional perpendicular shocks will reform. At higher Mach numbers whistlers should be excited, and the shock should become stabilised in two dimensions. \index{process!suppression of reformation} 

This surprising result suggests that sufficiently high Mach numbers are needed in order to excite whistlers; on the other hand, when the Mach number will become  large (\cite{Hellinger2007} investigated only the range of Mach numbers $<5$)   
then other effects should set on, and the shock should become non-stationary with whistlers becoming unimportant and reformation becoming possible again. 
 
\section{Ion Dynamics}\noindent
Until now we have avoided discussing the behaviour of particles in the simulations. The mere idealised reflection process we have already discussed, as far as this could be done analytically. The complicated geometry and dynamics of particle motion in shocks necessitates to return to simulations.

We have mentioned that reflected particles are forming a foot on the shock and may contribute to the reformation of the shock. We have, however, not yet gone into detail and into the investigation of the relation of the particle distributions in phase and real space observed in the simulations and their relation to the shock dynamics. This will be done in the present section for the ions on the basis of simulation studies.

\subsection{Ion dynamics in shock reformation}\noindent\index{process!reformation}\index{shocks!reformation}
The two plots on the right-hand side of  Figure\,\ref{chap4-fig-bis1} show phase space representations of ions in the early simulations of shock formation by \cite{Biskamp1972} at $t\omega_{ci}=3$ close to the end of the simulation. The first box is the dependence of the ion velocity $v_x$ on position $x$ (direction of inflow), the second plot the transverse ion velocity component $v_y$. Each macroparticle present in the box is identified by a dot. Hence high particle phase space density is reflected by accumulation of many such dots. The plasma enters the central region of the box from left at high positive speed $v_x$ and $v_y=0$ and leaves the central region to the right at high density and very low speed $v_x$. the inflowing plasma was extended relatively narrow in $v_x$ indicating low temperature. The shocked plasma distribution is broader thus having higher temperature. Moreover, some of the ions are seen to be smeared out from large positive up to large negative velocities in $v_x$ indicating that strong particle scattering has occurred and that there are particles of high velocity in both directions being equivalent to heated plasma. In the foot region the closed circle in the particle distribution signifies the presence of the reflected trapped ions which are accelerated in both direction. their presence decelerates the inflow as can be seen by the drop in the velocity $v_x$ just before entering the closed loop trapped ion region. The reflected particles are seen on the left of this closed loop having negative speeds in $v_x$ therefore being directed upstream.  The $(v_y,x)$-box show that the reflected and trapped ions gyrate and are accelerated into $y$-direction. Also many of the ions downstream of the shock possess non-zero $v_y$ velocity components, some of the large, indicating acceleration along the shock front. 

\begin{figure}[t!]
\hspace{0.0cm}\centerline{\includegraphics[width=1.0\textwidth,clip=]{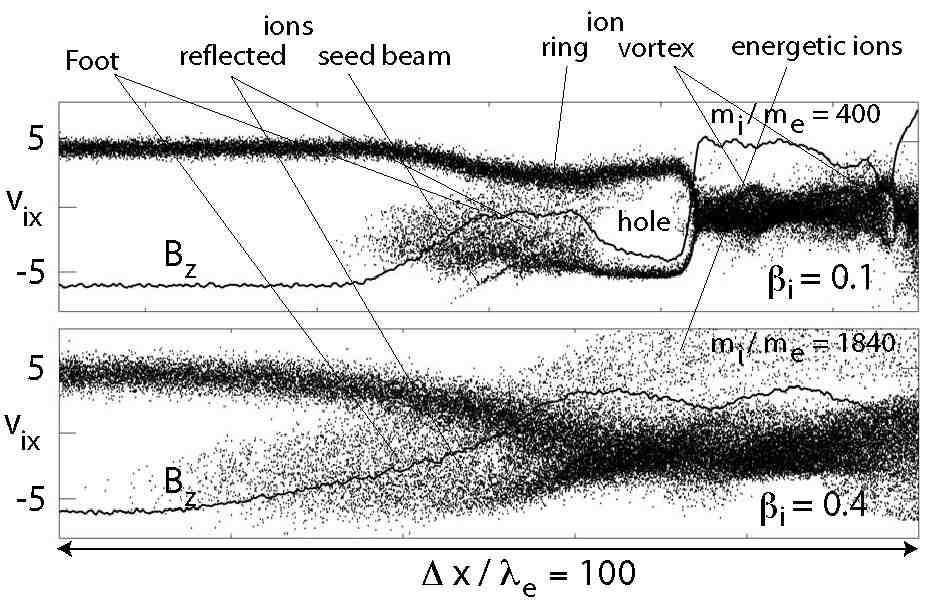} }
\caption[Ion phase space in supercritical simulations of quasi-perpendicular shocks]
{\footnotesize Ion phase space for two simulations of supercritical quasi-perpendicular shock formations for same Mach number and $\beta_e=0.2$ but different mass ratios and $\beta_i$ \citep[after][]{Scholer2003}. {\it Top}: Foot formation and shock reformation in a low-$\beta_i$ shock simulation. The reflected ion beam accumulates in the shock foot forming a ring-vortex ion distribution which causes scattering of the reflected and upstream ions and generates a high foot magnetic field. The hole of the vortex corresponds to low magnetic filed, the ring to large magnetic fields. When the foot field further increases it takes over and becomes the new ramp. At this time the ramp position jumps from the current position to that of the foot edge thereby reforming the shock. Such former reformation cycles can be recognised in the downstream distribution as remains of ion vortices (holes) in the otherwise hot downstream distribution. {\it Bottom}: The same simulation with realistic mass ratio but large $\beta_i=0.4$ No reformation and no ion vortex is observed. Instead, the foot ions because of their high temperature smear out the entire gap between the original inflow and the reflected ion beams. Note also the occurrence of a large number of downstream diffuse energetic ions in this case.}\label{chap4-fig-scho-3}
\end{figure}

Even though it seems clear from this figure already that the reflected ions are involved in shock reformation and the dynamics of the shock profile, the process remains to be unclear until higher resolution simulations and simulations at much higher mass ratio can be performed. Figure\,\ref{chap4-fig-scho03-1} gave already a much clearer idea of the ion dynamics. The corresponding ion phase space plots are shown in Figure\,\ref{chap4-fig-scho-3} for the two mass ratios $m_i/m_e=200$ (top) and 1840 (bottom)  respectively. Both plots show just the enlarged shock foot transition region over the same scale of $100 c/\omega_{pe}$. Also the electron $\beta_e=0.2$ has been kept constant in both simulations, while the ion $\beta_i$ is been changed. Also these plots show only the normal component of the ion velocity for the nearly perpendicular supercritical shock. In both plots the magnetic field $B_z$ has been drawn as the thin line showing the magnetic shock profile over the spatial distance $\Delta x$.  

The upper low-mass-ratio low-$\beta_i$ panel shows the cold dense ion inflow at velocity $v_{ix}\sim 5$ (in units of the upstream Alfv\'en velocity) being retarded to nearly Mach number 1 already when entering the foot. This retardation is due to its interaction with the intense but cold (narrow in velocity space) reflected ion beam which is seen as the narrow negative $v_{xi}$-velocity beam originating from the shock ramp. This reflected ion beam needs a certain distance to interact with the upstream plasma inflow. This distance is the lengths the beam-beam excited waves need to grow.  But once the interaction becomes strong enough, the reflected ions are scattered in addition to being turned around by gyration. Both effect cause a reduction in the velocity $v_x$ of the reflected ions which turn to flow in $y$ direction, causing the magnetic bump that develops in this region of the foot. It is interesting to remark that in the $(v_x,x)$-plane the reflected ions close with the upstream flow into an ion ring distribution just in front of the ramp. Behind the ramp, which is the point of bifurcation of the ion distribution, a broad hot ion distribution arises which at some locations shows rudimentary remains of rings (ion vortices) from former reformation cycles. Their magnetic signatures are dips in the magnetic field. The next reformation cycle can the be expected to completely close the ion vortex in the foot and to transform the ramp from its current position to the position of the foot. the first sign of this process is already seen in the foot ion distribution, which shows the birth of a faint new reflected ion beam at high negative speeds. This beam is not participating in the formation of the ring but serves as the seed of the newly reflected population.

The same behaviour is found in large mass ratio simulations as long as $\beta_i$ is small. This is obvious from the large mass-ratio magnetic field shown in Figure\,\ref{chap4-fig-scho03-1}. As long as $\beta_i$ remains to be small, the shock undergoes reformations also for realistic mass ratios. In other words, as long as the plasma is relatively cool the real shocks found in nature should develop feet which at a later time quasi-periodically become the shock ramp. 

This changes, when $\beta_i$ increases as is suggested from consideration of the lower panel in Figure\,\ref{chap4-fig-scho-3}. There a realistic mass ratio has been assumed, but $\beta_i=0.4$, and no reformation is observed, at least not during the possible simulation times.  Instead, the shock develops a very long foot region that is extended twice as far into the upstream region as in the low-$\beta_i$ case. Clearly, the high ion temperature smears out the reflected ion population over the entire gap region between the upstream and reflected beam regions, and no vortex can develop. This implies that the foot remains smooth and does not evolve into a ramp. 
\begin{figure}[t!]
\hspace{0.0cm}\centerline{\includegraphics[width=1.0\textwidth,height=0.4\textheight,clip=]{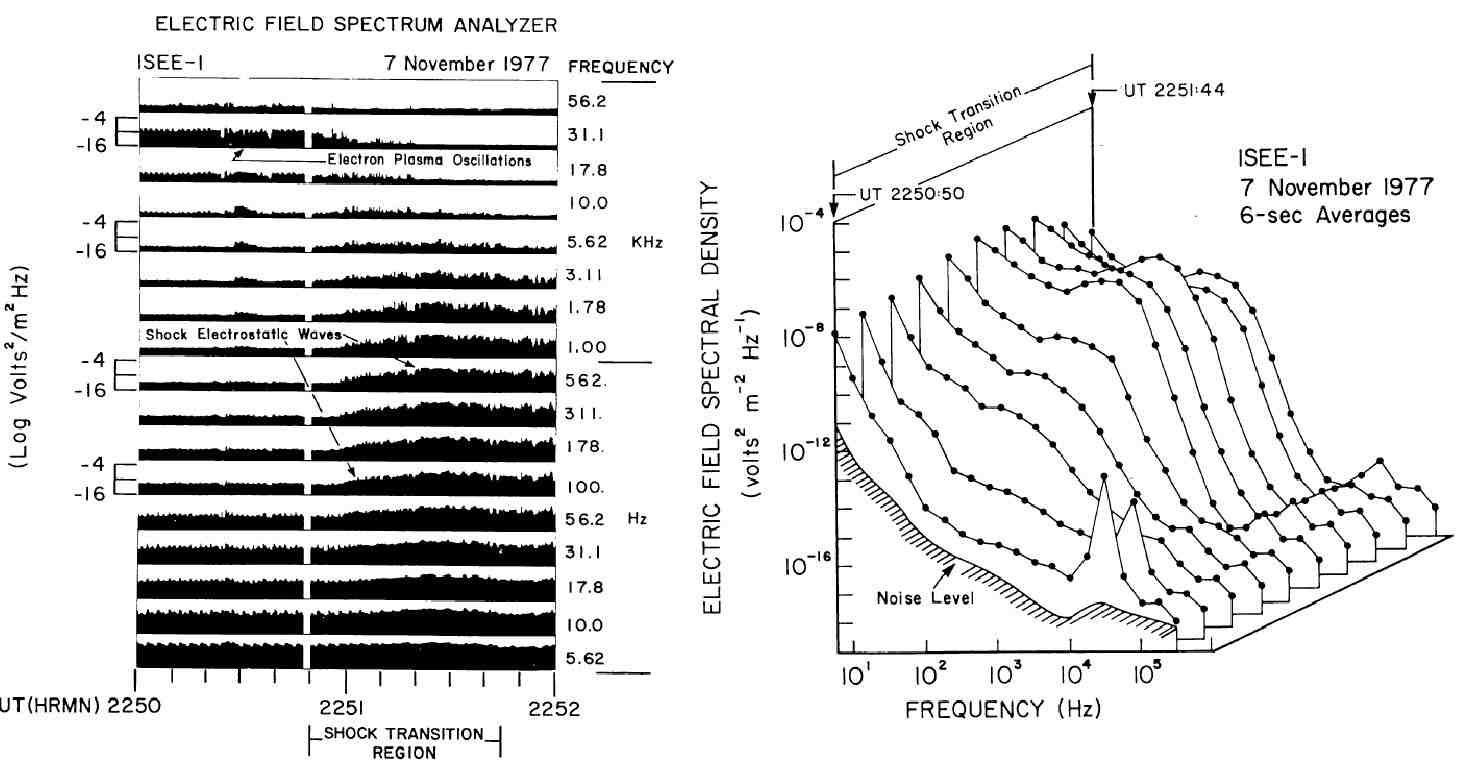} }
\caption[Wave spectra during a spacecraft shock crossing]
{\footnotesize Electric wave spectra measured during the spacecraft crossing of an interplanetary shock \citep[after][]{Gurnett1979}. {\it Left}: Power spectra (in V$^2$/m$^2$Hz) with respect to time in a number of frequency channels. The spacecraft approaches the shock from left and crosses over it. The increase in power is well documented from low to high frequencies when coming into the shock transition region. {\it Right}: A sequence of shock electric spectra during this crossing given as power spectral density with respect to frequency. The dramatic increase of the low frequency wave power is seen when the spacecraft approaches and crosses over the shock. Behind the shock the power remains high but lower than in the transition region. The Bump around a few 100 Hz is the most interesting from the point of view of instability. These waves are excited by electron-ion instabilities discussed in the next section. }\label{chap4-fig-gurn-qpaw}
\end{figure}
However, inspecting the panels of Figure\,\ref{chap4-fig-scho-3} it becomes immediately clear that  suppression of reformation is a relative process. Reformation will be suppressed only when the thermal speed $v_i$ of the ions is large enough to bridge the gap between the reflected and incoming ion beams, i.e. large enough to fill the hole. semi-empiricall one can establish a condition for shock reformation as $v_i< \alpha V_{n1}$ when taking into account that the normal speed of incoming ions is simply specularly turned nagative. Since this is never exactly the case, the coefficient will roughly be in the interval $1.5<\alpha<2$. This condition for reformation to occur can be written as
\begin{equation}\label{chap4-eq-refcond1}
{\cal M}_A> \beta_i^\frac{1}{2}/\alpha
\end{equation}
where the Alfv\'enic Mach number is defined on the the normal upstream velocity $V_{1n}$.
The larger the Mach number becomes the less suppression of reformation will play a role, and at really high Mach numbers one expects that either reformation becomes a normal process or that other time-dependent processes set on which lead to a non-stationary state shock transition probably being of chaotic unpredictable behaviour. As we have already argued earlier this is quite normal as the shock is  thermodynamically and thermally in a non-equailibirum state: it is a region where electrons and ions have violently different temperatures, it is not in pressure equilibrium, upstream and downstream temperatures are different, and it hosts a number of non-Boltzmannian phase space distributions all concentrated in a small volume of real space. Under such conditions stationary states will occur only exceptionally. \index{process!nonstationarity}
\begin{figure}[t!]
\hspace{0.0cm}\centerline{\includegraphics[width=0.6\textwidth,clip=]{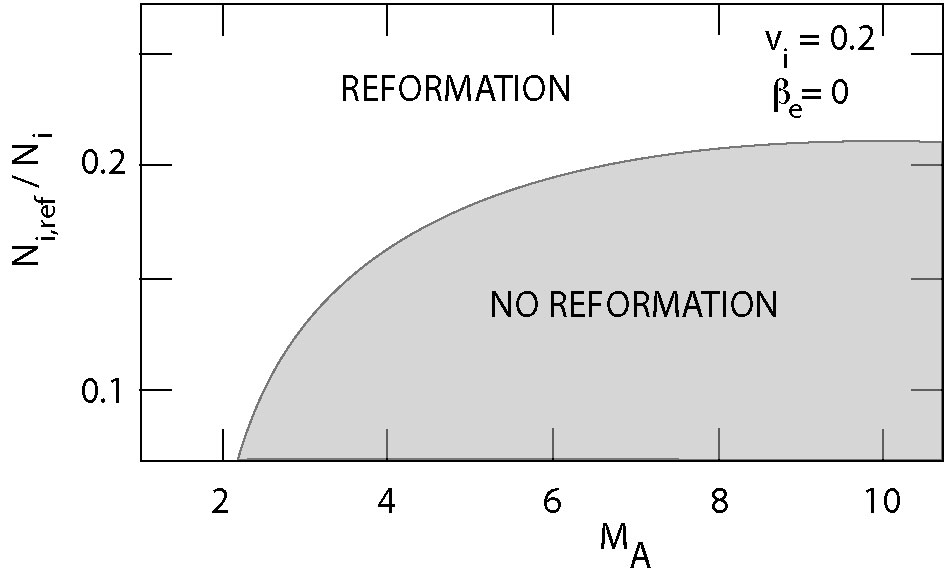} }
\caption[Wave spectra during a spacecraft shock crossing]
{\footnotesize Parametric determination of the fraction of shock-reflected ions in the foot of a quasi-perpendicular shock as function of Alfv\'enic Mach number for the special case of $\beta_e=0$ and ion thermal velocities $v_i=0.2$ measured in $\omega_{pe}$d, where d is the numerical grid spacing \citep[after][]{Hada2003}. }\label{chap4-fig-hada1}
\end{figure}
\cite{Hada2003} have recently attempted to semi-empirically determine the fraction of reflected ions needed for reformation to occur. They performed a large parametric search based on hybrid simulations by changing the Mach number and determined the fraction of reflected ions in the foot when reformation occurred. Their result is shown in Figure\,\ref{chap4-fig-hada1} for cold electrons $\beta_e=0$ (assuming that electrons do not contribute to reformation and reformation being exclusively due to ion-viscosity, and for fixed thermal velocity of upstream and reflected ions $v_i/\omega_{pe}{\rm d}=0.2$, where d is the spatial simulation-grid spacing.

\subsection{Ion instabilities and ion waves}\noindent
So far we have described the field and particle properties as have been observed in simulations of quasi-perpendicular supercritical shocks. It has become obvious that in the different regions of the shock transition the particle distributions carry free energy. This is true for the foot region, the shock transition and overshoot as well as the downstream region. And it is true for both species, electrons and ions. This free energy is the source of a number of instabilities which excite waves of different kinds in the various shock transition regions which can be measured. Figure\,\ref{chap4-fig-gurn-qpaw}, taken from \cite{Gurnett1979}, shows an example of such measurements when the ISEE-1 spacecraft\index{spacecraft!ISEE} crossed an interplanetary shock\index{shocks!interplanetary} travelling outward in the solar wind\index{solar wind}. The passage of the shock over the spacecraft is seen in the wave instrument in the various channels as a steep increase of the power spectral density of the electric field which is highest in the crossing of the ramp and at medium frequencies of a few 100 Hz under the conditions of the crossing. After the crossing took place the wave power in the downstream region of the shock still remained high but was lower than during the shock crossing. In the right part of the figure the different electric wave power spectra are shown in their time sequence as a stack plot. From this plot the dramatic increase of the power in the medium frequencies during the crossing is nicely visible. Most of the waves excited in this frequency range are caused by electron-ion instabilities which we will discuss in the next section.

The free energy source of the instabilities is less the temperature anisotropy than the direct differences in bulk flow properties of the different species components. We therefore ignore the temperature anisotropy differences even though such instabilities may arise, in particular when ion-whistlers are excited of which we know that they can be driven by a temperature anisotropy. To some extent the occurrence of two (counterstreaming in direction $x$ of the shock normal) ion beams already fakes a bulk temperature on the ion component thus generating some relationship between a two-beam situation and a temperature anisotropy. Similarly transversely heated reflected ions superposed on a low perpendicular temperature inflowing ion background fakes a perpendicular temperature anisotropy. For our purposes, however, the bulk flow differences are more interesting and have, in fact, been more closely investigated right from the beginning \citep{Forslund1970,Forslund1970a,Papadopoulos1971,Wu1984}. 

\subsubsection{Foot region waves}
\begin{figure*}[t!]
\hspace{0.0cm}\centerline{\includegraphics[width=1.0\textwidth,clip=]{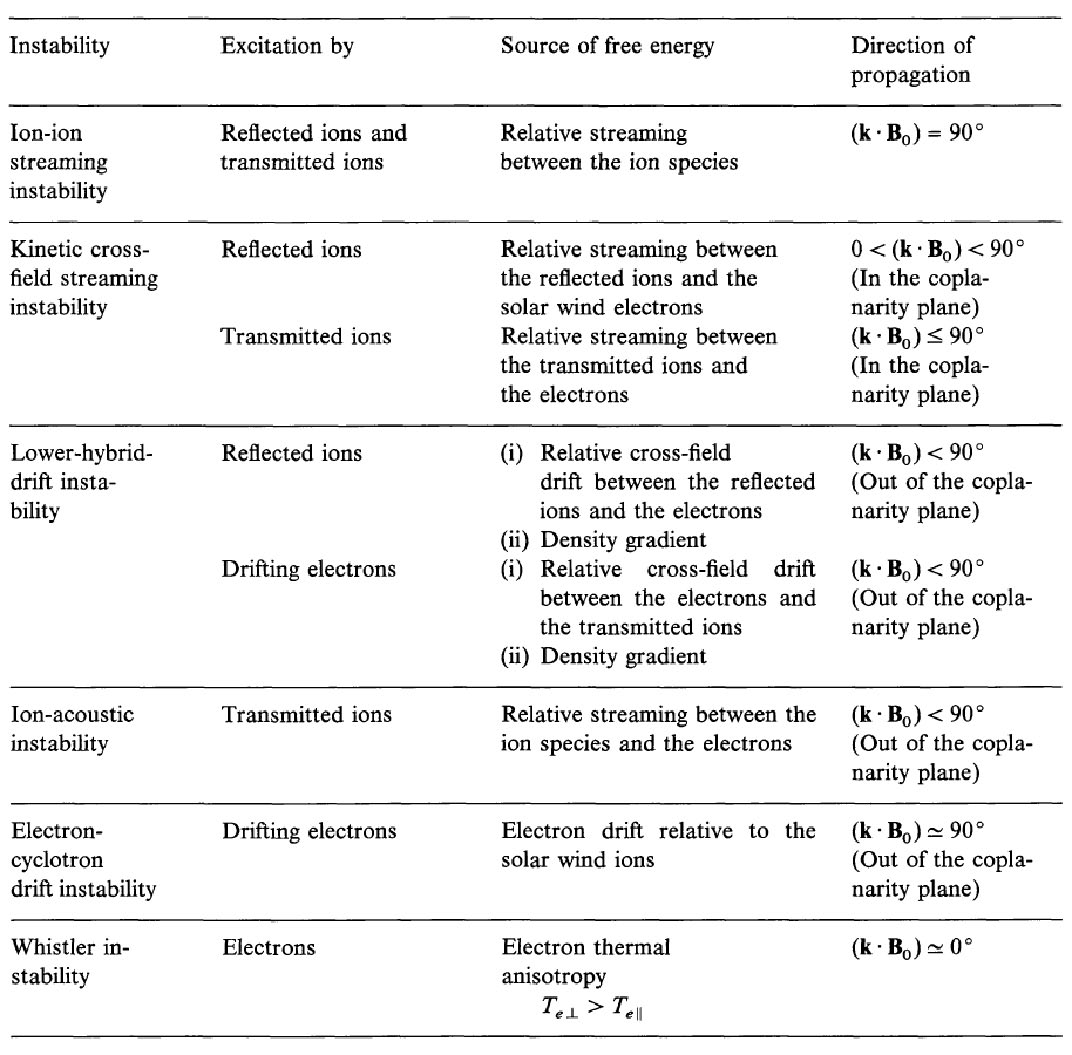} }
\caption{Instabilities in the Foot Region \citep[copied from][]{Wu1984}.}\label{Table1}
\end{figure*}

\noindent \index{waves!foot region}Let us first consider the foot region. The free energy sources here are the relative drifts between the incoming electrons and reflected ions and the incoming electrons and incoming ions. The presence of the reflected ions causes a decrease of the ion bulk velocity in the foot region. This implies that the incoming electrons are decelerated so that the current in shock normal direction is zero, i.e. the flow is current-free in normal direction. However, this has the consequence that a relative bulk velocity  between electrons and reflected ions or electrons and incoming ions arises. These differences will contribute to the excitation of instabilities.
Electrons are not resolved in hybrid simulations, however. In this section we will restrict to ion instabilities leaving the essentially more interesting ion-electron instabilities for the next section. A list of the most important instabilities in the foot region is given in Figure\,\ref{Table1}. This list has been complied by \cite{Wu1984}. It is interesting to note that only a few of these instabilities have ever been identified in actual observations and in the simulations even though theoretically they should be present. This can have several reasons, too small growth rates, too strong Landau damping, for instance, in the presence of hot ions, convective losses or very quick saturation due to heating effects, competition with other waves or wave-wave interaction and so on. 

The ion-ion instability \citep{Papadopoulos1971,Wu1984} generates waves in the whistler/lower hybrid frequency range $\omega_{ci}\ll\omega\omega_{ce}$\index{waves!whistler}\index{waves!lower hybrid} as we have discussed in Chapter 2. Its energy source is the beam-beam free energy of two counter streaming ion beams, one the reflected ion beam, the other the inflow. As long as the wavelengths re shorter than the reflected ion beam gyroradius the instability is relatively high frequency and electrostatic close to perpendicular and at the lower hybrid frequency. However, at slightly longer wavelength the magnetisation of the ions comes into play and the instability generates electromagnetic  whistlers. These are the whistlers which are observed at angles larger than the critical whistler angle mentioned earlier and probably also in the two-dimensional simulation case by \cite{Hellinger2007} for the parameters used there. 
The most important ion-driven instability in the foot region of quasi-perpendicular supercritical shocks has is the whistler instability which we have mentioned already several times. It is related either to the reflected ion beams or to the assumed temperature anisotropies \citep[as assumed to exist -- though never been confirmed by observations -- by][]{Wu1984}, or to result from diamagnetic density-gradient drifts in the lower-hybrid band as the electromagnetic branch of the lower-habrid/modified two stream instability. the electrostatic part of which we will discuss in the next section on electronic instabilities and electron dynamics. 

\cite{Hellinger2007}, in their two-dimensional hybrid simulations\index{simulations!two-dimensional hybrid} noted above, have seen the evolution of whistlers without identifying their sources. Recently \cite{Scholer2007} performed an extensive  parametric search for the whistler waves in the foot region of oblique shocks between $60^\circ\leq\thetabn\leq80^\circ$, the region where in one-di\-men\-si\-o\-nal PIC simulations intense whistlers should become excited theoretically when reflected ions are present. This is the case for the more oblique but still quasi-perpendicular supercritical shocks. A wide range of Alfv\'enic Mach numbers was used, and strong excitation of whistlers in the parametric range was found. We will discuss these observations here in more detail as they are the currently existing best available simulation results representing the current state of the art in the field of whistler excitation in connection with the formation, stability and time dependence of supercritical shocks at the time of writing this review.

Before discussing their results we briefly review the physics involved in the importance of whistlers in shock foot stability as had already been anticipated by \cite{Biskamp1972} following a suggestion by \cite{Sagdeev1966}. As we have mentioned earlier, a linear whistler wave precursor can stand in front of the quasi-perpendicular shock as long as the Mach number ${\cal M}<{\cal M}_{\rm wh}$ is smaller than the critical whistler Mach number
\begin{equation}
{\cal M}_{\rm wh}=\frac{1}{2}\left(\frac{m_i}{m_e}\right)^{\!\!\frac{1}{2}}\!\!|\cos\thetabn|\quad {\rm or~nonlinearly} \quad \frac{{\cal M}_{\rm wh, nl}}{{\cal M}_{\rm wh}}=\sqrt{2}\left[1-\left(\frac{27\beta}{128{\cal M}_{\rm wh}}\right)^{\!\!\frac{1}{3}}\right]
\end{equation}
The second expression result when including the nonlinear growth of the whistler amplitude during the steepening process \citep[for the derivation of this expression see, e.g.,][]{Kazantsev1963,Krasnoselskikh2002}. The nonlinear critical whistler Mach number ${\cal M}_{\rm wh, nl}$ is larger by a factor of $\sqrt{2}$ than the whistler Mach number ${\cal M}_{\rm wh}$. It depends weakly on the plasma-$\beta$ which has a decreasing effect on it, slightly reducing the whistler range.  

For instance, with realistic mass ratio $m_i/m_e=1840$ and $\thetabn=87^\circ$ the whistler critical Mach number is quite small, ${\cal M}_{\rm wh}\simeq 1.14$. In all smaller Mach number simulations no standing whistlers can thus be expected. It has also been speculated that above the above critical whistler Mach number the shock ramp is replaced by a nonlinear whistler wave train with wavelength of the order of $\lambda_e$. Approximating such a train as a train of whistler solitons one realises that the amplitude of the solitons increases with Mach number. Hence, the critical whistler Mach number in this case must become dependent on the whistler amplitude \citep{Krasnoselskikh2002}. This leads to the slightly larger nonlinear whistler critical Mach number ${\cal M}_{\rm wh, nl}$ on the right in the above equation.\index{Mach number!nonlinear whistler} 

Between the two Mach numbers ${\cal M}_{\rm wh}\leq {\cal M}\leq{\cal M}_{\rm wh, nl}$ the nonlinear whistler soliton train can exist attached to the ramp. However, when the nonlinear whistler Mach number is exceeded this is not possible anymore, and the whistler wave should turn over due to a so-called gradient catastrophe leading to nonstationarity of the shock front, which we will discuss later.\index{soliton!whistler train}
In the simulations of \cite{Biskamp1972} the simulation range was in favour of the excitation of whistlers which have also been seen and by these authors had been attributed to a nonlinear instability between the two ion beams and the electric field of a standing whistler wave.\index{waves!phase locked whistlers}
\begin{figure}[t!]
\hspace{0.0cm}\centerline{\includegraphics[width=1.0\textwidth,height=0.4\textheight,clip=]{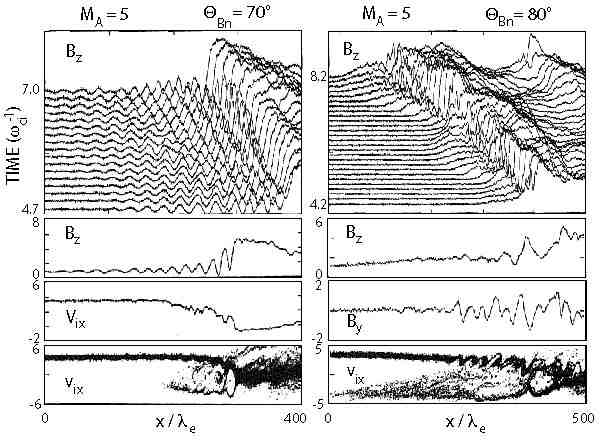} }
\caption[Simulation of whistler instability in shock foot]
{\footnotesize One-dimensional full praticle PIC simulations with realistic mass ratio for the same Mach number but different angle $\thetabn=70^\circ\,\,{\rm and}\,\,80^\circ$  \citep[after][]{Scholer2007}. {\it Left top}: Mach number ${\cal M}<{\cal M}_{\rm wh}$. The shock foot region is filled with waves of two polarizations, one of the the expected standing whistler waves which interfere with the other kind of waves. No substantial shock reformation is observed in this case.  {\it Right top}: Here the Mach number is in the range Mach number ${\cal M}_{\rm wh}<{\cal M}<{\cal M}_{\rm wh, nl}$. Two reformation cycles are visible during the simulation run in the magnetic field, and no whistler waves occur because the Mach number exceeds the first whistler Mach number. However there is also no nonlinearly steepened whistler in the shock front which is simply taken over by the foot after one reformation cycle. {\it Bottom three panels}: On the left shown the standing whistler oscillations in the magnetic field on the left, the decrease in the flow velocity when entering the foot due to the whistler scattering of the ions, and the particle phase space in $v_(ix)$ with the reflected beam and the scattered foot ions with little vortex formation such that the reformation is inhibited. On the right seen is the more irregular structure of the transition, the component $B_y$ in the magnetic field and the accelerated foot ions in the absence of whistlers. }\label{chap4-fig-scho-5}
\end{figure}

\cite{Scholer2007} perform PIC simulations with physical Mass ratio. For all shocks with $\thetabn\leq 83^\circ$ the whistler critical Mach number is well above the critical Mach number such that the shock is supercritical. This is in order to check the excitation of whistlers in the different regimes of ${\cal M}$. In the left part of  Figure\,\ref{chap4-fig-scho-5} the Mach number is below the critical whistler Mach number, and in the shock foot region a group of phase locked whistler waves is excited with increasing amplitudes towards the shock ramp. This is nicely seen on the left in both the magnetic stack plot as also in the time profile in the second panel on the left from top. The whistlers slow the incoming flow $V_{ix}$ down before it reaches the ramp. In the phase space plot the incoming and reflected beams are seen as is the scattering and trapping of the resonant ions in the whistler waves.
\begin{figure}[t!]
\hspace{0.0cm}\centerline{\includegraphics[width=0.8\textwidth,height=0.4\textheight,clip=]{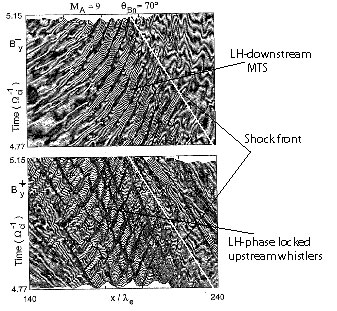} }
\caption[Different whistler helicities in shock reformation]
{\footnotesize One-dimensional full particle PIC simulations with realistic mass ratio for  Mach number ${\cal M}_A=9$ and $\thetabn=70^\circ$ in the non-reformation whistler regime  \citep[after][]{Scholer2007}. {\it Top}: Negative helicity waves $B_y^-$ propagating to the right are left-hand polarised short wavelength waves moving downstream toward the shock and being mostly absorbed in the shock transition.  {\it Bottom}: Positive helicity waves $B_y^+$. These waves move upstream and thus are also left-hand circularly polarised waves. They move at shock velocity which identifies them as the phase-locked standing whistler precursors in the shock-upstream region with decaying upstream amplitude and long wavelength. Some interference is seen on these waves. Their left-hand polarisation identifies them as ion-beam excited whistlers and not as electron temperature anisotropy exited whistlers. }\label{chap4-fig-scho-6}
\end{figure}

\indent The right part of the figure shows an identical simulation with Mach number above the critical whistler Mach number but below the nonlinear whistler Mach number. The stack plot shows two well developed reformation cycles with all signs of normal reformation. The magnetic field profile chosen in the second panel is at time $t\omega_{ci}=7.6$ in the second reformation cycle when the foot loop is well developed. The magnetic field signature shows the distortion due to the foot which is caused by a large amplitude non-phase-standing whistler wave that evolves nonlinearly. However, the reformation is not due to this whistler but due to the accumulation of gyrating ions at the foot edge as known from previous simulations. 

Near the ramp the ions become trapped in a large-amplitude whistler loop, as is seen in the phase space plot. The loop coincides with a minimum in the $B_z$-component of the magnetic field. The whistler, on the other hand, does practically not affect reformation, even though it structures the overshoot region. Reformation time is defined thus, as we know already, by the gyration time of ions in the foot, being of the order of a few ion-cyclotron periods. When the shock becomes more oblique, the whistler effect increases again at fixed Mach number as the Mach number enters the domain below critical whistler Mach number ${\cal M}_{\rm wh}$, and the shock transition becomes much more structured and reformation less important.

In order to see what kind of waves are excited during the whistler cycles, a separation of the magnetic wave spectrum $B_y$ into positive $B_y^+$ and negative $B_y^-$ helicity components has been performed for a ${\cal M}_A= 9$, $\thetabn = 70^\circ $ simulation run. Figure\,\ref{chap4-fig-scho-6} shows the result. The negative helicity waves $B_y^-$ propagate toward the shock, i.e. to the right. After correcting for the convection velocity which is also to the right, these waves are left-hand polarised waves. The lower panel shows positive helicity $B_Y^+$ waves propagating the the left, so they are upstream propagating waves and are also left-hand polarised. The positive helicity waves have longer wavelength than negative helicity waves. They propagate close to shock speed upstream. They are thus almost standing in the shock frame. These are the phase-locked upstream left-hand polarised (ion beam and not electron temperature anisotropy driven) whistlers.

The downstream propagating negative helicity waves are no whistlers. They are caused in quite a different way which is related to the electromagnetic modified two-stream instability\index{instability!modified-two-stream} which we will discuss in the next section on electron waves.
\begin{figure*}[t!]
\hspace{0.0cm}\centerline{\includegraphics[width=1.0\textwidth,clip=]{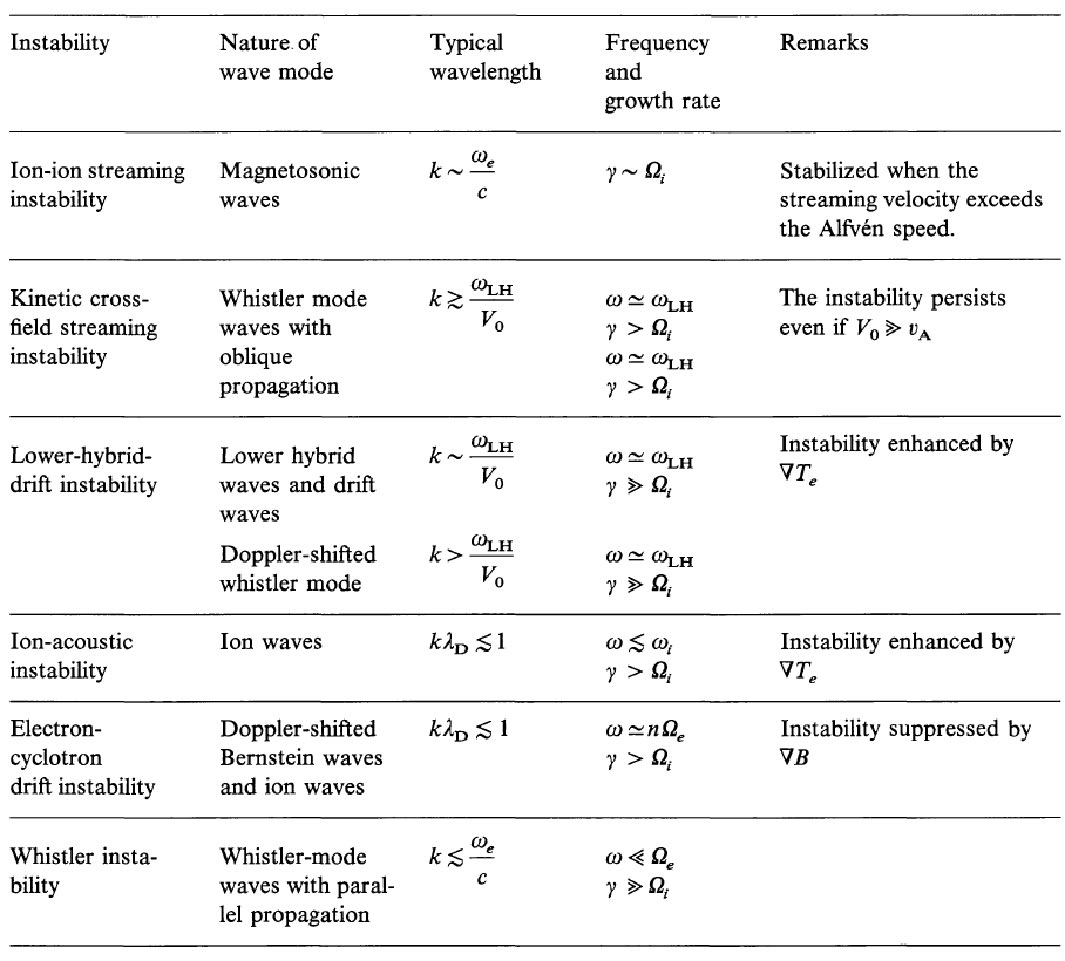} }
\caption{Instabilities in the Ramp Transition \citep[copied from][]{Wu1984}.}\label{Table2}
\end{figure*}

\subsubsection{Ramp transition waves}

\noindent Stability of the ramp is a question that is not independent of the stability of the foot as both are closely connected by the reformation process of the quasi-perpendicular shock front. Since the suggestion of \cite{Sagdeev1966}  is widely believed that the whistler waves excited in the foot are the main responsible for the stability and steepening of the foot. In fact, they might accumulate their, store energy in both magnetic and electric field, trap particles and excite different waves. These processes are still barely understood and not sufficiently investigated neither theoretically nor by simulation studies. A very early and to large extent out of time list is given in Figure\,\ref{Table2}  of the theoretical expectations for possible instabilities in the shock ramp region \citep{Wu1984}. As in the case of the foot region, the instabilities in the ramp which might be of real importance have turned out to not fit very well into the scheme of this listing. In both cases, the case of the foot and the case of the ramp, this disagreement reflects the weakness of theoretical speculation which is not supported by direct observations on the one hand and clear parametric searches in numerical experiments on the other. 
\begin{figure}[t!]
\hspace{0.0cm}\centerline{\includegraphics[width=0.75\textwidth,height=0.45\textheight,clip=]{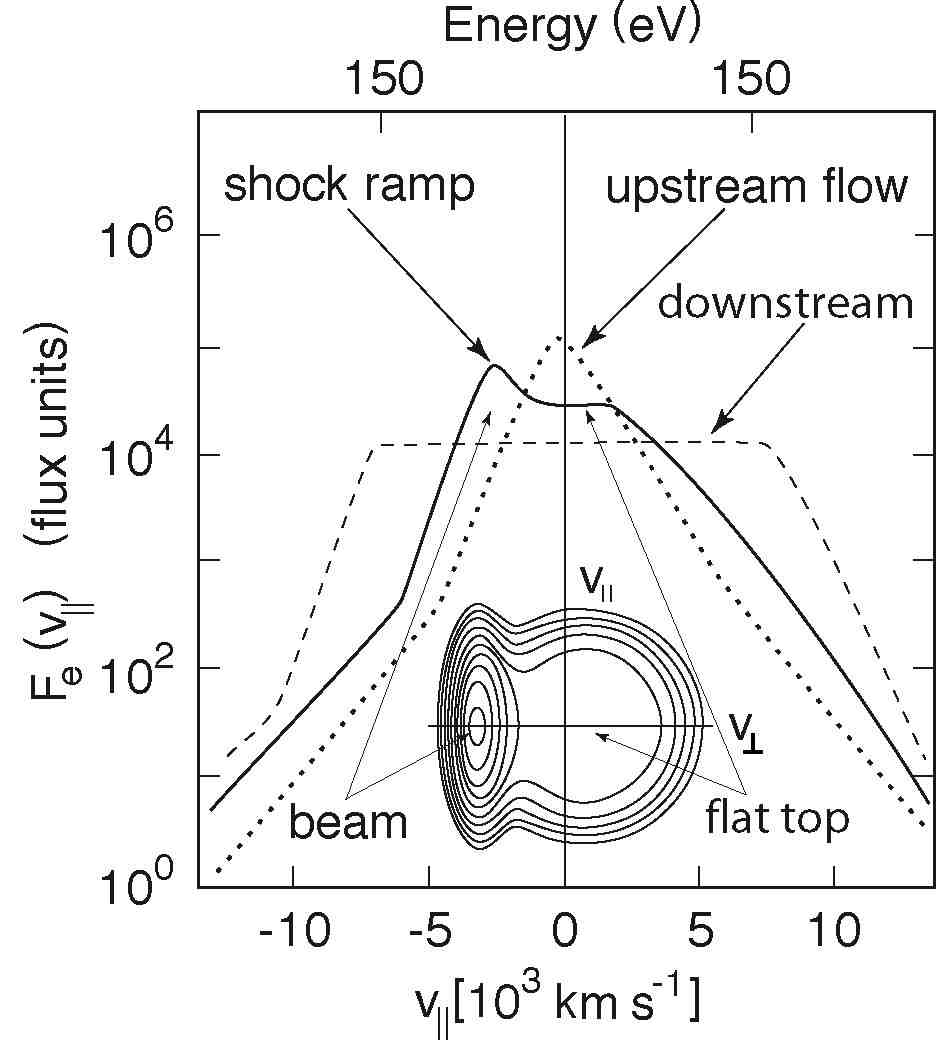} }
\caption[Shock electron distributions]
{\footnotesize Several successive reduced parallel electron distribution functions $F_e(v_\|)$ during the crossing of the supercritical bow shock of the Earth by ISEE 2 on December 13, 1977. The cuts through the distribution show the transition from the Maxwellian-plus-halo upstream flow distribution through the shock ramp distribution to the close to the shock downstream distribution. The shock ramp distribution is intermediate in evolving into a flat-top distribution of the kind of the downstream distribution but contains in its upstream directed part a well expressed shock-reflected electron beam of velocity of a few 1000\,km\,s$^{-1}$ which is sufficiently fast to destabilise the shock front and excite electron plasma waves  \citep[after][]{Gurnett1985}.}\label{chap4-fig-gurnfeld}
\end{figure}

Waves excited or existing in the ramp cannot be considered separate from the stability of the shock ramp. They are mostly related to electron instabilities and will to some extent been considered there in the next section. On the other hand they are also related to the non-stationarity of a shock ramp. We will therefore return to them also in the respective section on the time dependence of evolution of shock ramps and their stability. It is, however, worth mentioning that recently very large electric fields have been detected during passages of the Polar satellite across the quasi-perpendicular bow shock when the spacecraft was traversing the shock ramp. These observations showed that in the shock ramp electric fields on scales $\lesssim \lambda_e=c/\omega_{pe}$ exist with amplitudes of the order of several 100\,mV\,m$^{-1}$. These are amongst the strongest localised electric fields measured in space \citep{Bale2007}. Clearly, these localised fields are related to the electron dynamics in the shock and in particular in the shock overshoot/ramp region. Excitation of intense electron waves in the shock ramp has been expected for long time already since the observation of the (reduced) electron distribution across the shock from a drifting to a flat-topped distribution \citep{Feldman1983}. Figure {chap4-fig-gurnfeld}, taken from \cite{Gurnett1985}, shows this transition. The interesting point is that right in the ramp/overshoot region the reduced electron distribution function shows the presence of an electron beam in addition to an already quite well developed flat top on the distribution. Such a beam will almost inevitably serve as the cause of instability.

\section{Electron Dynamics}\noindent 
When talking about the dynamics of electrons in quasi-perpendicular, oblique, or quasi-parallel shocks, hybrid simulations cannot be used anymore. Instead, one must return to the more involved full particle PIC simulation codes or to Vlasov codes, which directly solve the Vlasov equation in the same way as a fluid equation, this time, however, for the ``phase space fluids" of ions and electrons. In both cases short time scales of the order of the electron gyro-period $\omega_{ce}^{-1}$ or even the electron plasma period $\omega_{pe}^{-1}$ must be resolved, and resolution of spatial scales of the order of the electron inertial scale $\lambda_e$ is required. It is thus no surprise that reliable simulations of this kind which also resolve the electron dynamics became available only within the last decade with the improved computing capacities. 
\begin{figure}[t!]
\hspace{0.0cm}\centerline{\includegraphics[width=0.95\textwidth,height=0.45\textheight,clip=]{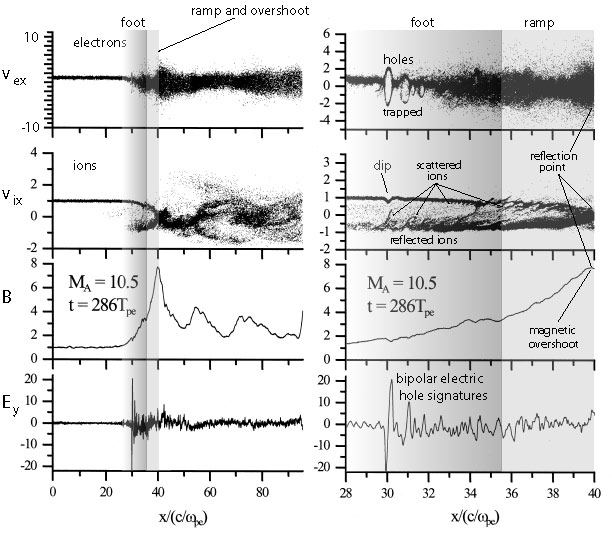} }
\caption[Electron heating in shocks(Shimada \& Hoshino (2000)]
{\footnotesize One-dimensional full particle PIC simulations with mass ratio 20 for  Mach number ${\cal M}_A=10.5$ and $\thetabn=90^\circ$ resolving the electron scales  \citep[after][]{Shimada2000}. {\it Left}: Simulation overview for electron and ion phase spaces, magnetic field and electric field. Ion reflection from the ramp and foot formation is seen in the second panel from top. The electrons are heated in the foot region. the heating coincides with large amplitudes in the electric field in the lowest panel. {\it Right}: Expanded view of the shaded foot and ramp regions on the left. The electron heating location turns out to be a site of electron hole formation. Three Buneman two-stream instability\index{instability!Buneman} holes are nicely formed on this scale with trapped electrons. The broadening of the distribution and thus heating is due to the holes. Retardation of the ions in interaction with the holes is seen in the second panel which is due to the retarding electric potential in the large amplitude electric field oscillations (lowest panel). Interestingly enough, ion reflection takes place in the very overshoot! The ion distribution is highly structured in the entire region which is obviously due to interaction with many smaller scale electron holes.}\label{chap4-fig-shim-1}
\end{figure}
\subsection{Shock foot electron instabilities}\noindent
\cite{Papadopoulos1988} proposed that in the foot region of a perpendicular highly supercritical shock the velocity differences between reflected ions and electrons from the upstream plasma inflow should be responsible for the excitation of the Buneman two-stream instability thus heating the electrons, generating anomalous conductivity and causing dissipation of flow energy which contributes to shock formation. 

\subsubsection{Buneman two-stream heating in strictly perpendicular shocks}\noindent \cite{Shimada2000} and \cite{Schmitz2002} building on this idea performed full particle PIC simulations in strictly perpendicular shocks discovering that the Buneman two-stream instability can indeed work in the foot region of the shock and can heat and accelerate the electrons. \cite{Shimada2000} initiated their one-dimensional simulations  for a small mass ratio of $m_i/m_e=20$, $\beta_i=\beta_e=0.15$, and Alfv\'enic Mach numbers $3.4\leq{\cal M}_A\leq 10.5$. 

Figure\,\ref{chap4-fig-shim-1} shows some of their simulation results. It is interesting to inspect the right part of the Figure which shows the (shaded) ramp and foot regions on the left in expanded view. The electron phase space shows the development of electron holes which are generated by the Buneman two stream instability in this strictly perpendicular shock simulation. The signature of the electrostatic field $E_y$ in the lowest panel shows the bipolar electric field structure the holes cause. The average field is zero, but in the hole it switches to large negative values, returns to large positive values and damps back to zero when passing along the direction normal to the hole. This is exactly the theoretical behaviour expected for both, solitons and electron holes\index{waves!BGK modes}\index{soliton!BGK holes} of the form of BGK modes. As known from simulations (see Chapter 2) such BGK-hole structures will trap electrons and heat them, they do, on the other hand, also accelerate passing electrons to large velocities. Both is seen here also in the simulations in the vicinity of the shock: Three such holes are completely resolved in the right high resolution part of the figure, with decreasing amplitude when located closer to the shock ramp. All of them contain a small number of trapped electrons over a wide range of speeds which on the gross scale on the left fakes the high temperature of the electrons they achieve in the hole. This is just the effect of heating by the two-stream instability. In addition the electron velocity shows two accelerated populations, one with positive velocity about 2-3 times the initial electron speed, the other reflected component with velocity almost as large as the positive component but in the opposite direction suggesting that the electron current in the holes is almost compensated by the electron distribution itself. 

Obviously the further strong heating of electrons in the ramp is caused by many smaller amplitude overlapping holes as is suggested by the structures in the inflowing and reflected ion distributions which do also strongly interact with the electric field of the holes. This is seen in the incoming ion component in the first hole as a dip in the velocity. The hole retards the incoming flow. It is also seen in the reflected ion component as strong distortion of their backward directed velocity when encountering a hole. The smaller speed ions are obviously retarded in their backward flow and are partially trapped in the negative electric field part of the hole. Very similar strong scattering of the incoming ion component is seen in the ramp region. This suggests that a large number of electric field structures are located in the ramp which scatter the incoming ions. These must be related to the highly fluctuating electric field component in the ramp seen on the right in the lowest panel. 

Two further observations which are related to the ion component are of considerable interest: The first is that the retardation of the incoming ion flow and the scattering of the reflected ions in the foot region cause a signal on the magnetic field component. The second is that the reflection of the main incoming ion beam, i.e. the incoming plasma takes place at the location of the magnetic overshoot and not in the shock ramp. Therefore, physically spoken, the shock ramp is also part of the foot, while it is the narrow overshoot region where the reflection occurs in a strictly perpendicular supercritical shock with cold ion inflow. The actual ramp region is much narrower than for instance shown in the figure. Its actual width is only of the order of $\Delta\sim (1-2) c/\omega_{pe}$. 
\begin{figure}[t!]
\hspace{0.0cm}\centerline{\includegraphics[width=0.95\textwidth,clip=]{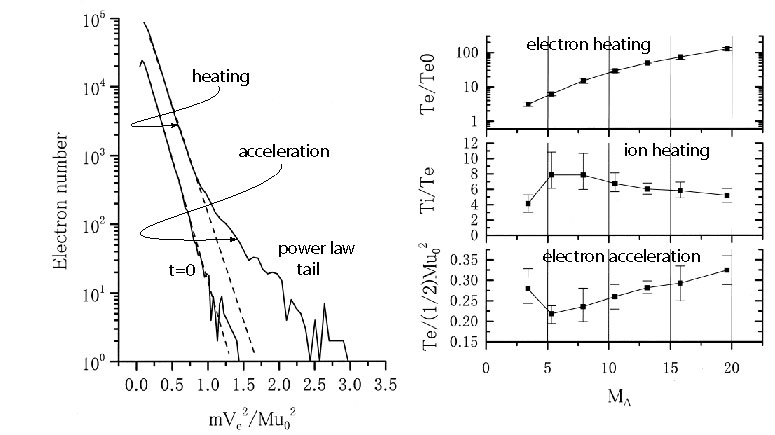} }
\caption[Electron Buneman distribution in shocks(Shimada \& Hoshino (2000)]
{\footnotesize  {\it Left}: The electron distribution in the shock arising from the action of the Buneman electron hole interaction. The interaction not only causes heating but also an energetic tail on the electron distribution function. This tail has the shape of a power law $F{\epsilon}\propto\epsilon^{-\alpha}$, with power $\alpha\approx 1.7$. Note that this power gives a very flat distribution close to marginal flatness $\alpha=\frac{3}{2}$ below that an infinitely extended distribution function has no energy moment.  {\it Right}: Evolution of the average electron temperature, ion temperature and ratio of electron temperature to initial kinetic energy in the simulations as function of Alfv\'enic Mach  number \citep[after][]{Shimada2000}. All quantities are in relative units of computation.}\label{chap4-fig-shim-2}
\end{figure}

\subsubsection{Electron heating and acceleration}\noindent
\cite{Shimada2000} followed the evolution of the electron vortices (holes) and showed that a hole once evolved distorts the ion and electron velocities in such a way that nonlinearly the velocity difference can increase and cause the generation of secondary vortices, which leads to excessive electron heating \citep[see also][]{Shimada2005}. The result is the generation of an extended electron tail on the electron distribution. This is seen from the left part of Figure\,\ref{chap4-fig-shim-2} in a log-lin representation of the electron number versus normalised electron energy. When plotting the data on a log-log scale (not shown) one realises that the newly produced tail of the electron distribution has a power law slope $F(\epsilon)\propto \epsilon^{-\alpha}$, notably with power $\alpha \approx 1.7$. (Note that this power is close to the marginally flattest power $\alpha=\frac{3}{2}$ below that an infinitely extended power law energy distribution has no energy moment more and thus ceases to be a distribution. In fact, any real nonrelativistic power law will not be infinitely extended but will be truncated due to the finite extent of the volume and loss of energetic  particles). In the right part of this figure the dependence of electron heating and ion cooling on Mach number for the investigated range of Alfv\'enic Mach numbers is shown.  The effect does not occur for small Mach numbers, too small for the Buneman two-stream instability to be excited. However, once excited, the heating increases strongly with ${\cal M}_A$. Over the range $5<{\cal M}_A<20$ the increase in electron temperature (electron energy stored mainly in the tail of the distribution)   is a factor of 40-50, which demonstrates the strong non-collisional but anomalous transfer of kinetic flow energy into electron energy via the two-stream instability. However, one should keep in mind that this result holds merely for a one-dimensional simulation of strictly perpendicular shocks.

At this place we should look again at real observations during crossings of real collisionless shocks in space. Recently, during passages of the Polar satellite\index{spacecraft!Polar} of the quasi-perpendicular Bow Shock of the Earth, very strong localised electric fields have been detected. These fields exist on scales $\lesssim \lambda_e=c/\omega_{pe}$, less  than the electron skin-depth, and reach enormous values of $\lesssim 100$\,mV\,m$^{-1}$ parallel and $\lesssim 600$\,mV\,m$^{-1}$ perpendicular to the magnetic field. They must naturally be related to the electron dynamics and should play a substantial role in the formation and dissipation processes of the quasi-perpendicular shock. They should also be of utmost importance in accelerating electrons and possibly also ions at shocks. Their nature still remains unclear, however, it is reasonable to assume that they are generated by some electron current instability via either the Buneman-two-stream instability, the modified two-stream instability which we discuss below, or the ion-acoustic instability, depending on the current strength. In any case they will turn out to belong to the family of Bernstein-Green-Kruskal modes which are encountered frequencly in collisionless plasmas. 

\subsection{Modified-two stream instability in quasi-perpendicular shocks}\noindent The Buneman two-stream instability works on scales short with respect to the gyroradii, in paritular on an electron scale $\leq \lambda_e=c/\omega_{pe}$. This condition is less easily satisfied in quasi-perpendicular shocks. However, here other instabilities can evolve which are relatives of the Buneman two-stream instablity. 

The condition that there is no current flowing in the shock normal direction during foot formation and reflection of ions at the shock requires that the electron inflow from upstream is decelerated when entering the foot region. This causes a difference in the flow velocities between the incoming ions and the incoming electrons. In a quasi-perpendicular shock where the configuration is such that the the wave vector ${\bf k}=(k_\|,{\bf k}_\perp)$ is allowed to have a component $k_\|$ parallel to the magnetic field, the velocity difference between ions and electrons can lead to the excitation of the modified two-stream (MTS) instability, which is a modification of the Buneman instability acting in the direction perpendicular to the magnetic field (see Chapter 2). This instability is an electromagnetic instability that couples the Buneman two-stream instability to the whistler mode. The waves generated propagate on the whistler mode branch with frequency $\omega_{\rm mtsi}\sim\omega_{lh}\ll \omega_{ce},\omega_{pe}$, close to the lower-hybrid frequency, but far below both the electron cyclotron and electron plasma frequencies, respectively. These waves, being electromagnetic whistler-like are thus capable to even modify the shock profile when being swept downstream from the foot region towards the shock ramp. In addition, the obliqueness of the shock generates a magnetic field aligned electric field component of the wave which leads to acceleration and trapping and eventually also pre-heating of the electrons in the shock foot parallel to the magnetic field. 
\begin{figure}[t!]
\hspace{0.0cm}\centerline{\includegraphics[width=0.65\textwidth,clip=]{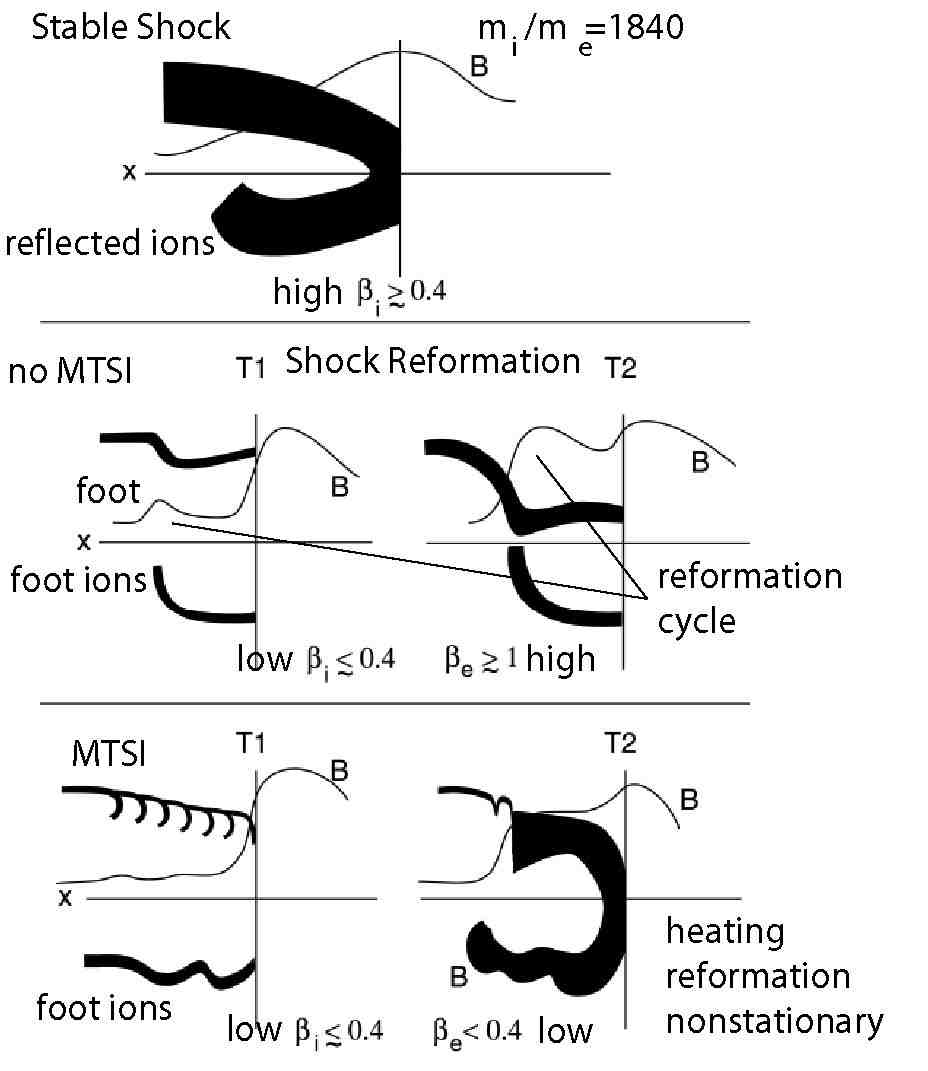} }
\caption[Schematic of shock dynamics without and with MTSI]
{\footnotesize  Schematic of the dependence of the shock structure on the combinations of $\beta_i, \beta_e$ for quasi-perpendicular supercritical but non-whistler shocks. For large $\beta_i$ the shock is stable even though ions are reflected. At small $\beta_i$, large $\beta_e$ the shock reforms due to accumulation of ions at the edge of the foot forming n reformation cycle. For small $\beta$ the MTSI evolves in the foot, strong heating and complicated dynamics evolves due to nonlinear interaction, heating and hole-vortex  formation \citep[after][]{Scholer2004}. }\label{chap4-fig-schoma04-1}
\end{figure}

\subsubsection{Relation to the Buneman Instability}\noindent
\cite{Scholer2004} investigated the transition from Buneman to modified two-stream (MTS) instabilities as function of mass ratio $m_i/m_e$ and for various $\be_i,\beta_e$ in the regime where no upstream standing whistlers exist, i.e. above the critical whistler Mach number ${\cal M}_A>{\cal M}_{\rm wh}$. This investigation is restricted to oblique shocks, however with ${\bf k}$-vectors being strictly perpendicular to the shock along the shock normal and for one-dimensional simulations only. This excludes any waves which could propagate along the magnetic field into the inclined direction. Nevertheless, this investigation is interesting in several respects. First it showed that for mass ratios $m_i/m_e\lesssim 400$ no modified two stream instabilities occur since their growth rates are small. The electron dynamics and the shock behaviour in this range are determined by the Buneman two-stream instability unless the electron temperature\index{instability!modified-two-stream} is large enough to inhibit its growth in which case ion-acoustic instability should (or could) set on (but has not been observed or has not been searched for). For larger mass ratios (and particularly for the realistic mass ratio) the Buneman two stream instability ceases to be excited. Instead, the modified two-stream instability (MTSI) takes over which is strong enough to completely determine the behaviour of the electrons. A summary of their results is schematically given in Figure\,\ref{chap4-fig-schoma04-1}.  
\begin{figure}[t!]
\hspace{0.0cm}\centerline{\includegraphics[width=0.9\textwidth,clip=]{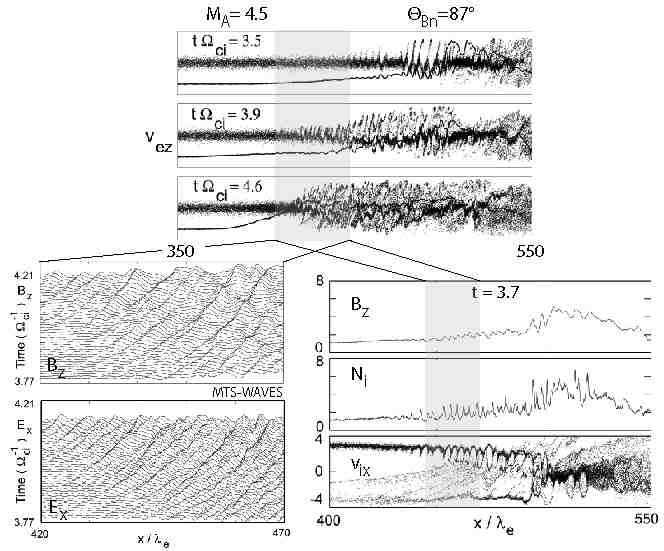} }
\caption[Schematic of shock dynamics without and with MTSI]
{\footnotesize  {\it Top}: Electron phase space evolution showing the distortion of the electrons until thermalisation during the modified two-stream instability. {\it Lower left}: Magnetic and electric wave components of MTSI waves present in the grey shaded area in space during part of the time shown and are moving toward the shock ramp in the left-hand mode as discussed earlier. These waves steepen when reaching the shock front. Distortion of the ion distribution is the result as shown in {\it Lower right}.  \citep[after][]{Scholer2004}. }\label{chap4-fig-schoma04-2}
\end{figure}

Figure\,\ref{chap4-fig-schoma04-2} shows the evolution of the MTS-waves \index{interactions!wave-wave}as seen in the full particle real mass ratio simulations. Three instants of time in electron phase space are shown on the top. In the shaded area and during an interval out of this time the wave spectrum has been determined from the magnetic and electric field components, which is shown in the bottom parts of the figure. Large amplitude waves of left-hand polarisation are propagating toward the shock during this reformation cycle. The waves are similar to those we have already discussed, but this time it is seen that these waves are related to the electron dynamics. They are excited by the modified two-stream instability in the foot in interaction between the retarded electrons and the fast ions. 

\subsubsection{Modified two-stream instability and quasi-perpendicular shock reformation}\noindent They cause reformation of the shock, but in a different way than it is caused for low mass ratio  by the Buneman-instability. There the reformation was the result of accumulation of ions at the upstream edge of the foot,  while here it is caused by participation of the foot ions in the MTSI all over the foot and particularly close to the shock ramp and presumably also at the ramp itself. Phase mixing of the ions leads to bulk thermalisation and formation of a hot retarded ion component in the foot region which has similar properties like the downstream population and, when sufficiently compressed takes over the role of the shock ramp. This can be seen from the lower right part of Figure\,\ref{chap4-fig-schoma04-2} which is a snapshot at time $t\omega_{ci}=3.7$  showing the magnetic profile, the density profile with its strong distortions, and the evolution of the ion distribution which evolves into large thermalised vortices towards the front of the shock (note that the shading indicates here also the spatial domain where the wave spectra have been taken).
\begin{figure}[t!]
\hspace{0.0cm}\centerline{\includegraphics[width=1.0\textwidth,clip=]{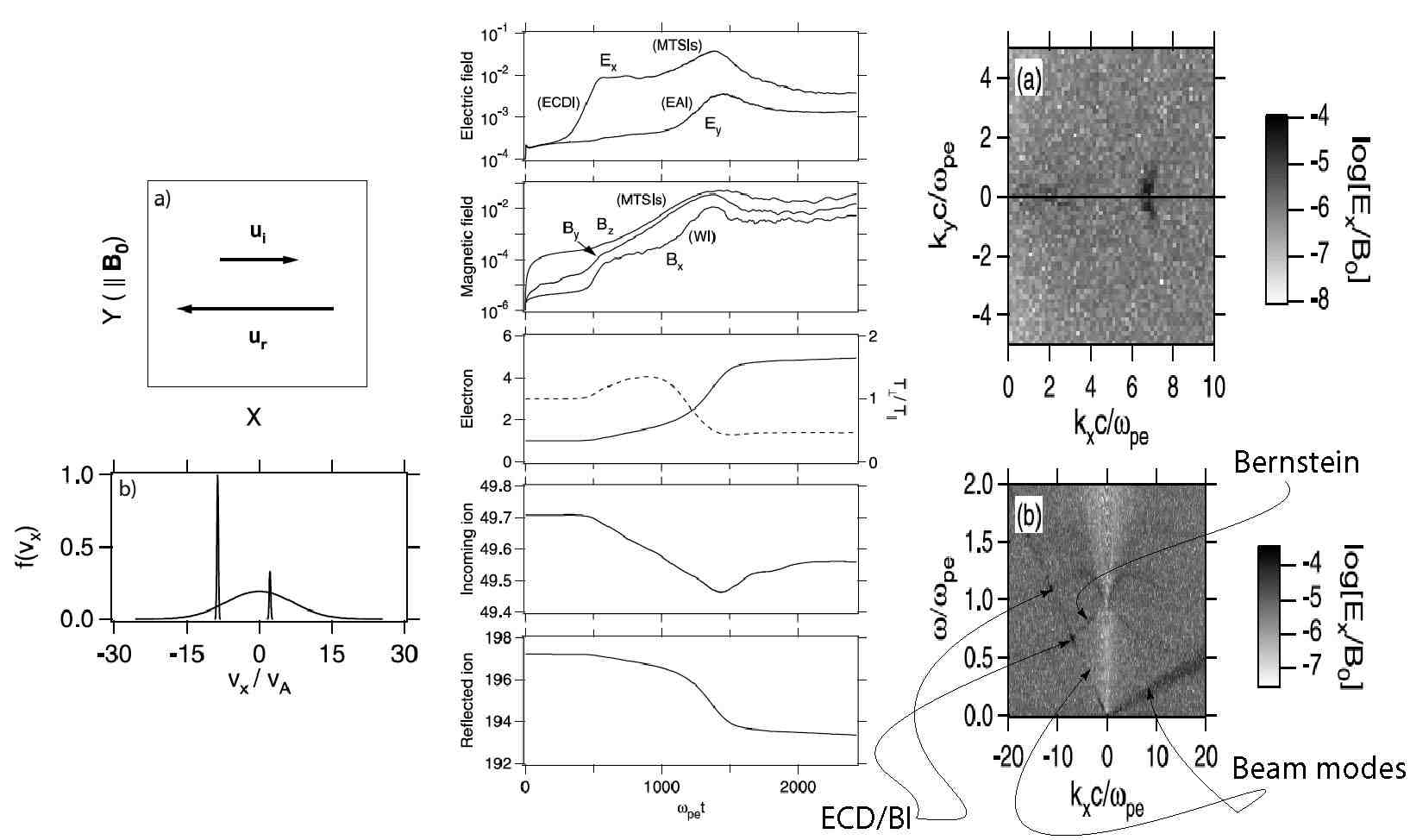} }
\caption[Simulation of MTSI]
{\footnotesize  {\it Left}: The phase space distribution set-up for the simulation. The original magnetic field is in $z$-direction. The upper panel shows the incoming and reflected ion beams. The lower panel shows the two cold ion distributions, incoming and reflected, and the hot electron distribution, shifted slightly in order to satisfy the zero-current condition in shock-normal direction. {\it Centre}: Time histories of the energy densities of the simulation quantities: electric and magnetic wave fields, electrons and the two ion components. {\it Right}: Wave power spectra in ${\bf k}$-space at early times $t\omega_{pe}<404.8$ showing the excited power in the Buneman mode in the upper panel. The lower panel shows the dispersion relation. The two straight lines correspond to the damped beam modes of the reflected (negative slope) and direct (positive slope) ion  beams. The enhanced power in the two dark spots is due to the ECD-instability, which is the Buneman mode which excited under these early conditions in the simulation as the interaction between the reflected ion beam mode and the first and second Bernstein harmonic waves  \citep[after][]{Matsukiyo2006a}. }\label{chap4-fig-matsu-1}
\end{figure}

The generation of MTS-waves by the modified two-stream instability has been investigated in depth theoretically and with the help of specially tailored one-dimensional numerical simulation studies by \cite{Matsukiyo2003}, and in two-dimensional simulations by \cite{Matsukiyo2006a,Matsukiyo2006b} which we are going to discuss in detail.

\begin{figure}[t!]
\hspace{0.0cm}\centerline{\includegraphics[width=1.0\textwidth,clip=]{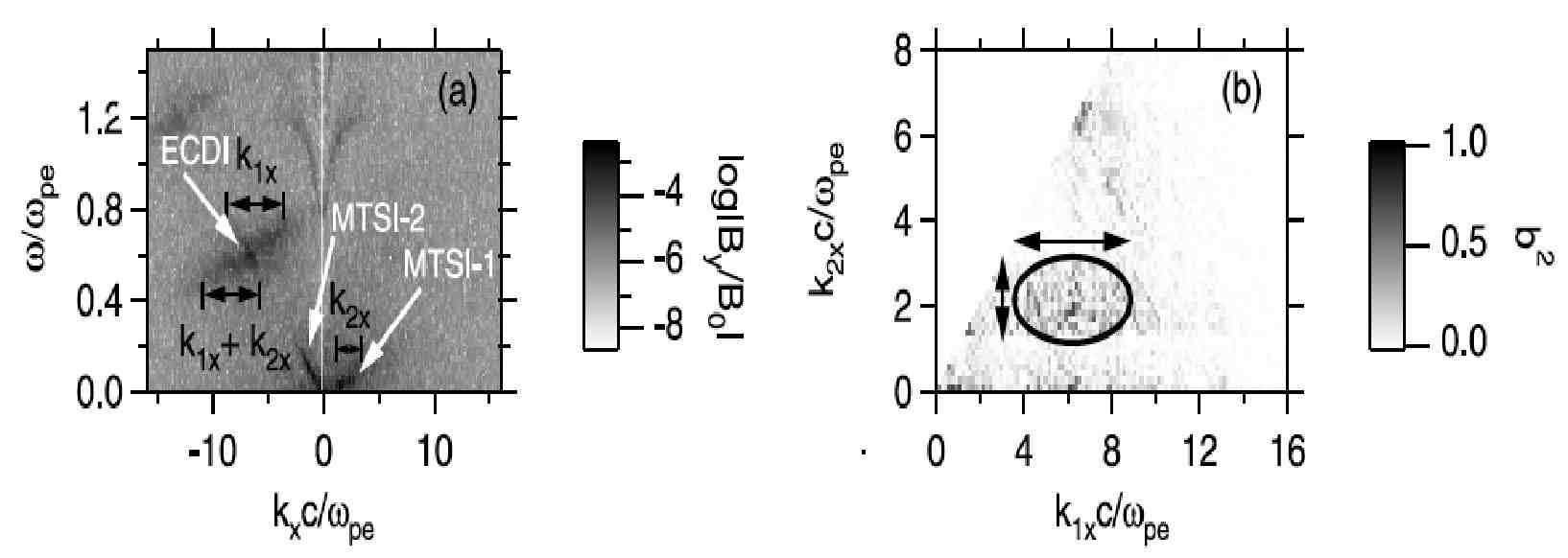} }
\caption[Dispersion relation from simulated MTSI]
{\footnotesize  {\it Top}: The dispersion relation for the time interval $607.2<t\omega_{pe}<1011.9$ showing the ECDI (Buneman mode), the original MTI-1 and the secondary MTSI-2 which is generated by wave-wave interaction. The corresponding reaction in $k_x$ numbers is indicated for the waves which participate in the three-wave process.. {\it Bottom}: The power spectral density in the $(k_x,k_y)$-plane. The ellipse indicates the wave numbers that contribute to the wave-wave interaction of the MTSI-1 and ECDI  \citep[after][]{Matsukiyo2006a}. }\label{chap4-fig-matsu-2}
\end{figure}

\paragraph{Modified two-stream generation mechanism: Tailored simulations.}
Figure\,\ref{chap4-fig-matsu-1} in its left-hand parts shows the set-up of the two-dimensional simulation and the resulting time histories of fields and particles. The incoming and reflected ion velocities are shown for time zero in the $(x,y)$-plane where the co-ordinate $y$ is about parallel to the magnetic field. The phase space at time zero contains the three distributions of inflow and reflected ions and hot incoming electrons. The slight displacement between the latter and the incoming ions accounts for zero normal current flow in presence of reflected ions. Clearly this configuration is unstable causing instabilities between the ion beams and electrons (in addition to the slowly growing ion-ion instabilities discussed earlier). 
The basic physics of the instability can be readily identified from the time histories of the fields and particles in the middle of Figure\,\ref{chap4-fig-matsu-2}. The first exponential growth phase of the $E_x$-component for times $\omega_{pe}t<500$ is due to the Electron-Cyclotron-Drift instability (ECDI) which we have omitted in our theoretical analysis in Chapter 2 \citep[cf., also,][]{Muschietti2006}. This instability is driven by the ion beam when it interacts with obliquely propagating electron-Bernstein waves (electron-cyclotron waves). In fact, this instability, in the present case is nothing else but the Buneman instability (BI) which\index{instability!electron-cyclotron-drift} for the given set-up is initially unstable (as is seen from the bulk velocity difference between the ion and electron phase space distributions on the left of the figure) due to the interaction of the ion beam mode with the lowest order electron-cyclotron mode. Initially there is some growth also in the magnetic field which is strongest in $B_z$ and much weaker in $B_y$ and $B_x$. However, until the MTSI sets on the magnetic field energy does not grow substantially. This changes with onset of the MTSI when all components increase with $B_y,B_z$ dominating and being of equal intensity, showing that due to the magnetic wave field of the MTSI the instantaneous magnetic field develops a transverse component.  

The MTSI does not grow in this initial state because its growth rate is small for these conditions. So during the saturation phase the ECDI does still dominate in the flat regime until the MTSI takes over and causes further growth of the already large amplitude electric field fluctuations. This stage  after $\omega_{pe}t>10^3$ is characterised by a growth phase also in $E_y$ (which is due to the electron acoustic instability EAI which can be excited in presence of both a cold and a hot electron component) and, surprisingly, the normal component $B_x$. This latter component might be caused by the Weibel instability (WI) when a substantial anisotropy is generated. Such an anisotropy exists for the ions, in fact, in our case as they propagate solely in $\pm x$-direction at grossly different speeds. The growth rate of the instability, neglecting the magnetic field, i.e. setting $\omega_{ce}=0$, is $\gamma_{\rm WI}=(V_i/\lambda_i)(1+1/k^2\lambda_e^2)^{-\frac{1}{2}}$ \citep{Weibel1959}, \index{instability!Weibel}where $\lambda_{e,i}=c/\omega_{pe,i}$ are the ion and electron  inertial lengths, respectively. When the magnetic field is not neglected but the ions are taken as non-magnetised, as is the case in the shock foot, then
\begin{equation}
\gamma_{\,\,\rm WI}=\frac{V_i}{\lambda_i}\left(1+\frac{1}{k^2\lambda_i^2}\right)^{-\frac{1}{2}} \qquad{\rm for}\quad \left(1+\frac{1}{k^2\lambda_i^2}\right)\frac{\omega_{ce}^2}{\omega_{pi}^2}\gg\frac{V_i^2}{c^2}\left(1+\frac{1}{k^2\lambda_e^2}\right)
\end{equation}
At short wavelengths the growth rate of this instability can be quite large. Its maximum is assumed for $k^2\to\infty$ when it becomes the order of $(\gamma_{\,\,\rm WI}/\omega_{ci})_{\rm max}\sim V_i/V_A\simeq {\cal M}_A$. 

At the expected wavenumber $k\lambda_i\sim 1$ is is just a factor $\sqrt{2}$ smaller than its maximum value and decreases rapidly towards longer wavelengths. One may thus expect that large Mach number shocks generate magnetic fields by the Weibel instability, in which case the field becomes non-coplanar, and small-scale stationary magnetic structures appear in the shock foot and ramp. Still, this is a little speculative. However, if the Weibel instability exists it will generate many small-scale magnetic structures in the shock. This is, in itself, sufficiently interesting to be noted. The simulations show the presence of $B_x\neq 0$, suggesting that the magnetic field becomes three-dimensional since the Weibel instability has zero frequency and thus produces a steady normal field component. 

The right outermost part of the figure shows two power spectra of the electric field in $(k_x,k_y)$-space at times $t\omega_{pe}=253$ (top) and $t\omega_{pe}<404.8$ (bottom). In the top panel the power of the waves concentrates at $(k_x\lambda_e,k_y\lambda_e)=(6.8,0)$. These waves propagate nearly perpendicular to the ambient magnetic field. 

The lower panel shows the dispersion relation $\omega(k_x)$ for these waves in a grey scale representation. The two straight dark lines with negative and positive slopes belong to the damped ion beam modes for the reflected (negative slope) and incoming (positive slope) ion beams. 

There are two dark specks on the reflected beam mode where the intensity of the electric field (which is shown here only) is enhanced. These specks are separated by about the electron cyclotron frequency in frequency. They belong to the crossings of the reflected ion beam mode and the two lowest harmonics of the electron Bernstein modes which is the ECD-instability which in this case is also the Buneman instability (BI).  This mode has been investigated also by \cite{Muschietti2006} in one-dimensional PIC simulations and has been shown to be present in the foot region. Since we now know that it is the Buneman mode, it is no surprise to find it in the early stage also here in two-dimensions, when the conditions are favourable for the Buneman mode to be excited and the initial situation is still one-dimensional. It is, however, important to note that the instability is excited by the Buneman two-stream mechanism resulting from the large difference in bulk speeds between electrons and reflected ions. In the later stages, as the existence of the electromagnetic left-hand polarised negative helicity waves seen in Figure\,\ref{chap4-fig-schoma04-2} confirm, the ECDI/Buneman mode\index{instability!Buneman} ceases to exist and is replaced by the MTS-instability which generates oblique, nearly perpendicularly propagating large amplitude electromagnetic waves which also form hole structures and heat electrons and ions.
\begin{figure}[t!]
\hspace{0.0cm}\centerline{\includegraphics[width=0.95\textwidth,clip=]{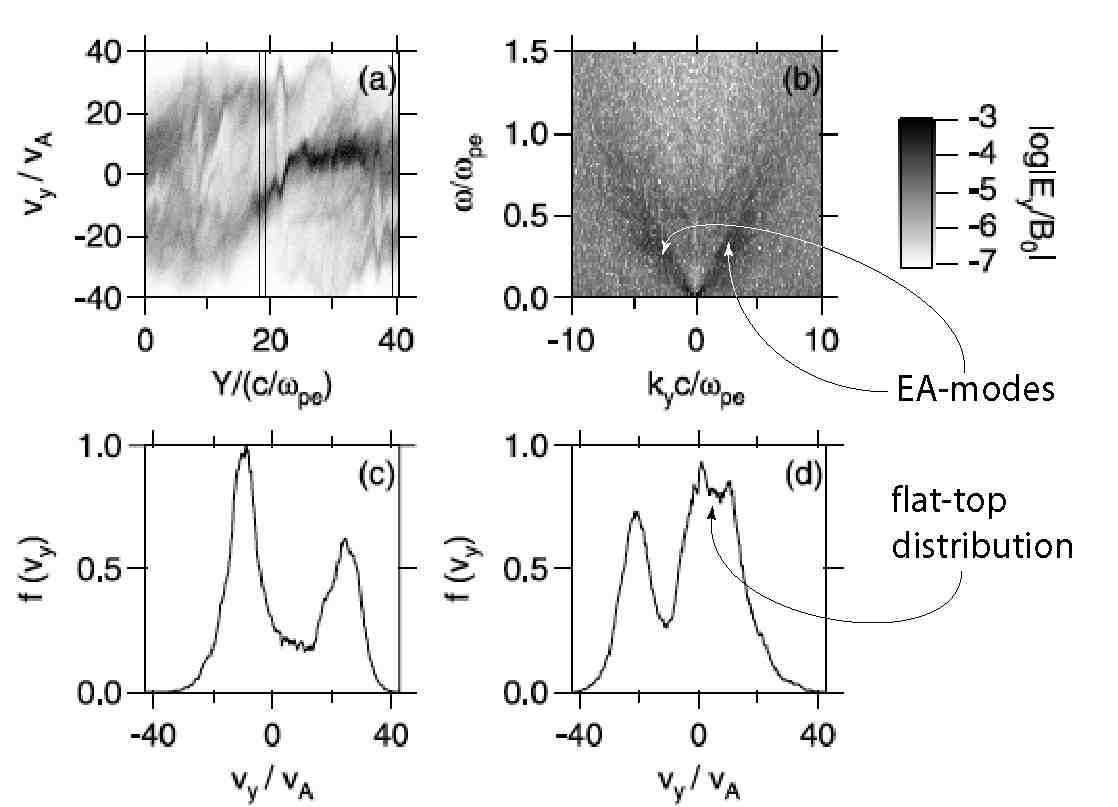} }
\caption[Density and field profiles from MTSI]
{\footnotesize {\it Top}: Electron phase space plot (left) at $19.4<x/\lambda_e<21$, and (right) dispersion relation $\omega(k_y)$ for the period $910.7<t\omega_{pe}<1315.5$ as obtained from $E_y$. This dispersion relation shows the occurrence of EA-waves with strictly linear dispersion and frequency below $\omega_{ea}\lesssim\omega_{pe}$ while propagating in both directions. These are generated in presence of the two-electron component structure seen in the distribution function below. They are responsible for the subtle fine-structuring of the electron distribution in the phase space as is seen in the upper left panel representation of $v_y$ versus $y$ which exhibits trapping and scattering of electrons on very small scales. {\it Bottom}:  Two electron distribution functions at $17.6<y/\lambda_e<19$ and $39.1<y/\lambda_e<40.5$ at $t=1000$, as indicated in the upper left panel by the vertical lines, showing the large electron hole distribution that is generated by the MTSI and some smaller substructures.  \citep[after][]{Matsukiyo2006a}. }\label{chap4-fig-matsu-3}
\end{figure}

Figure\,\ref{chap4-fig-matsu-2} shows the next time slot in the presentation of the dispersion relation(left).  During this time the dispersion relation becomes considerably more complicated than before because the waves have evolved to large amplitudes, large enough to cause various interactions among the waves which react on the wave and particle distributions and, in addition, because of the nonlinearity of the plasma state when the wave energy saturates. 

Firstly, the ECDI is clearly seen as a broad spot now smeared out over a considerable frequency-wavenumber domain. The MTSI is the nearly straigh short line at low frequencies and small positive $k_x$ (indicated as MTSI-1 in the figure). Remember that in the wave spectrogram these waves moved towards the shock ramp. This means their slope is positive in the dispersiogram! 

\paragraph{Secondary modified two-stream instability: Wave-wave interaction.} In additon to these modes another negatively moving low frequency wave appears. This is also an MTSI, but it is a secondary one, which \cite{Matsukiyo2006a} have shown to arises in a three wave process when the ECDI-BI and the MTSI-1 interact causing a wave with  $k_x=k_{\rm BI}+k_{\rm MTSI-1}=k_{\rm MTSI-2}$. The right part of the figure shows the enhanced wave power for this process extracted from the data on the way of a bi-spectral analysis and represented in the $k$-plane. The ellipse encircles the wavenumbers which are involved into the three-wave interaction, the original ECD-wave, and the resulting MTS-2-wave. Clearly, a whole range of waves participates in the interaction because the ECD-spectrum has broadened when saturating, and many combinations of ECD and MTS-1 waves satisfy the nonlinear three-wave interaction condition. 

The top-left panel of Figure\,\ref{chap4-fig-matsu-3} shows the evolution of the electron velocity $v_y$ during this interval and averaged over a range of $x$-values along the normal. This velocity is about perpendicular to the magnetic field; its dynamic range of variation is impressive. The panel at the lower left shows the electron phase space distribution. Two electron beams are seen to propagate at counter streaming velocities. These beams can already be identified in the upper panel. Due to the interaction with the unstable waves the region between the beams is partially filled. These distributions have been taken in the interval $17.6<y/\lambda_e<19$ as indicated in the top panel. 
\begin{figure}[t!]
\hspace{0.0cm}\centerline{\includegraphics[width=0.75\textwidth,clip=]{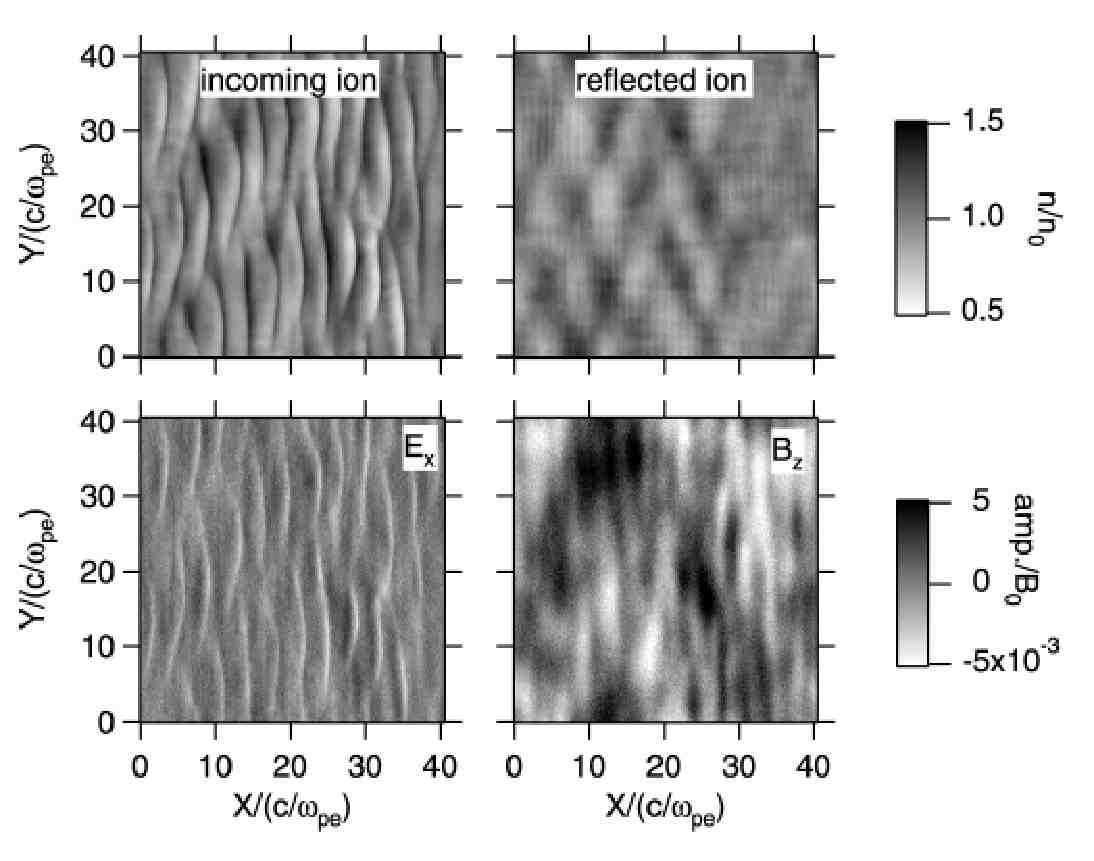} }
\caption[Density and field profiles from MTSI]
{\footnotesize {\it Top}: Incoming (left)  and reflected (right) ion densities at the late time  $t\omega_{pe}=2023.9$ shown in the ($x,y$)-plane. This time, in terms of the ion cyclotron period, corresponds to $t\omega_{ci}=0.55$, i.e. about half an ion-cyclotron period. At later times the ion magnetisation would come into play as well.  {\it Bottom}:  The corresponding electric field $E_x$ and magnetic field$B_z$ profiles.  One observes that the ECD-waves (Bernstein modes) have decayed by feeding their energy into electron heating. The two MTS-modes are still visible as the wavy variations in the incoming and reflected ion beams. The original MTS-1 wave modulates the incoming beam, which is seen in its downward propagation towards the shock. The secondary MTS-2 mode modulates, in addition, the reflected ion beam causing the interference pattern seen in the reflected beam density. The electric field is modulated by the MTS-2 wave, while the magnetic field contains signatures of both, MTS-1 and MTS-2 \citep[after][]{Matsukiyo2006a}. }\label{chap4-fig-matsu-4}
\end{figure}

Another distribution a little further in the interval $39.1<y/\lambda_e<40,5$, at the rear end of the top panel, is shown in the lower right panel. Here the distribution has evolved into a totally different combination of two electron populations, one top-flat and hot, the other one narrow, i.e. cold, but of same height indicating the retardation of one beam and heating of the other. Altogether the electron plasma has been heated to high temperature. Returning to the  upper left panel the complicated structure of the distribution is nicely seen with several sub-beams evolving and also with electron trapping in some vortices being visible for instance in the upper part around $y\simeq 22\lambda_e$. 

Coming now to the upper right panel, which shows the dispersion relation in the $y$-direction, one recognises two low-frequency linear wave modes propagating in positive and negative $y$-directions. These waves are electron-acoustic (EA) modes which are excited in the presence of the two electron distributions, the hot top-flat distribution and the cold beam distribution. They have strictly linear dispersion and frequencies below the electron plasma frequency $\omega_{ea}\lesssim\omega_{pe}$. Because they interact strongly with the electron distribution, they are responsible for the fine-structuring in the electron distribution function seen in the top-left panel of Figure\,\ref{chap4-fig-matsu-3} where they cause electron trapping and scattering  which results in electron heating and electron acceleration.

We close this section by presenting ion densities of the reflected and incoming beams and the corresponding modulations of the electric  $E_x$ and magnetic variation fields $B_z$, respectively, in Figure\,\ref{chap4-fig-matsu-4} in two-dimensional grey-scale representation in the $(x,y)$-plane. One observes that the ECD-waves (Bernstein and Buneman-modes) have decayed. They have been feeding their energy into electron heating, creating electron holes, trapping electrons and shaking them, as we have discussed above. The two MTS-modes are still visible. They dominate the ion density structure being visible as the wavy variations in the incoming and reflected ion beams. The original MTS-1 wave only  modulates the incoming beam. This is recognised from its long-wavelength downward propagation towards the shock. The secondary shorter scale MTS-2 mode modulates, in addition, the reflected ion beam causing the interference pattern seen in the reflected ion-beam density. The electric field is merely  modulated by the short wavelength MTS-2 wave. 

\paragraph{Weibel instability caused effects.} On the other hand, the magnetic field contains signatures of both, MTS-1 and MTS-2 thus exhibiting a more irregular structure than the electric field. Here, probably, also the small-scale structures of the Weibel instability do contribute. We have noted its effect already above,  but it would be rather difficult to extract them from the figure as they should appear as stationary vortices, which are convected downstream towards the shock front with the speed of the average bulk  flow. Their dynamics remains to be unresolved, i.e. it is not clear what will happen to them when encountering the shock front.  One possibility would be that they accumulate there and generate a non-coplanar magnetic  component. Nevertheless, the possibility for the Weibel instability to evolve in supercritical quasi-perpendicular shocks is of interest as Weibel vortices could, if confirmed by observations, cause an irregular fine structuring of the magnetic field in the shock ramp transition,  which would have consequences for the particle dynamics, trapping, scattering, reflecting and acceleration of particles from the shock front. It could, moreover, also lead to small scale reconnection in the shock front,  which so far has not been believed to exist in the shock, including the various side-effects of reconnection. Weibel vortices could also pass into the downstream region where they might contribute to the downstream magnetic turbulence where they would occur as magnetic nulls or holes for which the shock would be the source\index{turbulence!magnetic}. 

\section{The Problem of Stationarity}\noindent\index{shocks!non-stationary}
In this last section of the present chapter we will be dealing with the time-dependence of quasi-perpendicular shocks. Since in the previous sections we have frequently dealt with time variations, there is little new about time-dependences of shocks. Nevertheless, in the past few years the so-called problem of shock non-stationarity has brown into an own field of shock research. So what does it mean that shocks can be or even are non-stationary? One kind of non-stationarity is shock reformation. This is a periodic or better quasi-periodical process in which the shock ramp for the time of foot formation remains about stationary, i.e. the shock ramp moves only very slowly ahead. In fact, during reformation the shock everything else but stationary: from one reformation cycle to the next the ramp flattens and broadens while the new shock foot grows and steepens. And towards the end of the reformation cycle the shock ramp suddenly jumps ahead from its old position to the edge position of the shock foot. Could  one follow this evolution over very many reformation cycles\index{shocks!reformation}, one would find that there is no real periodicity but that the process of reformation, i.e. the time sequence of final forward jumps of the shock ramp would form a quasi-periodic or even chaotic time series. unfortunately, computer capacities do not yet allow to simulate more than a few reformation cycles such that this conjecture cannot be proved yet. But it is simple logic that reformation cannot be strictly cyclic; there are too many processes involved into it, too many instabilities cooperate, and the particle dynamics is too complicated for a strictly periodic process to be maintained over longer times than one or two ion-gyroperiods. In addition, once the shock is considered two-dimensional -- or even three-dimensional -- the additional degrees of freedom introduced by the higher dimensions multiply and the probability for the shock of becoming a stationary or even cyclic entity decreases rapidly. This is particularly true for high-Mach number shocks even under non-relativistic conditions. Hence, we may expect that a realistic high-Mach number, i.e. supercritical shock will necessarily be non-stationary. 

Principally, stationarity is a question of scales. On the macroscopic scale, the scale of the macroscopic obstacle and the macroscopic flow a shock will be stationary as long as the flow and the obstacle are stationary. For instance, such stationary shocks are the planetary bow shocks that stand in front of the planetary magnetospheres or ionospheres. On the scales of the magnetospheric diameters their variation is of the same order as the variation of the solar wind or -- if the magnetospheres themselves behave dynamically -- the time and spatial scales of their variation is of the same order as the time and spatial scale of the magnetospheric variation, for instance the diurnal precession of the Earth's magnetic axis which causes a strictly periodic flapping of the magnetosphere and thus a strictly periodic variation of the position and shape of the Earth's bow shock. On spatial scales of the order of the ion gyroradius and temporal scales of the order of the ion gyroperiod there is little reason to believe in stationarity of a collisionless supercritical shock wave. The whole problem of stationarity reduces to the investigation of instabilities and their different spatial and temporal scales and ranges, their evolution, saturation, being the sources of wave-wave interactions and nonlinear wave-particle interactions and so on. These we have already discussed in the former section as far as the current state of the investigations do allow. What, thus, remains is to ask how a shock surface can become modulated in higher dimensions and what reasons can be given for such modulations.   
 
\subsection{Theoretical reasons for shocks being non-stationary}\noindent
That collisionless shock waves might exhibit a non-stationary behaviour was suggested early on from the first laboratory experiments on collisionless shocks \citep{Auer1962,Paul1967}.  \cite{Morse1972} were the first to definitively conclude from their one-dimensional shock full particle PIC simulations that collisionless shocks seem to be non-stationary on the scale of the ion gyroperiod. Afterwards, time variations in the behaviour and evolution of collisionless shocks have been recovered permanently in shock simulations \citep[e.g.,][]{Lembege1987a,Lembege1987b}.

This is no surprise as we have mentioned several times already. The {\it principal reason} is that shocks, and in particular supercritical shocks which are not balanced by collisional dissipation, are in thermal non-equilibrium and are thermodynamically not in balance. Hence, locally they are longing for any opportunity to escape this physically unpleasant situation in order to achieve balance and thermal equilibration. However, as simple as this reason might look, as difficult is it to find out what under certain given condition will actually happen and which way a shock will locally direct itself for a try to escape non-equilibrium and to achieve equilibrium. Even though when it is permanently driven by an unchangeable flow and a stationary obstacle it will chose any kind of irregularity, fluctuation or detuning to drive some kind of instability, cause dissipation; and when driving will become too hard in any sense, it will overturn and break and in this way it will maximise dissipation if it is not possible to achieve in any smoother way. 

Non-stationary behaviour of quasi-perpendicular shocks has been anticipated theoretically, following \cite{Sagdeev1966}, by \cite{Kennel1985} who noted the existence of the critical whistler Mach number ${\cal M}_{\rm wh}$, which we have discussed above in comparison to numerical simulations. \cite{Galeev1988} tried to give a theoretical account for reasons of the anticipated non-stationary character of supercritical shocks. They investigated the role of whistlers in the nonlinear domain at the ramp, finding that whistlers for flow speeds sufficiently above the Alfv\'en speed do not possess soliton solutions and thus do not sustain the steady state of a shock. This means very simply that neither dissipation nor dispersion can sustain the nonlinear steepening of the waves, and therefore the waves should cause breaking of the flow and lead to non-stationary behaviour of the ramp and crest. this process is called by them `gradient catastrophe'. These authors also dealt with quasi-electrostatic waves of frequencies close to the lower-hybrid frequency $\omega_{lh}$ to which they also attributed responsibility for wave breaking. 

Simulations by \cite{Quest1986}, \cite{Lembege1992}, \cite{Savoini1994} and \cite{Hada2003} for low mass-ratios have attempted to illuminate some aspects of this non-stationary behaviour. \cite{Lowe2003}  and \cite{Burgess2006} have investigated two-dimensional rippling of the shock surface in hybrid simulations and its consequences. Full particle simulations up to realistic mass ratios have been performed by \cite{Scholer2004}, \cite{Matsukiyo2006a,Matsukiyo2006b} and \cite{Scholer2007}. We will return to these attempts. Here we first follow the analytical and simulational attempts of \cite{Krasnoselskikh2002} to advertise the general non-stationarity of quasi-perpendicular shocks. We should, however, note thta in principle there is not reason for a shock to behave like we wish, i.e. to behave stationary. It might, if necessary, break and overturn or mike not; the only requirement being that it follows the laws of physics.

\subsubsection{Nonlinear whistler mediated non-stationarity}\noindent\index{waves!whistler}
\cite{Krasnoselskikh2002} rely on a method developed by \cite{Whitham1974} to describe the nonlinear breaking of simple waves by adding to the simple wave evolution equation a nonlocal term that takes care of the accumulating short wavelength waves. The Whitham equation reads
\begin{equation}
\frac{\partial v}{\partial t}+v\nabla_x v +\int_{-\infty}^\infty\!\! {\rm d}\xi \,K(x-\xi)\nabla_\xi v(t,\xi)=0, \qquad K(x)=\frac{1}{2\pi}\int_{-\infty}^\infty\!\! {\rm d}k\frac{\omega(k)}{k}{\rm e}^{ikx}
\end{equation}
If this new term is purely dispersive, it reproduces the Korteweg-de Vries equation, if it is dissipative it reproduces Burgers' equation. In general, stationary solutions $\partial v/\partial t \to 0$ peak for $K(x)\sim |x|^{-\alpha}$ for $x\to 0$, and $\alpha>0$. \cite{Krasnoselskikh2002} use a dispersion relation
\begin{equation}
\frac{\omega}{kV_A}=\left(\frac{m_i}{m_e}\right)^{\!\!\frac{1}{2}}\frac{|\cos\theta|\,k\lambda_e}{1+k^2\lambda_e^2}, \qquad {\rm for}\quad k\lambda_i\gg 1,~~ \cos^2\theta\gg \frac{m_e}{m_i}
\end{equation}
describing low frequency whistlers at oblique propagation which, when inserted into the above integral for $K(x)$ asymptotically for $|x|\to 0$ yields $K(x)\sim\pi^{-1}\sqrt{m_i/m_e}|\cos\theta|[{\textsf C}+\ln|x| +\dots]$. Here ${\textsf C}= 0.577\dots$ is Euler's constant. Since $|x|^\alpha\ln|x|\to 0$ for all positive $\alpha>0$ and $|x|\to 0$, nonlinear low-frequency whistler waves will necessarily break by the above condition. Thus, when whistlers are involved into shock steepening, and when $\alpha>0$, they will necessarily break as their dispersion does not balance the nonlinear steepening. This happens when the Mach number exceeds the nonlinear whistler Mach number ${\cal M >M}_{\rm wh,nl}$. Both, whistler dispersion and dissipation by reflected ions cannot stop the whistlers from growing and steepening anymore, then. 

In order to prevent breaking, another mechanism of dissipation is required. Still based on the whistler assumption, \cite{Krasnoselskikh2002} argue that the shock ramp would radiate small wavelength whistler trains upstream as a new dissipation mechanism. This works, however, only as long as the Mach number remains to be smaller than another critical Mach number ${\cal M}<{\cal M}_{\rm wh, g}$ that is based on the whistler group velocity, $\partial \omega/\partial k$, 
\begin{equation}
{\cal M}_{\rm wh, g}=\left(\frac{27\, m_i}{64\, m_e}\right)^{\!\!\frac{1}{2}}| \cos\thetabn |\lesssim 19.8,
\end{equation}
since for larger ${\cal M}$ the whistler-wave energy will be confined to the shock and cannot propagate upstream. The right-hand estimate holds for an electron-proton plasma and $\cos\thetabn\sim 0.707$, i.e. at the largest shock-normal angle $\thetabn=45^\circ$ of quasi-perpendicular shocks.   \index{Mach number!whistler group}

In other words, since the nonlinear whistler Mach number is always larger than the whistler-group Mach number,  whistler energy will leave the shock upstream only in a narrow Mach number range ${\cal M}_{\rm wh}<{\cal M}<{\cal M}_{\rm wh, g}<{\cal M}_{\rm wh, nl}$, corresponding to $15<{\cal M}/|\cos\thetabn|<19.8<21.3$.  Corresponding one-dimensional full particle PIC simulations with realistic mass ratio $m_i/m_e=1840$ have been performed by \cite{Scholer2004} and \cite{Scholer2007} confirming that whistlers affect the stationary or non-stationary behaviour of nearly perpendicular shocks. 

At larger ${\cal M}>19.8|\cos\thetabn|$ the whistler energy is again confined to the shock and will  be swept downstream towards the shock when transported by the passing though continuously retarded flow. In the region of the foot and ramp where the energy accumulates it will cause different instabilities  some of which propagate downstream. Such processes can be wave-wave interactions driven by the high whistler energy, as has already originally been envisaged by \cite{Sagdeev1966}, or nonlinear wave-particle interactions. In addition to causing anomalous resistivity and anomalous dissipation, these processes should lead to emission of plasma waves from the shock, preferably into the direction downstream of the shock, as only there ${\cal M}\lesssim 1$, and the wave group and phase velocities can exceed the speed of the flow.

This whole discussion refers only to whistler waves and follows the traditional route. We have, however, seen in the previous sections that the foot of the shock is capable of generating waves of another kind, electromagnetic Buneman modes, modified-two-stream waves, and possibly even Weibel modes. These waves are highly productive in generation of electron heating; they cause magnetic disturbances that move towards the shock or also upstream. The role in dissipation and dispersion these waves play has not yet been clarified and is subject to further investigation. It is, however, clear that their excitation and presence in the shock foot produces electron heating, retarding of ions and ion heating as well and will thus provide an efficient dissipation mechanism. Whether this can prevent shock breaking and overturning at very large Mach numbers is not known yet.
\begin{figure}[t!]
\hspace{0.0cm}\centerline{\includegraphics[width=0.75\textwidth,clip=]{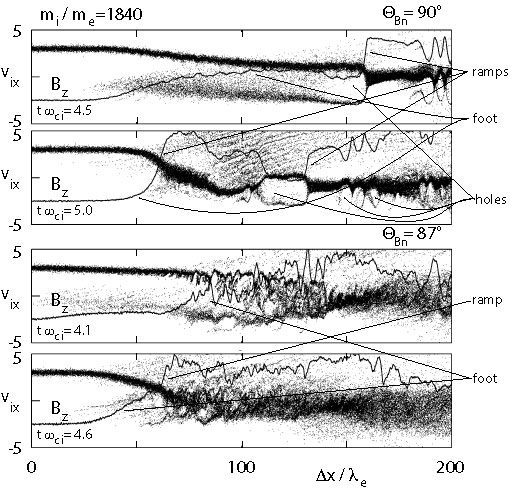} }
\caption[Phase space at 90 and 87 degrees.]
{\footnotesize The completely different reformation behaviour of shocks in one-dimensional PIC simulations with realistic mass ratio of 1840 for strictly perpendicular and oblique quasi-perpendicular shocks at exactly same parameter settings and scales. Shown is the magnetic field $B_z$ and the ion phase space at two subsequent  simulation times for each of the respective simulations. Since the evolution is different in both cases the $x$-coordinate is given as a relative scale not in $x$ but for the same interval lengths in $\Delta x$ for the instance when reformation takes place in both cases \citep[compiled from][]{Matsukiyo2006a}.  {\it Top}: Reformation at $\thetabn=90^\circ$ at two times showing the evolution of the foot in the magnetic field and the taking-over of the ramp by the foot while a new foot evolves. This process is governed by the Buneman two-stream instability. Large holes evolve on the ion distribution. Note the correlation of the ion holes with depressions in the magnetic field, In the second panel the old ramp is still visible as the boundary of the large ion hole. Farther downstream many holes are seen, each of them corresponding to a magnetic depression, and the regions between characterised by magnetic overshoots.  {\it Bottom}:  The corresponding evolution at $\thetabn=87^\circ$. High variability of the shock profile is observed which is identified as being due to the large amplitude MTS-waves travelling into the shock. The foot region is extended and very noisy both in the magnetic field and ion distributions, the latter being highly structured. The foot is extended much longer than in the perpendicular case. The two bottom panels might also show signatures of wave breaking in the ion velocities when groups of ions appear which overturn the main flow in forward downstream direction.}\label{chap4-fig-matsu-5}
\end{figure}
\subsubsection{Shock variability as a consequence of two-stream and modified two-stream waves}
\noindent Variability of the quasi-perpendicular shock has been demonstrated from numerical full particle PIC simulations in one and two dimensions to come about quite naturally for a wide range of -- sufficiently large -- Mach numbers. While in the low Mach number range whistlers are involved in the variability, reformation and non-stationarity, the simulations have clearly demonstrated that at higher Mach numbers the responsible waves are the Buneman two-stream mode and the modified two-stream instability. This has been checked \citep[cf.,][]{Matsukiyo2006a,Matsukiyo2006b} by shock-independent simulations where the typical electron and ion phase space distributions have been used which occur in the vicinity of supercritical shocks during particle reflection events. So far the importance of these waves over whistlers has been investigated only for the shock-foot region. The shock ramp and overshoot are more difficult to model because of the presence of density and field gradients, their electrical non-neutrality, and the fuzziness of the particle phase-space distribution functions.

Differences were also  found between strictly perpendicular and quasi-perpendicular shocks. The former are much stronger subject to the Buneman two-stream instability that completely rules the reformation process in one and two dimensions in this case, causing phase-space holes to evolve and being responsible for quasi-periodic changes in the positions, heights and widths of the shock front and foot regions, respectively. We have already put forward arguments that an investigation of the long-term behaviour of this quasi-periodic variation should reveal that this process is irregular in a statistical sense. Even under apparently periodic reformation conditions the shock will presumably not behave stationary on the short time and spatial scales. This, however, can be checked only with the help of long-term simulations which so far are inhibited if done with sufficiently many macro-particles, realistic mass ratios and in more than one-dimension.

Figure\,\ref{chap4-fig-matsu-5} provides an impression of the variability of shock reformation in the two cases of a strictly perpendicular, highly supercritical shock, and the case of an oblique supercritcial shock at $\thetabn=87^\circ$, when whistler excitation is absent. The settings of the simulations are otherwise identical, but the evoultion of the two simulations is compleetely unrelated. This is because the perpendicular shock does not allow, in these one-dimensional simulations, for the modified two-stream instability to grow. So only the Buneman two-stream instability grows. It reforms the foot in the way we have already described, forming large holes and letting the shock ramp jump ahead in time-steps of the order of roughly an ion gyro-period. The shock foot acts decelerating on the flow, and already during reformation begins to reflect ions and to form a new foot. Most interestingly is that the holes survive quite a while downstream while being all the time related to magnetic depressions. At their boundaries large magnetic walls form which can be interpreted as magnetic compressions (or otherwise signatures of current vortices).

The oblique case looks different. It is highly variable both in time and in space. The magnetic profile is more irregular, and the ion-phase space exhibits much more structure than in the perpendicular case. This has been identified to be due to the combined action of the Buneman two-stream and the modified two-stream instabilities with the two-stream instability being important only during the initial state of the reformation process, while the modified two-stream instability dominates the later nonlinear evolution. Both, foot and ramp, are extended and vary strongly. It is quite obvious, that in this case one can speak of a stationary shock front only when referring to the long-term behaviour of the shock, much longer than the irregular reformation cycle lasts. At the scale of reformation and below there is no stationarity but variability and evolution, which can be attributed to the growth and interaction times of the MTSI and the various secondary processes caused by it.

To complete this section, we note in passing that the low-mass ratio two-dimensional full particle PIC simulations with small particle numbers performed by \cite{Lembege1992,Lembege2002} and \cite{Savoini1994,Savoini1999} also showed non-stationary behaviour of the quasi-perpendicular -- or perpendicular -- shock leading to so-called ``rippling" of shocks, which we will briefly describe in the next paragraph.

\subsection{Formation of ripples}\noindent\index{shocks!ripples}
One-dimensional theory and one-dimensional simulations implicitly treat the shock as an infinitely extended plane surface. In addition they allow only for instabilities to evolve in the direction of the shock normal at an angle relatively close to $90^\circ$ such that any waves along the shock surface are completely excluded and waves parallel to the magnetic field have very small wave numbers $k_\|=k_x\cos\thetabn\ll k_x$ corresponding to very long parallel wavelengths. To be more realistic, two-dimensional PIC simulations have been performed to investigate the effect of the additional freedom given by the second spatial dimension which allows instabilities to evolve in other than the shock normal direction. The cost of these simulations is being restricted to low mass ratios only. In the simulation of \cite{Lembege1992,Lembege2002} and \cite{Savoini1994,Savoini1999} the mass ratio has been taken as $m_i/m_e=42$ which implies from comparison to the high-mass ratios in one-dimensional simulations that the modified two-stream instability will presumably be excluded. The structure of the shock front in these simulations has been shown in Figure\,\ref{chap4-fig-lemb1}, the left-hand side of which shows the cyclic reformation of the shock -- which at these low mass ratios is clearly expected to occur -- at a period comparable to the ion-gyro period of the reflected ions in the foot of the shock. The right-hand side shows a pseudo-threedimensional profile of the shock in the two spacial dimensions, in the top part at the tie when the foot is fully developed, in the bottom part when the ramp has just reformed, i.e. when the foot has taken over to become the ramp. 
\begin{figure}[t!]
\hspace{0.0cm}\centerline{\includegraphics[width=1.0\textwidth,clip=]{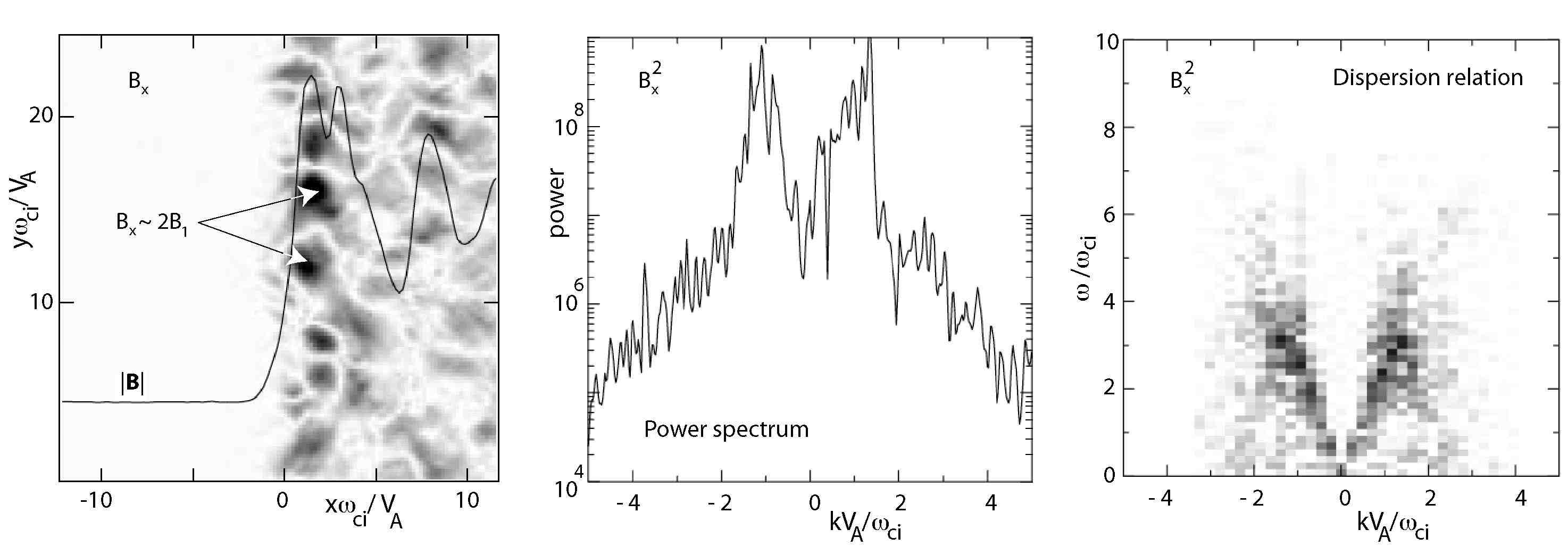} }
\caption[Rippling fluctuations]
{\footnotesize {\it Left}: Spatial distribution of the $B_n=B_x$-component of the magnetic field in the hybrid simulations of \cite{Lowe2003} with Mach number ${\cal M}_A=5.7$, and for $\thetabn=88^\circ$. The $B_x$-component is not zero; it reaches values twice the upstream magnetic field $B_1$ and shows quite structured behaviour along the shock surface which indicates that the surface is oscillating back and forth and that waves are running along the surface. These waves are interpreted as surface waves.   {\it Centre}:  The power in the presumable surface waves as determined from the simulations. Obviously the power concentrates around the ramp. {\it Right}: Apparent dispersion relation $\omega(k)$  of the fluctuations \citep[after][]{Lowe2003}. }\label{chap4-fig-lowe}
\end{figure}

We have already discussed this paper in connection with the reformation problem. What interests us here is that the shock ramp surface is by no means a smooth plane in the direction tangential to the shock. It exhibits large variations both in space and time which are correlated but not directly in phase with the presence of reflected ions in the foot. The overshoot, steepness and width of the ramp and ramp position oscillate at a not strictly periodical time-scale. In addition, the structure of the ramp also exhibits shorter scale fluctuations. 

These so-called ripples on the shock front \citep{Lembege2002} have been further investigated by \cite{Lowe2003}, and \cite{Burgess2006a, Burgess2006b} and \cite{Burgess2007} using hybrid-simulations in two dimensions. The simulations by \cite{Lowe2003} use $\thetabn=88^\circ$ for the quasi-perpendicular shock, and ${\cal M}_A=5.7$. since this hybrid-set-up satisfies the Rankine-Hugoniot conditions with ${\bf n\cdot B}=B_n$ constant, fluctuations in $B_n$ are attributed to rippling on the shock surface. Figure\,\ref{chap4-fig-lowe} shows the two-dimensional distribution of the component $B_n=B_x$ (left), the power $B_x^2$ (centre), and the `dispersion relation' $\omega(k)$ of the fluctuations in $B_x$ determined from the simulations. The fluctuations in the normal component of the magnetic field are far from being negligible; in fact, they are of the same value as the main component of the magnetic field $B_z$ reaching maximum values of twice the upstream magnetic field $B_1$. They are concentrated in the ramp, foot and overshoot. The dispersion relation is about linear and low frequency but exceeds the ion-cyclotron frequency for shorter wavelengths. there is no mode known which corresponds to these waves, so they are attributed to surface waves flowing in the shock transition. Maximum wavelengths are a few ion inertial lengths. The lesson learned is, however, quite simple: the shock exhibits structure along its surface which can presumably be attributed to waves running along the shock front and modulating it temporarily and spatially. The caveat of these simulations is however, their hybrid character which does not account for the full dynamics of the particles and therefore it cannot be concluded about the nature of the waves. 
\begin{figure}[t!]
\hspace{0.0cm}\centerline{\includegraphics[width=0.7\textwidth,height=0.3\textheight,clip=]{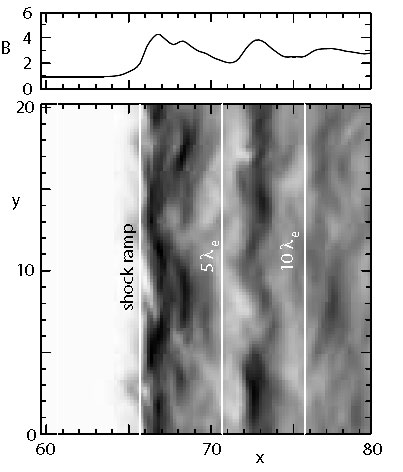} }
\caption[Rippling fluctuations 1]
{\footnotesize Two-dimensional structure of the surface waves \citep{Burgess2007}. {\it Top}: Magnetic field average across the shock.  {\it Bottom}: Grey scale plot of the surface waves. The three white lines show the presumable location of the nominal shock and two distances from it downstream \citep[after][]{Burgess2006b}. }\label{chap4-fig-lowe1}
\end{figure}

\cite{Burgess2007} have extended these investigations to infer about the driver of these waves. They find that it is the reflected ion component in the shock foot which flows along the shock surface and at large Mach numbers becomes unstable. Figure\,\ref{chap4-fig-lowe1} shows a grey scale plot of the two-dimensional structure of the surface wave oscillation. Its growth is proportional to the Mach number, i.e. it must therefore be proportional to the number of reflected ions, their velocity and to the upstream convection electric field that accelerates the ions. Presumably it is some variant of Kelvin-Helmholtz instability along the shock surface, which is driven by the reflected ion flow along the shock surface causing undulations or vortices at the shock and which, in the magnetic field, appear as ripples.\index{shocks!surface waves}\index{instability!Kelvin-Helmholtz}

It should be clear, however, that a long-term full particle simulation must be performed at real mass ratio before any reliable conclusion can be drawn about the existence of surface waves. We have seen that part of these waves is nothing but the exchange between the foot and the ramp during the reformation process. This applies to the long wave part. In addition is is indeed possible that the strong and fast flow of reflected ions along the shock surface can generate a Kelvin-Helmholtz instability of the shock front. However, it is not clear whether these oscillations are the sole action of the modulation of the shock surface in two or three dimensions. The only conclusion we can safely draw is that the shock surface even under ideal non-curved and quiet upstream conditions at high Mach numbers will not remain to be a quiet stable shock surface but will exhibit fluctuations in position, structure, overshoot amplitude and width on the scales of the ion inertial length and the ion cyclotron period.

\section{Summary and Conclusions}\noindent
Quasi-perpendicular supercritical shocks are among the best investigated collisionless shocks. Theory has predicted that they reflect ions, form feet and ramps and possess either whistler precursors are trails. Theory also predicted that whistlers could be phase-locked and stand in front of the shock ramp only for a limited range of Alfv\'enic Mach numbers. We have reviewed here the theoretically expected shock structure, the relevant scales, the most relevant particle simulations for perpendicular and quasi-perpendicular supercritical shocks, the shock-reformation process and its physics as far as it could be elucidated from one-dimensional and to a certain part also from two-dimensional simulations. The most relevant instabilities generated in the shock foot have been identified as the whistler instability for nearly perpendicular supercritical but low-Mach number shocks, leading to foot formation but not being decisive for feet, as it has turned out that feet in this Mach number and schock-normal angle ranges are produced by accumulation of gyrating ions at the upward edge of the feet. 

More important than whistlers have turned out the Buneman and modified two-stream modes, the former dominating shocks at perpendicular angles, the latter growing slowly but dominating at more oblique angles and at later simulation times with the effect of completely restructuring both the shock feet and ramps. Both instabilities generate phase space holes which during reformation survive and are added to the downstream plasma and, in addition, being responsible for low magnetic field values. 

Most interestingly, the plasma state just downstream of the shock is nothing else but the collection of the old shock ramps which have been left over from former reformation cycles and move relative to the shock frame in the direction downstream of the shock. 

The modified-two-stream instability in addition generates waves which flow into the shock ramp where they contribute to the dynamics of the ramp. Wave-wave interaction and wave particle interaction lead to the generation of secondary waves and to particle heating. Finally, simulations show that the shock front in more than one dimension is not a plane surface but exhibits a strong variability in time and space. This can be explained as surface waves of the shock front which might be driven by the reflected ion current flow along the surface similar to a Kelvin-Helmholtz instability. This question is still open to investigation. 
\begin{figure}[t!]
\hspace{0.0cm}\centerline{\includegraphics[width=1.0\textwidth,clip=]{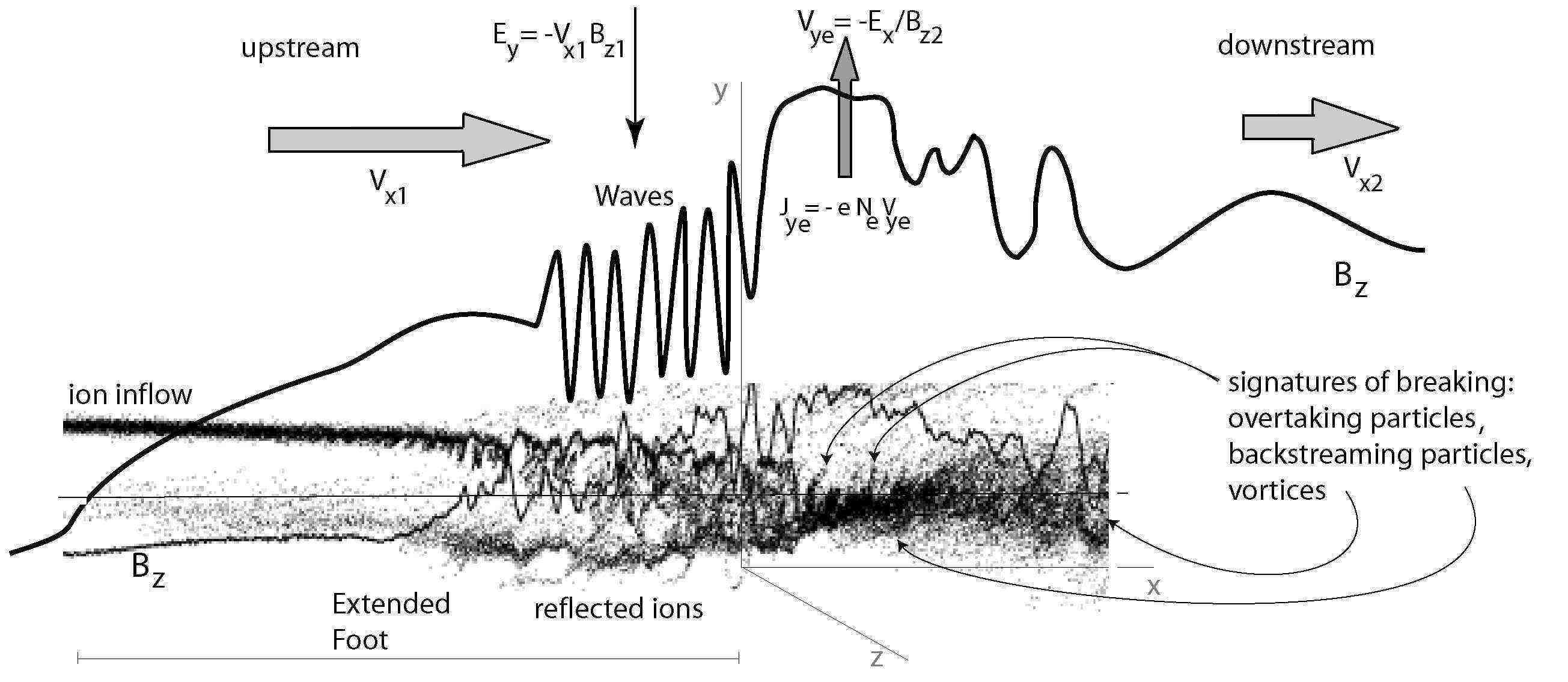} }
\caption[Sketch of shock dynamics with breaking]
{\footnotesize Schematic of the profile of a highly supercritical shock with waves just before shock reformation and signatures of beginning wave breaking. The sketch has been completed with a copy of the ion phase space from the simulations of \cite{Matsukiyo2006b} showing the structure of the ions in the ramp with the signatures of overtaking ions and backstreaming ions as well as ion vortices, all an indication of onset of breaking. }\label{chap4-fig-final}
\end{figure}
So far the evolution of the shock ramp, its stability and time variation as well as the physics of the region just downstream of the ramp is not yet well explored. It is, however, clear from the available intelligence that any serious investigation must be based on full particle simulations and experimenting with appropriate sets of distribution functions suggested by the simulations in order to investigate the instabilities and interactions between the waves and particles as well as between waves and waves in order to understand the physics. This has to a certain degree already been achieved for the foot region. In the shock ramp and in the strongly disturbed region behind the ramp it is more difficult as the conditions there are less clean and the definition of the responsible distributions is more difficult. Moreover, plasma and field gradient must be taken into account in this region, and the electric charge separation field that is partially responsible for ion reflection cannot be neglected as well.  With the further increase of computing capacity and the refinement of the models one expects that within the next decade also the physics of the shock ramp will become more transparent.

One particularly interesting question is related to the stability of shocks\index{shocks!stability}. In a certain range of small Mach numbers shocks can be balanced by generation of anomalous resistance. However, above a critical Mach number they become supercritical and reflect ions in order to generate an ion viscosity that helps dissipating the excess energy. This dissipation goes via the above mentioned instabilities and less on the way of ion viscosity in the classical sense of fluid theory. However, for even larger Mach numbers these processes will also not suffice to stabilise a quasi-perpendicular shock. This poses the question what will then happen? It has been suggested that strongly nonlinear processes driven by whistlers will set on and lead to non-stationarity of the shock ramp. This might be the case. However, only simulations at high Mach numbers and full mass ratios in large enough systems can answer this question. 

We can, however, state that the problem of stationarity or cyclic behaviour of the shock is not the problem of the shock; rather it is the problem of our understanding. For the shock does nothing else count than dissipating in some way the excess inflow energy and momentum of the inflow. If it cannot do this by either anomalous dissipation, shock reflection, foot formation, precursors and early retarding the flow, by generating various instabilities, then it will not care but will break and turn over as this will be the only way for reducing the scale to the microscopic dimension and in this way produce violent heating and energy dissipation. 

At the time of writing it remains unclear whether and how such breaking\index{shocks!breaking} proceeds. In particular, in the magnetic field breaking-off of field lines is inhibited by Maxwell's equations. Magnetic field lines cannot break-off as this would require the existence of magnetic monopoles; magnetic field lines can kink but otherwise must always remain to be simply connected. Hence, any breaking that is going on can either take place only in the particle density or bulk velocity. Breaking is always connected with formation of vortices in the particle velocity and density. In this sense the appearance of phase space vortices at high Mach numbers resembles already a tendency for shock breaking. In the light of this discussion the lower two panels in Figure\,\ref{chap4-fig-matsu-5} can also be interpreted as breaking and overturning of the quasi-perpendicular ($\thetabn=87^\circ$) realistic mass-ratio supercritical shock. In particular during the phase before reformation (third panel from top) the magnetic field behaves very irregularly, and both the incoming and reflected beam form many partial vortices prior to the reflection point (at $\Delta x\sim 140\,\lambda_e$). Behind the reflection point the ion velocity shows formation of bursts of ions which run away in forward direction, which is just what we expect when breaking occurs. A sketch of the dynamical processes is shown in Figure\,\ref{chap4-fig-final}.

Breaking also requires rearranging of the magnetic field, i.e. high diffusion on short scales or reconnection, or it requires generation of small scale magnetic structures like magnetic bubbles. The former needs reconnection to take place in high Mach number shocks, the latter becomes possible if the Weibel instability is strong enough to generate many small-scale magnetic bubbles. 

As long as the shock can prevent breaking it will produce instabilities and fast particles, however for too large Mach numbers it will not care whether we understand its dynamics or not. It will confront us with the reality of its own choice and will let us wonder and argue what this reality is.  


\end{document}